\documentclass[a4paper,11pt]{article}

\usepackage{amsmath,amstext,amssymb}
\usepackage{bm}
\usepackage{cite}
\usepackage{psfrag,graphicx,subfigure,xcolor}
\usepackage[small,bf]{caption}
\usepackage{calc}

\newcommand{\N}{\mathcal{N}}
\newcommand{\wn}{\mathfrak{w}}
\newcommand{\mn}{\mathfrak{m}}

\newcommand{\R}{\mathfrak{R}}

\newcommand{\dd}{\mathrm{d}}
\renewcommand{\Re}{\mathop\mathrm{Re}}
\renewcommand{\Im}{\mathop\mathrm{Im}}

\newcommand{\eqlabel}[1]{\label{#1}}
\def\cale{{\cal E}}
\def\calf{{\cal F}}
\def\call{{\cal L}}
\def\calo{{\cal O}}
\def\caln{{\cal N}}
\def\cals{{\cal S}}
\def\calv{{\cal V}}
\def\calw{{\cal W}}
\def\ie{{i.\,e.\ }}
\def\eg{{e.\,g.\ }}
\def\dt{\tilde d}
\def\ft{\tilde f}
\def\Xt{\tilde{X}}
\def\mut{\tilde{\mu}}

\def\vrho{\varrho}

\def\del{\partial}
\def\diag{\mathrm{diag}}
\def\str{\mathrm{Str}}
\def\ee{{\mathrm e}}
\def\ii{{\mathrm i}}

\def\vol{\text{vol}}
\def\Z4{{\mathbb Z_4}}

\newcommand{\dslash}[1]{#1\hskip-7pt \diagup} 
\newcommand{\dSlash}[1]{#1\hskip-11pt \diagup} 

\numberwithin{equation}{section}

\title{\vspace{-3em}\parbox{\textwidth}{\small\hfill MPP--2008--65}\\[3em] 
	\bf Finite baryon and isospin chemical potential in AdS/CFT with
	flavor}
\author{Johanna Erdmenger\footnote{jke@mppmu.mpg.de},
	Matthias Kaminski\footnote{kaminski@mppmu.mpg.de},
	Patrick Kerner\footnote{pkerner@mppmu.mpg.de},
	Felix Rust\footnote{rust@mppmu.mpg.de}
	\\[\bigskipamount]
	\small %
	Max-Planck-Institut f\"ur Physik (Werner-Heisenberg-Institut)\\
	\small %
	F\"ohringer Ring 6, 80805 M\"unchen, Germany
}
\date{}

\begin{document}
\maketitle
\thispagestyle{empty}

\vfill

\begin{abstract}
	\noindent
	We investigate the thermodynamics of a thermal field theory in presence of
	both a baryon and an isospin chemical potential. For this we consider a
	probe of several D$7$-branes embedded in the AdS-Schwarzschild black hole
	background.
	
	We determine the structure of the phase diagram and calculate the relevant
	thermodynamical quantities both in the canonical and in the grand canonical
	ensemble. We discuss how accidental symmetries present reflect themselves in
	the phase diagram: In the case of two flavors, for small densities, there is
	a rotational symmetry in the plane spanned by the baryon and isospin density
	which breaks down at large densities, leaving a $\mathbb Z_4$ symmetry.
	
	Finally, we calculate vector mode spectral functions and determine their
	dependence on either the baryon or the isospin density. For large densities,
	a new excitation forms with energy below the known supersymmetric spectrum.
	Increasing the density further, this excitation becomes unstable. We
	speculate that this instability indicates a new phase of condensed mesons.
	
\end{abstract}

\vfill

\vfill

\vfill

\setcounter{tocdepth}{2}

\newpage
\tableofcontents

\clearpage

\section{Introduction}

Generalizations of the AdS/CFT correspondence
\cite{Maldacena:1997re,Aharony:1999ti} have proven to be very useful in
describing strongly coupled QCD-like theories. In particular, the
AdS-Schwarzschild black hole background provides a gravity dual for
four-dimensional $\mathcal{N}=4$ $SU(N)$ SYM theory at finite temperature
\cite{Witten:1998zw}. This is expected to be relevant for describing some
aspects of the strongly coupled quark-gluon plasma. Hydrodynamic methods based
on gauge/gravity duality have been applied very successfully to describing
transport processes \cite{Son:2007vk,Policastro:2001yc,Policastro:2002se,
Policastro:2002tn,Herzog:2002fn,Herzog:2003ke,Kovtun:2003wp,Buchel:2003tz,
Kovtun:2004de,Buchel:2004di,Benincasa:2005iv,Mas:2006dy,Son:2006em,Maeda:2006by,Herzog:2006se,CasalderreySolana:2006rq,Buchel:2008wy,Myers:2008yi}.

A further important ingredient for obtaining gravity duals of field theories
similar to QCD is the addition of flavor degrees of freedom via the embedding of
D$7$-brane probes \cite{Karch:2002sh}. A D$7$-brane probe embedded in the AdS
Schwarzschild background displays an interesting first-order phase transition
between embeddings which do not reach the black hole horizon and those who do
\cite{Babington:2003vm,Kirsch:2004km,Mateos:2007vn,Kruczenski:2003uq}. The first kind of
embeddings is often said to be a ``Minkowski embedding'' while the second is
called a ``black hole embedding''. The phase diagram is modified in presence of
a $U(1)$ baryon chemical potential given by a non-zero vev for the time
component of the gauge field on the brane \cite{Kobayashi:2006sb,Mateos:2007vc,Nakamura:2006xk,Nakamura:2007nx}. The effect of a finite baryon chemical potential is also studied in the $dS_4$ background, \eg\cite{Ghoroku:2007re}, and in the Sakai-Sugimoto model, \eg\cite{Kim:2006gp,Horigome:2006xu}.

The spectral functions for the D$7$ system, calculated according to the
holographic prescription of \cite{Son:2002sd,Herzog:2002pc}, also display a
rich structure, as investigated for instance in \cite{Hoyos:2006gb,Hoyos:2007zz,   
Amado:2007yr,Myers:2007we, Mateos:2007yp,Myers:2008cj,Mas:2008jz}. In particular, as shown in
\cite{Erdmenger:2007ja}, in presence of a baryon chemical potential, the vector
quasiparticle resonances approach the supersymmetric meson mass spectrum of
\cite{Kruczenski:2003be} for large values of the quark mass or small values of
the temperature. At small values of the quark mass or large values of the
temperature, the frequencies of the resonance peaks increase with rising
temperature. Moreover it was shown in \cite{Erdmenger:2007ja} that in presence
of an $SU(2)$ isospin chemical potential given by a non-zero vev for the time
component of the gauge field on a probe of two coincident D$7$-branes, the
vector spectral function displays a triplet splitting of the resonance peaks.

In the present paper we determine the structure of the phase diagram in presence
of both a $U(1)$ baryon and an $SU(2)$ isospin chemical potential, of the form
\begin{equation}
\label{eq:isoIntro}
	\mu =   \mu^B \left( \begin{array}{cc}1 & 0 \\ 0 &  1\end{array}\right)
	      + \mu^I \left( \begin{array}{cc}1 & 0 \\ 0 & -1\end{array}\right).
\end{equation}
We thus restrict ourselves to an isospin chemical potential in the $\sigma^3$
direction in flavor space. This restriction was suggested and used also in
\cite{Yamada:2006rx}. The values of $\mu^B$ and $\mu^I$ are given in terms of
fields which depend on the holographic radial direction. Some features of this
setup are generalized to a general small number $N_f\ll N_c$ of flavor probe
branes. For the discussion of the phase structure we restrict to $N_f=2$.

Using a suitable linear combination of the flavor components, the action
determining the brane embedding splits into a sum of Abelian contributions.
Moreover the choice of~\eqref{eq:isoIntro} leads to a number of accidental
symmetries which we list in detail. We discuss the case of arbitrary $N_f$ as
well as $N_f=2$. In addition to a reflection and a permutation symmetry,
there is also an approximate $O(N_f)$ symmetry if the baryonic and isospin
densities, parametrized by $\tilde d^B$ and $\tilde d^I$, are small.

We investigate the phase diagram and calculate thermodynamic quantities such as
the free energy (grand potential), entropy, energy and speed of sound in the
canonical (grand canonical) ensemble. Generically, a special case arises when
the normalized baryon and isospin densities coincide, $d^I=d^B$. The relation
$|d^I|=|d^B|$ determines fixed lines of the permutation and reflection symmetry
transformations.

In the canonical ensemble at vanishing isospin density, there is a phase
transition between black hole embeddings%
\footnote{Note that as discussed in
\cite{Mateos:2007vc}, this is really a transition between a black hole and a
mixed phase, since the lower energy black hole phase has been shown to be
unstable.}
below a critical baryon density \cite{Kobayashi:2006sb}. At both finite baryon
and isospin density, the according critical point becomes a critical line which
displays the approximate $O(2)$ symmetry up to a small derivation which we
quantify. We proceed by investigating a new first order phase transition in the
density plane in the limit of large quark mass over temperature ratio. The phase
diagram clearly displays the symmetry fixed lines $|d^I|=|d^B|$ at which the
chemical potentials develop a discontinuous gap, which marks the first order
phase transition. This is similar to what is observed in two-color QCD
\cite[fig.~1]{Splittorff:2000mm}. In the four different regions of the phase
diagram, the larger of the two densities measures the degrees of freedom in our
setup and contributes linearly to the free energy. The smaller density
contributes nonlinearly and may be interpreted as a electrostatic-like charge
induced by the non-Abelian charges of the quarks.

In the grand canonical ensemble, the sum $\mu^B+\mu^I$ is relevant. For
$\mu^B+\mu^I<M_q$ there is a black hole/Minkowski phase transition which is
absent for $\mu^B+\mu^I>M_q$, with $M_q$ the bare quark mass. As we discuss in
detail, a new phase becomes visible in the grand canonical phase diagram. This
is related to the fact that it is not possible to find a bijective map between
the canonical and the grand canonical ensemble in the limit of large mass over
temperature ratio and of chemical potentials of the order of the quark mass.
This implies an instability, \eg \cite[chap.~3]{0034-4885-50-7-001}.

This motivates us to calculate the spectral function for the gauge field
fluctuations in the canonical ensemble at either finite baryon or finite isospin
particle density. We identify resonance peaks of the spectral functions with
mesonic quasiparticles. For finite baryon density, it was shown in
\cite{Erdmenger:2007ja} that for $\mu^B\approx M_c$, with $M_c$ the renormalized
quark mass in the plasma, the spectrum of these particles approaches the
supersymmetric spectrum $M_n/(2\pi T)=m\sqrt{(n+1)(n+2)/2}$, with
$m=2M_q/(\sqrt{\lambda} T)$ found in \cite{Kruczenski:2003be}. The quasiparticle
resonances become very narrow in this case. Here, however, we investigate the
case of large baryon density and find that these resonances become wide, \ie
unstable, in this limit.

In the case of finite isospin density, we additionally find a new excitation.
This new resonance develops at an energy below $M=2\pi T m$. The new resonance
becomes the lowest vector mesonic excitation of the system. It becomes unstable
at a critical isospin chemical potential. We interpret this as vector meson
condensation, similar to \cite{Aharony:2007uu}. Moreover we discuss the
structure of the phase diagram in $(\mu^I,M_q/T)$ and identify the new phase. We
relate the appearance of this phase to the instability discussed above.

We also calculate the effective baryon diffusion coefficient $D$ within
hydrodynamics from the membrane paradigm. We find that $D$ displays the
hydrodynamic crossover reminiscent of the phase transition present at small
densities. The position of the crossover displays a $\mathbb Z_4$ symmetry in
the $(d^B,d^I)$-plane.

\medskip

The paper is organized as follows. In section~\ref{sec:holographicSetup} we
introduce the setup of probe D$7$-branes in presence of both a baryon and an
isospin chemical potential, paying particular attention to the accidental
symmetries. We obtain numerical results for the chemical potentials and the
condensate as functions of the quark mass. In section~\ref{sec:thermoCan} we
study the phase diagram and thermodynamic quantities such as free energy,
entropy and energy in the canonical ensemble. In section~\ref{grandcanonical} we
perform a similar analysis in the grand canonical ensemble, revealing the
possibility of an instability. The hydrodynamic diffusion constant is studied in
section~\ref{sec:hydrodynamics}.  In section~\ref{sec:mesons} we
study the spectral functions of mesonic quasiparticle resonances. For large
isospin chemical potentials, we find an instability and identify the
corresponding region in the phase diagram.

\section{Holographic setup} \label{sec:holographicSetup}

\subsection{Background and brane configuration} \label{bg}

	We consider asymptotically $AdS_5\times S^5$ space-time. The $AdS_5\times
	S^5$ geometry is holographically dual to the $\caln=4$ super-Yang-Mills
	theory with gauge group $SU(N_c)$. The dual description of a finite
	temperature field theory is an AdS black hole. We use the coordinates of
	\cite{Kobayashi:2006sb} to write the AdS black hole background in Minkowski
	signature as 
	\begin{equation}
	\eqlabel{AdSmetric}
		ds^2=\frac{1}{2}\left(\frac{\vrho}{R}\right)^2\left(-\frac{f^2}{\ft}\dd
		t^2+\ft
		\dd\vec{x}^2\right)+\left(\frac{R}{\vrho}\right)^2(\dd\vrho^2+\vrho^2\dd\Omega_5^2)\,,
	\end{equation}
	with $\dd\Omega_5^2$ the metric of the unit 5-sphere and
	\begin{equation}
		 f(\vrho)=1-\frac{\vrho_H^4}{\vrho^4},\quad
		 \ft(\vrho)=1+\frac{\vrho_H^4}{\vrho^4}\,, 
	\end{equation}
	where $R$ is the AdS radius, with
	\begin{equation}
		 R^4=4\pi g_s N_c\,{\alpha'}^2 = 2\lambda\,{\alpha'}^2\,.
	\end{equation}
	The temperature of the black hole given by (\ref{AdSmetric}) may be
	determined by demanding regularity of the Euclidean section. It is given by
	\begin{equation}
			 T=\frac{\vrho_H}{\pi R^2}\,. 
	\end{equation}
	In the following we may use the dimensionless coordinate
	$\rho=\vrho/\vrho_H$, which covers the range from the event horizon at
	$\rho=1$ to the boundary of the AdS space at $\rho\to\infty$.
	
	To include fundamental matter, we embed $N_f$ coinciding D$7$-branes into
	the ten-dimensional space-time. These D$7$-branes host flavor gauge fields
	$A_\mu$ with gauge group $U(N_f)$. To write down the DBI action for the
	D$7$-branes, we introduce spherical coordinates $\{r,\Omega_3\}$ in the
	4567-directions and polar coordinates $\{L,\phi\}$ in the 89-directions
	\cite{Kobayashi:2006sb}. The angle between these two spaces is denoted by
	$\theta$ ($0\le\theta\le\pi/2$). The six-dimensional space in the
	$456789$-directions is given by
	\begin{equation}
		 \begin{split}
			\dd\varrho^2+\varrho^2\dd\Omega_5^2=&\,\dd r^2+r^2\dd\Omega_3^2+\dd L^2+L^2\dd\phi^2\\
			=&\,\dd\varrho^2+\varrho^2(\dd\theta^2+\cos^2\theta\dd\phi^2+\sin^2\theta\dd\Omega_3^2)\,,
		\end{split} 
	\end{equation}
	where $r=\varrho\sin\theta$, $\varrho^2=r^2+L^2$ and $L=\varrho\cos\theta$.
	Due to the symmetry, the embedding of the D$7$-branes only depends on the
	radial coordinate $\rho$. Defining $\chi=\cos\theta$, we parametrize the
	embedding by $\chi=\chi(\rho)$ and choose $\phi=0$ using the $O(2)$
	symmetry in the 89-direction. The induced metric $G$ on the D$7$-brane
	probes is then
	\begin{equation}
	\eqlabel{inducedmetric}
		ds^2(G)=\frac{\vrho^2}{2R^2}\left(-\frac{f^2}{\ft}\dd
		t^2+\ft\dd\vec{x}^2\right)+\frac{R^2}{\vrho^2}\frac{1-\chi^2+\vrho^2(\del_\vrho\chi)^2}{1-\chi^2}\dd\vrho^2+R^2(1-\chi^2)\dd\Omega_3^2\,.
	\end{equation}
	The square root of the determinant of $G$ is given by
	\begin{equation}
		\sqrt{-G}=\frac{\sqrt{h_3}}{4}\varrho^3f\ft(1-\chi^2)\sqrt{1-\chi^2+\varrho^2(\del_\varrho\chi)^2}\,,
	\end{equation}
	where $h_3$ is the determinant of the 3-sphere metric. The embedding
	function $\chi$ will be determined numerically in section~\ref{sec:eomEmb}.
	It depends on the baryon and isospin chemical potentials we introduce next.

	\subsection{Introducing baryon and isospin chemical potentials}
	\label{sec:introchempot}
	
	We introduce baryon and isospin chemical potential as a non-vanishing time
	component of the non-Abelian background gauge field living on the
	D$7$-branes \cite{Kobayashi:2006sb},
	\begin{equation}
		\mu=\lim_{\rho\to\infty}A_0(\rho)\,. 
	\end{equation}
	First we consider the case of two different flavors, $N_f=2$. The generators
	of the gauge group $U(N_f=2)$ can be chosen as the three Pauli matrices
	$\sigma^a$ completed with the identity $\sigma^0$,
	\begin{equation}
		 \sigma^0=\begin{pmatrix}1 &
	0\\ 0&1\end{pmatrix}\,,\quad \sigma^1=\begin{pmatrix}0 &
	1\\1&0\end{pmatrix}\,,\quad \sigma^2=\begin{pmatrix}0 &
	-\ii\\\ii&0\end{pmatrix}\,,\quad\sigma^3=\begin{pmatrix}1 &
	0\\0&-1\end{pmatrix}\,. 
	\end{equation}
	For simplicity, we only consider the diagonal representations of the gauge
	group. This is equivalent to rotating the flavor coordinates until our
	chemical potential lies in the third isospin direction. Thus, the field
	strength tensor $F^{N_f}=\dd A^{N_f}+A^{N_f}\wedge A^{N_f}$ on the
	D$7$-branes may be written as
	\begin{equation}
	\eqlabel{U(2)expand}
		F^{N_f=2}_{\mu\nu}=F^B_{\mu\nu}\sigma^0+F^I_{\mu\nu}\sigma^3\,.
	\end{equation}
	The part of the field strength $F^B\sigma^0$ corresponding to the $U(1)$
	gauge group charges the branes equally and therefore induces a baryon
	chemical potential \cite{Kobayashi:2006sb}. However, the part $F^I\sigma^3$
	corresponding to the $SU(2)$ gauge group charges the branes differently
	inducing a isospin chemical potential \cite{Erdmenger:2007ap,Apreda:2005yz}.
	The definition above may be generalized to arbitrary $N_f>2$. Recall that
	there are $(N_f-1)$ diagonal generators of $SU(N_f)$ which form the Cartan
	subalgebra. Inspired by the interpretation that a diagonal generator of
	$SU(N_f)$ should charge one brane differently with respect to all others, we
	write the diagonal generators as
	\begin{equation}
		\lambda^i=\diag(1,\dots,\overbrace{-(N_f-1)}^{i\text{-th	position}},\dots,1)\qquad i=2,\dots,N_f\,. 
	\end{equation}
	For this choice of matrices the first flavor component is treated as the
	reference quantity. For arbitrary $N_f$ the generator of the baryonic part
	of the gauge group $U(1)$ is called $\lambda^1$. Thus, we can generalize
	(\ref{U(2)expand}) to
	\begin{equation}
		F^{N_f}_{\mu\nu}=F^B_{\mu\nu}\lambda^1+\sum_{i=2}^{N_f}F^{I_i}_{\mu\nu}\lambda^i=\sum_{i=1}^{N_f}F^{I_i}_{\mu\nu}\lambda^i
	\end{equation}
	where again $F^B=F^{I_1}$ induces the baryon and $F^{I_i}$ for $i\ge 2$ the
	isospin chemical potential for the $i$-th flavor component. In our setup,
        the only non-vanishing components of the background field strength are
        $F_{40}=-F_{04}=\del_\rho A_0$ since $A_0$ depends on $\rho$ only.

	\subsection{The DBI action and equations of motion}
	\label{eom} 
The DBI action determines the shape of the brane embeddings, i.e. scalar fields~$\Phi$, as well as the configuration of gauge fields~$A$ on these branes. We consider the case of $N_f=2$ 
coincident D7-branes for which the non-Abelian DBI action reads
\begin{equation}
 \label{eq:non-AbelianDBI}
 S_{\text{DBI}}=-T_{D7}\str\int\!\dd^8\xi\:\sqrt{\det
     Q}\sqrt{\det\left(P_{ab}\left[E_{\mu\nu}+E_{\mu
           i}(Q^{-1}-\delta)^{ij}E_{j\nu}\right]+2\pi\alpha'F_{ab}\right)} 
\end{equation}
with
\begin{equation}
 \label{eq:16}
 Q^i{}_j=\delta^i{}_j+\ii 2\pi\alpha'[\Phi^i,\Phi^k]E_{kj} \, ,
\end{equation}
where for a D$p$-brane in $d$ dimensions we have 
$\mu,\,\nu=0,\dots, 9$, $a,\,b=0,\dots, p$, $i,\,j = (p+1),\dots, d$,
$E_{\mu\nu} = g_{\mu\nu} + B_{\mu\nu}$. In our case we set $p=7$, $d=10$, $B\equiv 0$.
The action \eqref{eq:non-AbelianDBI} can be simplified significantly by using the spatial and gauge
symmetries present in our setup. 
First we make use of the spatial rotation symmetry in our 8,9-directions in
order to rotate to the frame in which $\Phi^9\equiv0$. In this particular
frame, all the commutators of our two scalar  
fields $\Phi^8,\,\Phi^9$ vanish and thus $Q^i_j = \delta^i_j$. However, the 
non-Abelian structure of embeddings and gauge fields is still manifest in the pullback appearing in the action~\eqref{eq:non-AbelianDBI} as
\begin{equation}
 \label{eq:pullback}
 \begin{split}
   P_{ab}[g_{\mu\nu}]=&g_{ab}+(2\pi\alpha')^2 g_{ij}
   \left(\del_a\Phi^i\del_b\Phi^j+\ii\del_a\Phi^i[A_b,\Phi^j]\right.\\
   &\left. +\ii[A_a,\Phi^i]\del_b\Phi^j-[A_a,\Phi^i][A_b,\Phi^j]\right) \, .
\end{split}
\end{equation}
The only terms coupling the embeddings $\Phi$ to the gauge fields $A$ are given by the commutator 
terms in equation~\eqref{eq:pullback}. These commutators vanish due to the following argument. 
We are free to define the $\tau^0$ flavor direction to be the direction parallel to 
the non-vanishing embedding $\Phi^8$ using the $U(1)\subset U(N_f)$ gauge symmetry. Now we {\it choose} our baryonic gauge field to have only a non-vanishing $A_0^0 \dd x^0 \tau^0$ component. 
In the dual field theory, this particular choice corresponds to the case in which the baryon charge representation and the mass representation are simultaneously diagonalizable. 
A different choice is possible and will most likely change the field theory phenomenology.
This applies also to the case of non-coincident D$p$-branes. 
Finally we use the remaining gauge symmetry $SU(N_f)\subset U(N_f)$ in order to restrict to 
an isospin gauge field $A$ along the third flavor direction only, i.e.~$A_0^3\dd x^0 \tau^3$, 
without loss of generality. 
Thus the only representations appearing in the background are the diagonal 
elements $\tau^0$ and $\tau^3$ of $U(N_f=2)$, which constitute the Cartan subalgebra. 
By definition these representation matrices commute with each other and thus all commutators $[\Phi,A]$ vanish. Therefore all terms coupling the scalars $\Phi$ to the gauge fields $A$ vanish, 
such that their equations of motion decouple as well.
In this way we can consistently truncate the non-Abelian DBI action to  	
		\begin{equation}
			\eqlabel{actiondiagNf=2} \begin{split}
		S_{\text{DBI}}&=-T_{D7}\int\!\dd^8\xi\;\str\left(\sqrt{|\det\left(G\sigma^0+2\pi\alpha'F^{N_f=2}\right)|}\right)\\
		&=-T_{D7}\int\!\dd^8\xi\;\sqrt{-G}\,\str\left(\sqrt{\sigma^0+(2\pi\alpha')^2
		G^{00}G^{44}\left(F_{40}^{N_f=2}\right)^2}\right)\,, \end{split}
		\end{equation}
		where in the second line the determinant is calculated. Next we
		determine the square of the non-vanishing components of the field
		strength tensor, which is by construction diagonal in the flavor space,
		\begin{equation}
			 \begin{split}
		\left(F_{40}^{N_f=2}\right)^2&=\left[\left(F^B_{40}\right)^2+\left(F^I_{40}\right)^2\right]\sigma^0+2F^B_{40}F^I_{40}\sigma^3\\
		&=\diag\left\{\left(F^B_{40}+F^I_{40}\right)^2,\left(F^B_{40}-F^I_{40}\right)^2\right\}\,,
		\end{split} 
		\end{equation}
		where we used $\left(\sigma^3\right)^2=\sigma^0$. Defining the new
		fields\footnote{To distinguish between the physical basis and the basis,
		in which the fields decouple, we give the quantities in the physical
		basis as $A^{B/I}$ an upper index and the quantities in the decoupled
		basis as $X_i$ a lower index.}
		\begin{equation}
			X_1=A_0^B+A_0^I\qquad\text{and}\qquad X_2=A_0^B-A_0^I\,, 
		\end{equation}
		the square of the field strength can be written as\footnote{Notice that
		only the time component of the gauge field $A^{B/I}$ does not vanish and
		depends only on the radial coordinate $\rho$: $F^{B/I}_{40}=\del_\rho
		A^{B/I}_0$}
		\begin{equation}
			 \left(F_{40}^{N_f=2}\right)^2=\diag\left\{\left(\del_\rho
			 X_1\right)^2,\left(\del_\rho X_2\right)^2\right\}\,. 
		\end{equation}
		Notice that the field $X_i$ is the gauge field living on the $i$-th
		brane. Inserting the metric components, the action \eqref{actiondiagNf=2} becomes
		\begin{equation}
			\begin{split}
				S_{\text{DBI}}=-T_{D7}\int&\!\dd^8\xi\;\frac{\sqrt{h_3}}{4}\varrho^3f\ft(1-\chi^2)\\
				\times\Bigg(&\sqrt{1-\chi^2+\vrho^2(\del_\vrho\chi)^2-2(2\pi\alpha')^2\frac{\ft}{f^2}(1-\chi^2)(\del_\vrho
				X_1)^2}\\
				&+\sqrt{1-\chi^2+\vrho^2(\del_\vrho\chi)^2-2(2\pi\alpha')^2\frac{\ft}{f^2}(1-\chi^2)(\del_\vrho
				X_2)^2}\Bigg)\,.
			\end{split} 
		\end{equation}
		The transformation to the fields $X_i$ decouple the two branes and we
		obtain a sum of two Abelian DBI actions. The Abelian actions are known
		from the pure baryonic case \cite{Kobayashi:2006sb} and we therefore can
		use the ideas given in the pure baryonic case to study our setup in
		which baryon and isospin charges are switched on simultaneously.
		
		Now we generalize this result to arbitrary $N_f$: Also for this case, we
		can decouple the non-Abelian gauge field into Abelian gauge fields on
		each brane. For simplicity we suppress the according index $N_f$ on the
		field strength tensor. We again start from the DBI action for the D$7$-branes,
		\begin{equation}
			\eqlabel{actiondiag}
		S_{\text{DBI}}=-T_{D7}\int\!\dd^8\xi\;\str\left(\sqrt{|\det(G\lambda^1+2\pi\alpha'F)|}\right)\,.
		\end{equation}
		Since this action is diagonal in flavor space, we are able to evaluate
		the square root and the trace directly (for more details see appendix
		\ref{appaction}). After a redefinition of the fields
		\begin{equation}
		\eqlabel{newfields}
			X_1=\sum_{j=1}^{N_f}A^{I_j}_0\,,\qquad
			X_i=\sum_{\substack{j=1\\j\not=i}}^{N_f}A^{I_j}_0-(N_f-1)A^{I_i}_0\,,\quad
			i=2,\dots,N_f\,, 
		\end{equation}
		where $X_i$ is the $i$-th flavor component of the gauge field $A^{N_f}$,
		the non-Abelian DBI action again becomes a sum of $N_f$ Abelian DBI actions,
		\begin{equation}
		\eqlabel{action}
		\begin{split}
			S_{\text{DBI}}=-T_{D7}\int&\!\dd^8\xi\;\frac{\sqrt{h_3}}{4}\varrho^3f\ft(1-\chi^2)\\
			&\times\sum_{i=1}^{N_f}\sqrt{1-\chi^2+\vrho^2(\del_\vrho\chi)^2-2(2\pi\alpha')^2\frac{\ft}{f^2}(1-\chi^2)(\del_\vrho
			X_i)^2}\,.
		\end{split} 
		\end{equation}
		Thus we may consider each brane separately with the gauge fields $X_i$
		treated as a $U(1)$ gauge field living on the $i$-th brane. The action
		for each brane is the same as the action for the pure baryonic case
		\cite{Kobayashi:2006sb}. In the following we apply the approach of
		\cite{Kobayashi:2006sb}. First we calculate the electric displacement
		$d_i$ on the $i$-th brane, which is a constant of motion and
		proportional to the flavor charge density,
		\begin{equation}
		\eqlabel{dX}
		\begin{split}
			d_i=\frac{\delta
			S_{\text{DBI}}}{\delta(\del_\varrho
			X_i)}=&(2\pi\alpha')^2T_{D7}\sqrt{h_3}\varrho^3\frac{\ft^2}{2f}\\
			&\times\frac{(1-\chi^2)^2\del_\varrho
			X_i}{\sqrt{1-\chi^2+\varrho^2(\del_\varrho\chi)^2-2(2\pi\alpha')^2\frac{\ft}{f^2}(1-\chi^2)(\del_\varrho
			X_i)^2}}\,.
		\end{split} 
		\end{equation}
		From the relations of the gauge fields \eqref{newfields}, we can read
		off the relations between the conjugate charge densities
		\begin{equation}
		\eqlabel{ds}
			d^B=d^{I_1}=\sum_{j=1}^{N_f}d_j\,,\qquad
			d^{I_i}=\sum_{\substack{j=1\\j\not=i}}d_j-(N_f-1)d_i\quad i=2,\dots,N_f\,,
		\end{equation}
		which for $N_f=2$ becomes 
		\begin{equation}
			d^B=d_1+d_2\qquad\text{and}\qquad d^I=d_1-d_2\,. 
		\end{equation}
		We now construct the Legendre transformation of the action
		(\ref{action}) to eliminate the fields $X_i$ in favor of the constants
		$d_i$,
		\begin{equation}
		\eqlabel{actiontrans}
		\begin{split}
			\tilde{S}_{\text{DBI}}&=S_{\text{DBI}}-\int\!\dd^8\xi\;\sum_{i=1}^{N_f}X_i\frac{\delta
			S_{\text{DBI}}}{\delta (\del_\vrho X_i)}\\ &
			\begin{split}
				=-T_{D7}\int\!\dd^8\xi\;&\frac{\sqrt{h_3}}{4}\vrho^3f\ft(1-\chi^2)\sqrt{1-\chi^2+\vrho^2(\del_\vrho\chi)^2}\\
				&\times\sum_{i=1}^{N_f}\sqrt{1+\frac{8d_i^2}{(2\pi\alpha')^2T_{D7}^2\vrho^6\ft^3(1-\chi^2)^3}}\,.
			\end{split}
		\end{split} 
		\end{equation}
		The gauge fields can be calculated from the Legendre transformed action by
                $\del_\vrho X_i=-\delta\tilde{S}_{\text{DBI}}/\delta d_i$.
		
		It is convenient to introduce the dimensionless quantity
                \footnote{The definition of $\dt$ in the pure baryonic case
		\cite{Kobayashi:2006sb} differs from our convention by a factor $N_f$ such that the
		dimensionful physical parameters $d^{I_i}$ have the same normalization.}
		\begin{equation}\eqlabel{dt}
			 \dt_i=\frac{d_i}{2\pi\alpha'\varrho_H^3T_{D7}}\,.
		\end{equation}

		\subsubsection{Accidental symmetries}
		\label{sym} 
			
			The DBI action (\ref{actiontrans}) has the following discrete
			symmetries: \begin{description} \item[Permutation:
			$d_i\leftrightarrow d_j$ for $i,j=1,\ldots,N_f$ :] Since $d_i$ is
			the electric displacement of the $i$-th brane and the coincident
			branes have the same embedding function by definition, the
			substitution $d_i\leftrightarrow d_j$ interchanges the two branes.
			\item[Reflection: $d_i\leftrightarrow-d_i$ for each $i=1,\dots,N_f$
			:]We consider an $U(1)$ gauge field with only the time component
			non-vanishing. Since it does not depend on the time, our setup is
			like an electrostatic situation. To change the sign of the electric
			displacement $\dt_i$, we must change the sign of the gauge field
			$X_i$ (see equation (\ref{dX})). The sign change in the gauge field
			may be induced by a sign change of the electric charge generating
			the gauge field $X_i$, which may be interpreted as an interchange
			between particle and anti-particle. The symmetry $d_i\leftrightarrow
			-d_i$ is therefore induced by the invariance of our system
			interchanging particles with anti-particles on the $i$-th brane.
			\end{description} If we expand the action (\ref{actiontrans}) for
			small $d_i$,
			\begin{equation}
			 \begin{split}
				\tilde{S}_{\text{DBI}}=-T_{D7}\int\!\dd^8\xi\;&\frac{\sqrt{h_3}}{4}\rho^3f\ft
				(1-\chi^2)\sqrt{1-\chi^2+\rho^2(\del_\rho\chi)^2}\\
				&\times\left(N_f+\frac{8\sum_{i=1}^{N_f}d_i^2}{(2\pi\alpha')^2T_{D7}\rho^6\ft^3(1-\chi^2)^3}+\cdots\right)\,,
			\end{split} 
			\end{equation}
			we observe an approximate $O(N_f)$ symmetry
			\begin{equation}
				\eqlabel{approxO(2)} (d_1,\dots,d_{N_f})\leftrightarrow
				\calo(d_1,\dots,d_{N_f})\,,\quad \calo\in O(N_f)\,. 
			\end{equation}
			The approximate $O(N_f)$ symmetry, which consists of the $SO(N_f)$
			group and reflections, is the continuous symmetry generated by the
			discrete symmetries since the $SO(N_f)$ generators are real
			skew-symmetric matrices which can be represented by reflection and
			permutation matrices.
			
			In the physical basis the effect of the symmetry transformation
			above is given by
			\begin{itemize}
				\item $d_i\leftrightarrow d_j$: 
					\begin{equation}
						\begin{aligned}
							\text{for} &\quad i,j\not=1: && d^{I_i}\leftrightarrow
							d^{I_j}\\ \text{for}&\quad i=1,\, j\not=1:\quad &&d^B\leftrightarrow
							d^B\qquad\text{and}\qquad d^{I_j}\leftrightarrow -\sum_{k=2}^{N_f}d^{I_k},
						\end{aligned} 
					\end{equation}
				\item $d_i\leftrightarrow -d_i$:
					\begin{equation}
						d^{I_i}\leftrightarrow
						-d^{I_i}+2\sum_{j\not=i}d_j\qquad\text{and}\qquad d^{I_j}\leftrightarrow
						d^{I_j}-2d_i\quad j\not=i\,. 
					\end{equation}
			\end{itemize}
			For $N_f=2$ we get
			\begin{equation}
				\begin{aligned}
					d_1&\leftrightarrow d_2: && d^B\mapsto d^B && \text{and} && d^I\mapsto-d^I\\
					d_1&\leftrightarrow-d_1:\qquad && 	d^B\mapsto-d^I\qquad&&\text{and}\qquad && d^I\mapsto-d^B\\
					d_2&\leftrightarrow-d_2: && d^B\mapsto d^I && \text{and} && d^I\mapsto d^B\,.
				\end{aligned}
				\eqlabel{symNf=2} 
			\end{equation}
			These symmetries are also present in 2-color QCD theories with
			degenerated quark mass \cite{Splittorff:2000mm}. Since the
			transformation matrix to the physical basis is proportional to an
			$O(2)$ matrix, there is also an induced approximate $O(2)$ symmetry
			in the physical basis. The results for the case $N_f=3$ are shown in
			the appendix \ref{app:symNf=3}.

		\subsubsection{Gauge fields}
		\label{sec:eomGaugeFields} 
			
			We now determine the gauge fields which will allow us to compute the
			chemical potentials. Therefore, we write down the asymptotic form of
			the gauge field close to the boundary,
			\begin{equation}
				A_0^{I_i}=
				\mu^{I_i}-\frac{\dt^{I_i}}{2\pi\alpha'}\frac{\vrho_H}{\rho^2}+\cdots\qquad\text{for}\qquad
				i=1,\ldots,N_f\,, 
			\end{equation}
			where the coefficients $\mu^{I_i}$ and $\dt^{I_i}$ are
			related to the baryon/isospin chemical potential and
			the baryon/isospin density $n_q^{I_i}$, respectively,
			\begin{equation}\label{eq:ndt}    
				\dt^{I_i}=\frac{2^{\frac{5}{2}}n_q^{I_i}}{N_fN_c\sqrt{\lambda}T^3}.
			\end{equation}
			Obviously, the chemical potential $\mu^{I_i}$ depends on the
			corresponding density $\dt^{I_i}$. To determine, this dependence it
			is convenient to use the dimensionless quantities
			\begin{equation}
				\Xt_i=\frac{2\pi\alpha'}{\vrho_H}X_i\,,\qquad
				\mut_i=\frac{2\pi\alpha'}{\vrho_H}\mu_i=\sqrt{\frac{2}{\lambda}}\frac{\mu_i}{T}\,.
			\end{equation}
			As described above, the transformed action (\ref{actiontrans}) may
			be used to calculate the gauge fields and therefore the chemical
			potentials (see \cite{Kobayashi:2006sb} for more details)
			\begin{equation}
			\eqlabel{mu}
				\mut_i=\Xt_i(\rho=\infty)=2\dt_i\int_1^\infty\!\dd\rho\;\frac{f\sqrt{1-\chi^2+\vrho^2(\del_\rho\chi)^2}}{\sqrt{\ft(1-\chi^2)[\rho^6\ft^3(1-\chi^2)^3+8\dt_i^2]}}\,.
			\end{equation}
			The connection to the physical basis is given by
			\begin{equation}
				 \mut_1=\sum_{j=1}^{N_f}\mut^{I_j}\,,\qquad
				 \mut_i=\sum_{j\not=i}\mut^{I_j}-(N_f-1)\mut^{I_i}\quad i=2,\dots,N_f\,,
			\end{equation}
			which for $N_f=2$ becomes
			\begin{equation}
				\eqlabel{muphysNf=2}
				\mut^B=\frac{1}{2}(\mut_1+\mut_2)\qquad\text{and}\qquad\mut^I=\frac{1}{2}(\mut_1-\mut_2)\,.
			\end{equation}
			Notice that in the Abelian case \cite{Kobayashi:2006sb} it is found
			that in the limit $m\to\infty$ the chemical potentials $\mu_i$
			approach $M_q$ for positive $\dt_i$. This allows us to calculate the
			physical chemical potentials in the limit $m\to\infty$,
			\begin{equation}
				\mu^B\to M_q\qquad\text{and}\qquad \mu^{I_i}\to0\quad i=2,\ldots,N_f,
			\end{equation}
			if $\dt_j>0$ for $j=1,\ldots,N_f$. In the case $\dt_i<0$, using
			the accidental symmetry of the action $\dt_i\leftrightarrow-\dt_i$
			presented in section \ref{sym}, we get also a sign change in the
			chemical potential $\mu_i\leftrightarrow-\mu_i$. Therefore, if we change the sign of
			just one $\dt_i$, there is at least one isospin chemical potential
			$\mut^{I_i}$ which does not vanish any more. For $N_f=2$ we 
			explicitly get
			\begin{equation}
			\eqlabel{eq:mulargem}
			\begin{aligned}
				\mu^B & \to M_q\quad	&& \text{and}\quad &&\mu^I\to 0\qquad && \text{for }\dt_1,\,\dt_2>0      && \text{\ie}\quad 0\le\dt^I<\dt^B\\
				\mu^B & \to	0         	&& \text{and}       &&\mu^I\to M_q     && \text{for }-\dt_1<\dt_2<0\quad && \text{\ie}\quad 0\le\dt^B<\dt^I\,.
			\end{aligned} 
			\end{equation}
			If we set one density to zero in the case $N_f=2$, \eg $\dt_2=0$,
			only the first brane is charged. Therefore, the chemical potential
			$\mu_1$ approaches $M_q$ and $\mu_2$ approaches zero in the limit
			$m\to\infty$. Using equation \eqref{muphysNf=2}, this implies
			\begin{equation}
			\eqlabel{eq:discontidB=dI}
				\mu^B\to
				\frac{M_q}{2}\qquad\text{and}\qquad\mu^I\to\frac{M_q}{2}\qquad\text{for }
				\dt_1>0,\,\dt_2=0\quad\text{\ie}\quad 0<\dt^B=\dt^I\,. 
			\end{equation}
			This discontinuous step suggests that there is a phase transition
			between the regions $0\le \dt^I<\dt^B$ and $0\le\dt^B<\dt^I$. We
			also expect a similar phase transition for arbitrary $N_f$. In
			section \ref{canlargem} we discuss this phase transition further.

		\subsubsection{Embeddings}
		\label{sec:eomEmb}
			
			The equation of motion for the embedding function $\chi$ can be
			derived from (\ref{actiontrans}) as
			\begin{equation}
			\eqlabel{eomchi}
			\begin{split}
				&\del_\rho\left[\rho^5f\ft(1-\chi^2)\frac{\del_\rho\chi}{\sqrt{1-\chi^2+\rho^2(\del_\rho\chi)^2}}\sum_{i=1}^{N_f}\sqrt{1+\frac{8\dt_i^2}{\rho^6\ft^3(1-\chi^2)^3}}\right]\\
				=&-\frac{\rho^3f\ft\chi}{\sqrt{1-\chi^2+\rho^2(\del_\rho\chi)^2}}\Bigg[\left[3\left(1-\chi^2\right)+2\rho^2(\del_\rho\chi)^2\right]\sum_{i=1}^{N_f}\sqrt{1+\frac{8\dt_i^2}{\rho^6\ft^3(1-\chi^2)^3}}\\
				&-\frac{24}{\rho^6\ft^3(1-\chi^2)^3}\left(1-\chi^2+\rho^2(\del_\rho\chi)^2\right)\sum_{i=1}^{N_f}\frac{\dt_i^2}{\sqrt{1+\frac{8\dt_i^2}{\rho^6\ft^3(1-\chi^2)^3}}}\Bigg]\,.
			\end{split} 
			\end{equation}
			In addition to the case without isospin density
			\cite{Kobayashi:2006sb}, sums over the $N_f$ different densities
			appear in the equation of motion for the embedding function. Due to
			these sums, the equation of motion cannot be written down in a more
			compact form, as done in the Abelian case \cite{Kobayashi:2006sb}.
			
			To relate the gravity field $\chi$ to the dual gauge field
			parameters we consider the asymptotic form for the embedding
			function $\chi$ close to the boundary,
			\begin{equation}
				\eqlabel{asymbchi}
			\chi=\frac{m}{\rho}+\frac{c}{\rho^3}+\cdots\,. 
			\end{equation}
			Then the AdS/CFT dictionary relates the coefficients $m$ and $c$ to
			the bare quark mass $M_q$ and the quark condensate
			$\langle\bar{\psi}\psi\rangle$, respectively,
			\begin{equation}
			\eqlabel{dicmc}
				 m=\frac{2M_q}{\sqrt{\lambda}T}\,,\qquad
				 c=-\frac{8\langle\bar{\psi}\psi\rangle}{\sqrt{\lambda}N_fN_cT^3}\,.
			\end{equation}
			The initial conditions to solve equation of motion \eqref{eomchi}
                        numerically are $\chi(\rho=1)=\chi_0$ and $\del_\rho\chi(\rho=1)=0$.
                        We determine the parameter $m$ and $c$, which depend on $\chi_0$,
                        by fitting the numerical solution to the asymptotic form
			(\ref{asymbchi}).
			
\begin{figure}
				\centering
				\psfrag{w6}{\small $L(r)$}
				\psfrag{r}{\small $r$}
				\psfrag{dt0}{\scriptsize $\tilde d=0$}
				\psfrag{dtn0}{\scriptsize $\tilde	d=0.05$}
				\includegraphics[width=0.6\textwidth]{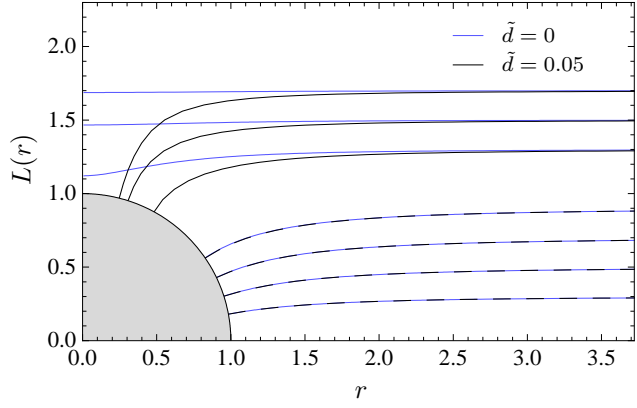}
				\caption{Embedding function $L(r)$ of D$7$-branes in the AdS
					black hole background, with the dimensionless
                                        coordinates $L=\rho\cos\theta=\rho\chi$ and
                                        $r=\rho\sin\theta=\rho\sqrt{1-\chi^2}$.}
				\label{fig:embedExample}
			\end{figure}

			Examples for embeddings are shown in fig.~\ref{fig:embedExample}. In
			the case of vanishing particle density we observe both black hole
			embeddings, ending on the black hole horizon, and Minkowski
			embeddings \cite{Babington:2003vm}. The latter embeddings do not
			touch the horizon. For any finite value of the particle
			density there are only black hole embeddings
			\cite{Kobayashi:2006sb}.
			
		\subsubsection{Strings from branes}
		\label{stringsbrane} 
			
			In the case of pure finite baryon density \cite{Kobayashi:2006sb} it
                        was shown that D$7$-branes representing heavy quarks develop
			a spike near the horizon, since they reach the horizon albeit their
			large value of $L$ at large $r$. This spike may be interpreted as a
                        bundle of strings stretching between the D$7$-branes and the black
                        hole. This bundle of strings is necessary in the presence of a
                        non-vanishing gauge field on the brane. Since in our case we
                        have a similar setup, we expect to have this spike, too. To formalize
                        this intuition, we analyze the near horizon limit of the brane
                        embedding in more detail as in \cite{Kobayashi:2006sb}. We start by
                        writing the Legendre-transformed action as
			\begin{equation}
				\tilde{S}_{\text{DBI}}=-\frac{T_{D7}}{\sqrt{2}}\int\!\dd^8\xi\;\frac{f}{\sqrt{\ft}}\sqrt{1+\frac{\vrho^2(\del_\vrho\chi)^2}{1-\chi^2}}\sum_{i=1}^{N_f}\sqrt{\frac{d_i^2}{(2\pi\alpha')^2T_{D7}}+\frac{\vrho^6\ft^3(1-\chi^2)^3}{8}}\,.
			\end{equation}
			Notice that $\chi=\cos\theta$, which becomes $\chi\simeq 1$ if the
			embedding is very near to the axis. Therefore, the second term in
			the square roots can be neglected and we get
			\begin{align}
				\tilde{S}_{\text{DBI}}&=-\frac{V_3\vol(S^3)}{2\pi\alpha'}\left(\sum_{i=1}^{N_f}d_i\right)\int\!\dd
				t\dd\varrho\;
				\frac{f}{\sqrt{2\ft}}\sqrt{1+\frac{\varrho^2(\del_\varrho\chi)^2}{1-\chi^2}}\nonumber\\
				&=-\frac{V_3\vol(S^3)}{2\pi\alpha'}\left(\sum_{i=1}^{N_f}d_i\right)\int\!\dd
				t\dd\varrho\;\sqrt{-g_{tt}(g_{\varrho\varrho}+g_{\theta\theta}(\del_\varrho\theta)^2)}\,,
			\end{align}
			where $V_3$ is the Minkowski space volume and $\vol(S^3)$ the volume
			of the 3-sphere. Recognize the fact that the result above can be
			written as the Nambu-Goto action for a bundle of strings stretching
			in $\varrho$ direction but free bending in the $\theta$ direction,
			\begin{equation}
				\tilde{S}_{\text{DBI}}=-V_3\vol(S^3)\left(\sum_{i=1}^{N_f}d_i\right)S_{NG}\,.
			\end{equation}
			Therefore just one of the densities $d_i$ must be non-zero to make
			the branes develop a spike close to the axis. This is the expected
			result. According to the discussion below equation~\eqref{eq:non-AbelianDBI}
we consider a consistent case in which the $N_f$ branes are coincident. 
                        Therefore they have the same embedding function, such that the complete
                        stack of $N_f$ D$7$-branes develops the spike discussed above.

	\subsection{Numerical results for the background fields\\ at constant baryon
	and isospin density}
	\label{bgcan} 

	\begin{figure}
		\centering
		\psfrag{chi0}{$\chi_0$}
		\psfrag{m}{$m$}
		\includegraphics[width=0.5\textwidth]{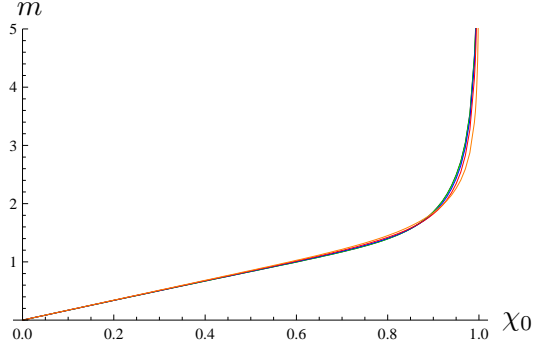}
		\caption{The dimensionless mass parameter $m$ as defined in equation
			\eqref{dicmc} versus the horizon value $\chi_0=\lim_{\rho\to 1}\chi$ of
			the embedding at baryon density $\dt^B=0.5$ for the case $N_f=2$.
			The five different curves correspond to isospin density $\dt^I=0$
			(black), $\dt^I=\frac{1}{4}\dt^B$ (green), $\dt^I=\frac{1}{2}\dt^B$
			(blue), $\dt^I=\frac{3}{4}\dt^B$ (red) and $\dt^I=\dt^B$ (orange).}
		\label{fig:bgmcan}
	\end{figure}
	
	\begin{figure}
		\centering
		\psfrag{m}{$m$}
		\psfrag{c}{$c$}
		\subfigure[]{\includegraphics[width=0.3\textwidth]{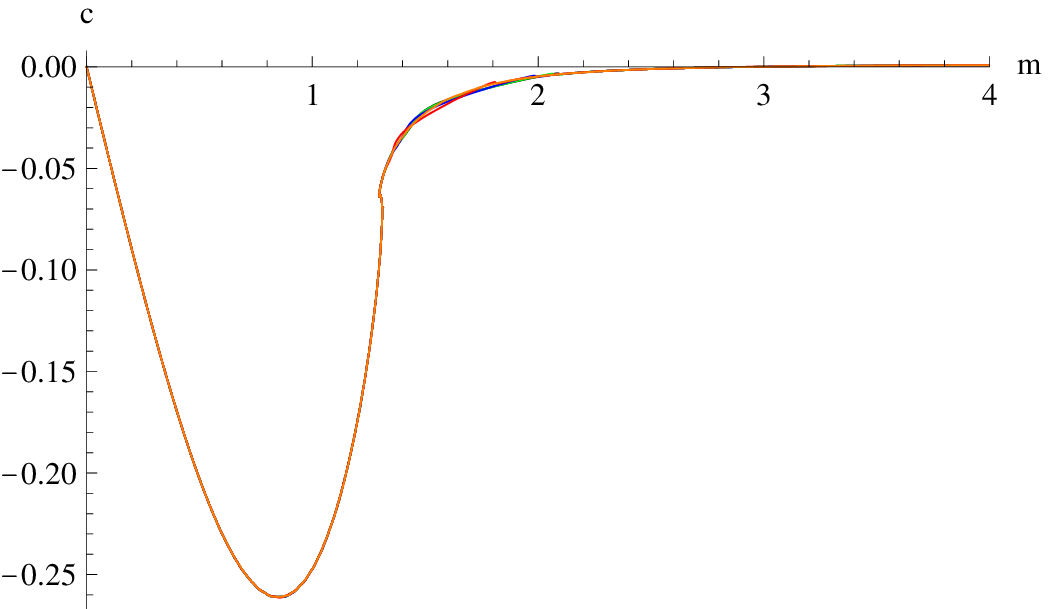}}
		\hspace{8pt}
		\subfigure[]{\includegraphics[width=0.3\textwidth]{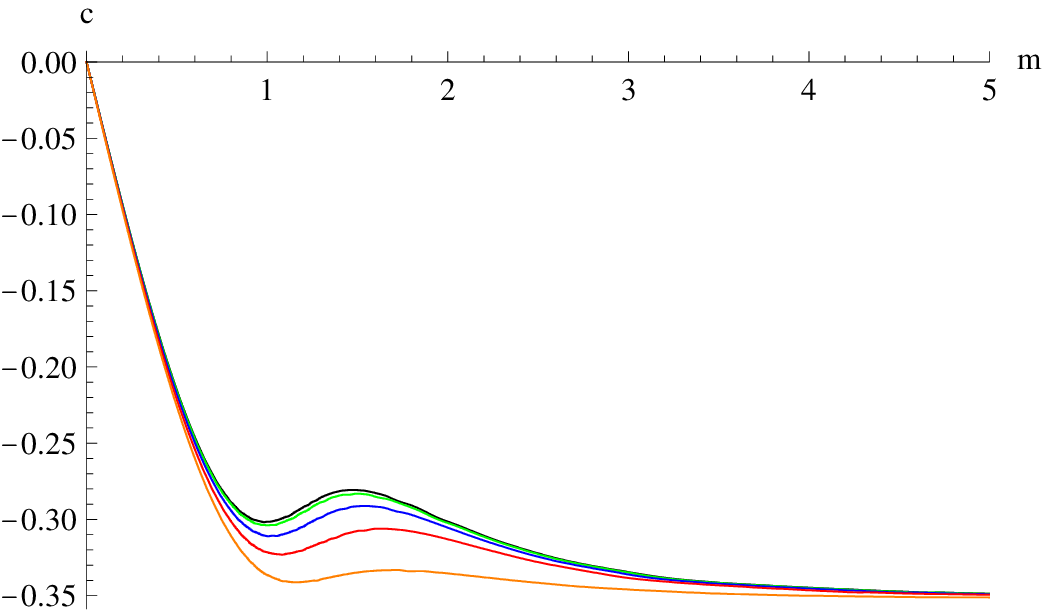}}
		\hspace{8pt}
		\subfigure[]{\includegraphics[width=0.3\textwidth]{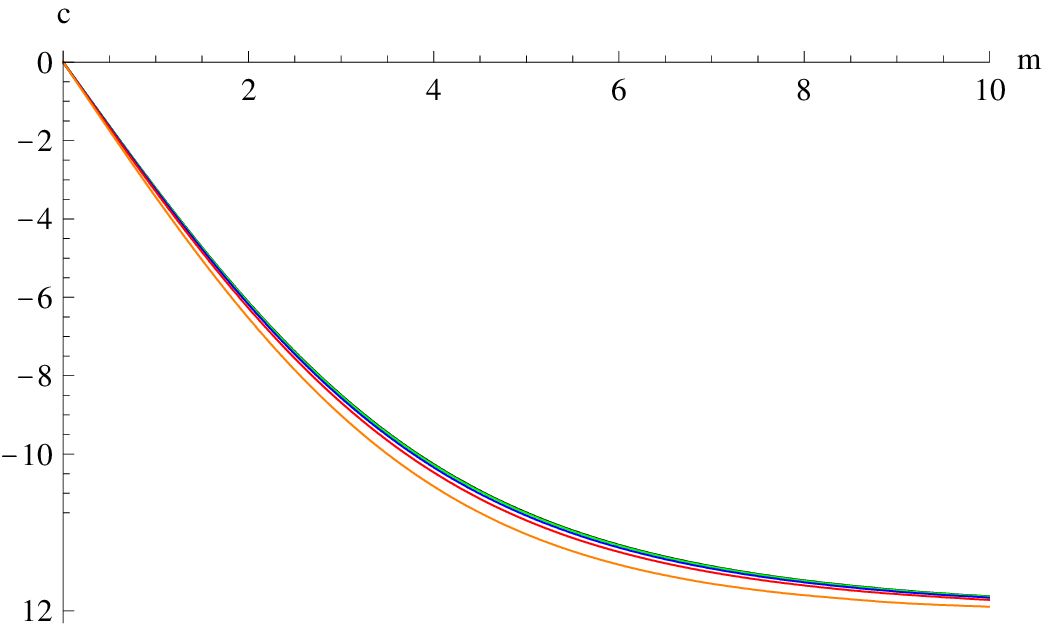}}
		\caption{The dimensionless chiral condensate $c$ versus the mass parameter
			$m$ as defined in equation \eqref{dicmc} at baryon density
			$\dt^B=5\cdot 10^{-5}$ (a), $\dt^B=0.5$ (b) and $\dt^B=20$ (c) for
			the case $N_f=2$. The five different curves in each figure
			correspond to isospin density $\dt^I=0$ (black),
			$\dt^I=\frac{1}{4}\dt^B$ (green), $\dt^I=\frac{1}{2}\dt^B$ (blue),
			$\dt^I=\frac{3}{4}\dt^B$ (red) and $\dt^I=\dt^B$ (orange).}
		\label{fig:bgccan}
	\end{figure}
	
	\begin{figure}[t]
	\centering
	\psfrag{m}{$m$}
	\psfrag{muB}{$\mu^B/M_q$}
	\psfrag{muI}{$\mu^I/M_q$}
	\subfigure[]{\includegraphics[width=0.45\textwidth]{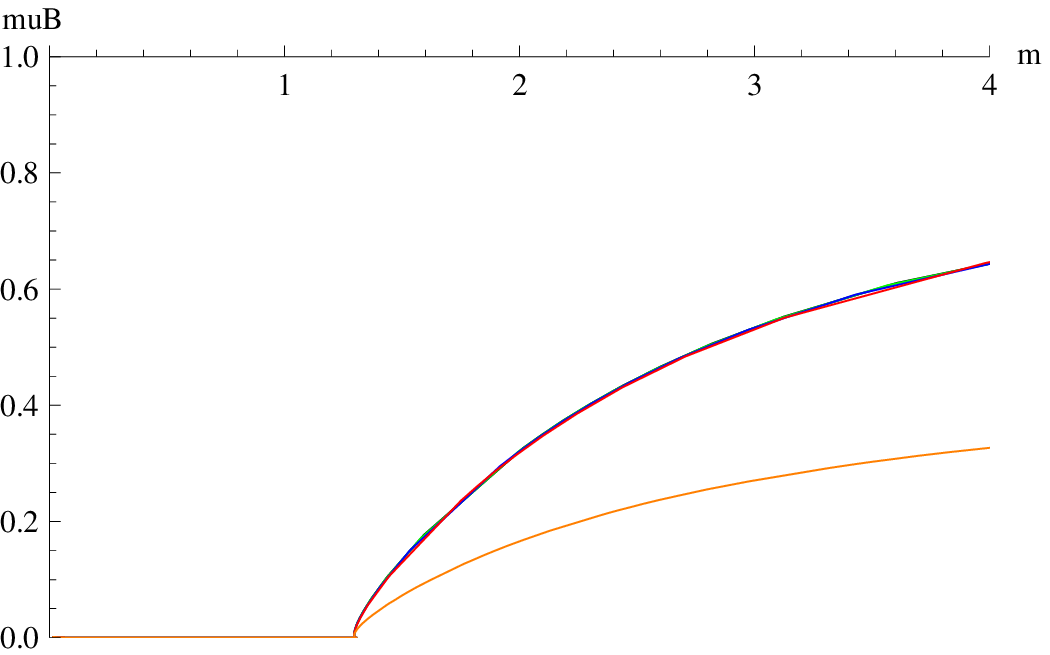}}
	\hspace{8pt}
	\subfigure[]{\includegraphics[width=0.45\textwidth]{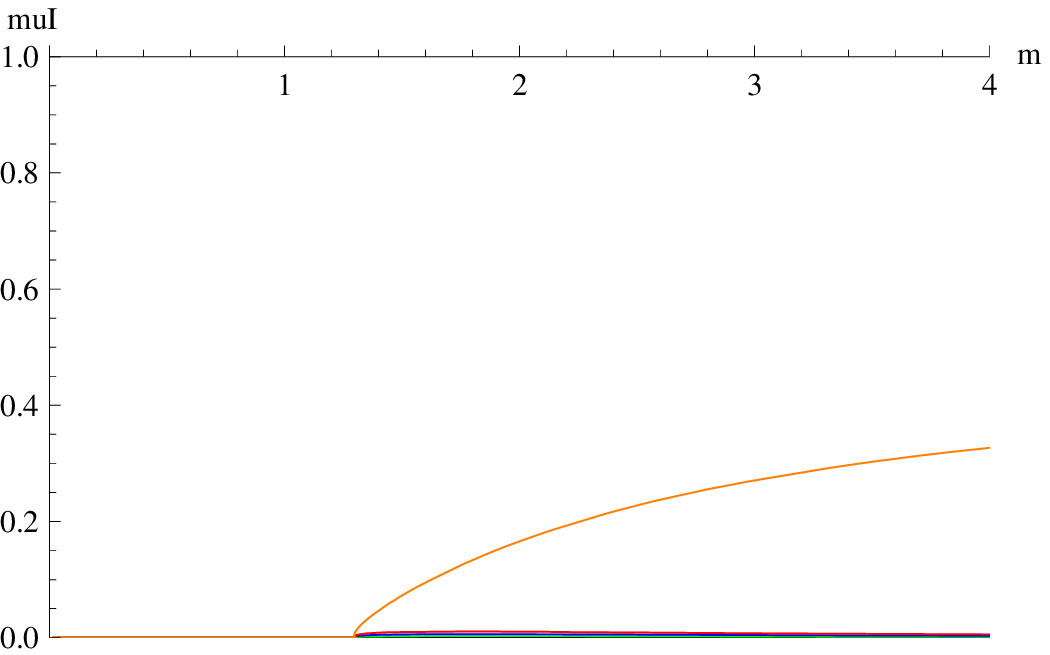}}
	\subfigure[]{\includegraphics[width=0.45\textwidth]{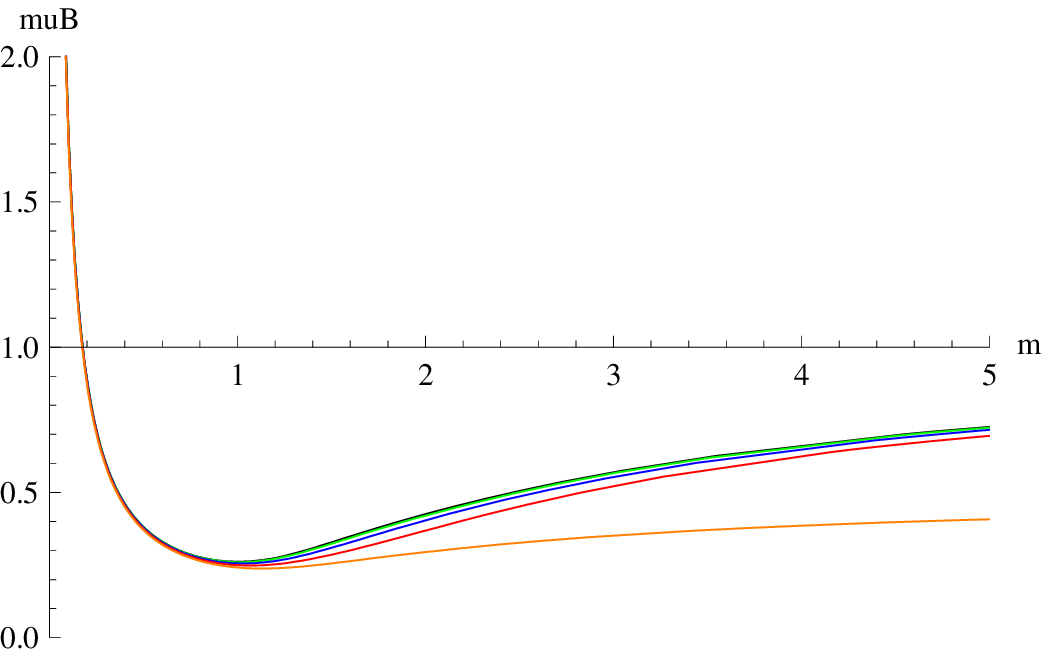}}
	\hspace{8pt}
	\subfigure[]{\includegraphics[width=0.45\textwidth]{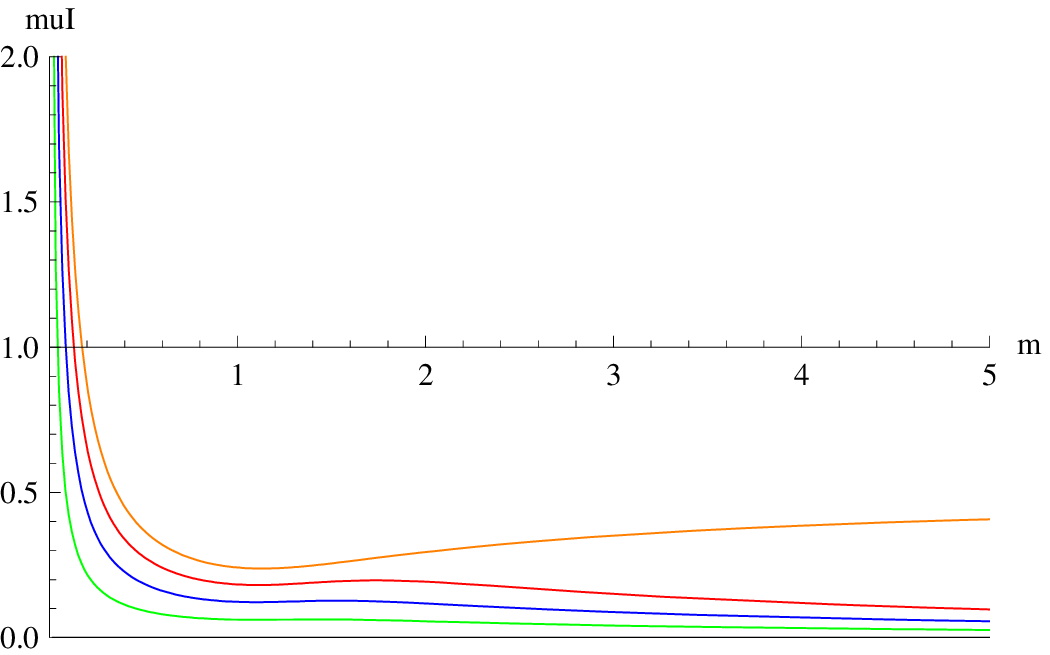}}
	\caption{The baryon (left) and isospin (right) chemical potential divided by
		the bare quark mass $M_q$ versus the mass parameter $m$ as defined in
		equation \eqref{dicmc} at baron density $\dt^B=5\cdot 10^{-5}$ in
		the upper figures and $\dt^B=0.5$ in the lower ones for the case
		$N_f=2$. The five different curves in each figure correspond to
		isospin density $\dt^I=0$ (black), $\dt^I=\frac{1}{4}\dt^B$ (green),
		$\dt^I=\frac{1}{2}\dt^B$ (blue), $\dt^I=\frac{3}{4}\dt^B$ (red) and
		$\dt^I=\dt^B$ (orange).}
	\label{fig:bgmucan}
	\end{figure}
	
	In this section we present numerical results for the background fields at
	constant densities, which will later be used to determine the thermodynamics
	in the canonical ensemble. We first analyze the embedding of the
	D$7$-branes. Fig.~\ref{fig:bgmcan} and \ref{fig:bgccan} show the dependence
	of the mass parameter $m$ on the horizon value of the embedding function
	$\chi_0=\lim_{\rho\to 1}\chi$ and of the chiral condensate $c$ on the mass
	parameter $m$ for the case $N_f=2$ at different baryon and isospin
	densities.
	
	Fig.~\ref{fig:bgmcan} shows the same divergent behavior of the mass
	parameter as found in \cite{Kobayashi:2006sb} for the pure baryonic case.
	Fig.~\ref{fig:bgccan} (a) shows that the low temperature black hole
	embeddings for small but non-zero densities mimic the behavior of Minkowski
	embeddings which describe the brane at zero densities. For small densities
	there is a first order phase transition between two black hole embeddings,
	which replaces the first order phase transition between black hole and
	Minkowski embeddings at zero densities \cite{Babington:2003vm,
	Mateos:2007vn}. Later in section \ref{phasediagramcan} we discuss that the
	phase transition between the two black hole embeddings must be replaced by a
	phase transition between black hole embeddings and an inhomogeneous mixture
	of black hole and Minkowski embeddings since the black hole embeddings alone
	are not the stable ground state of this theory. At a baryon density larger
	than a critical density, which value depends on the isospin density, the phase
        transition disappears and a local maximum appears in the chiral condensate
        (see fig.~\ref{fig:bgccan} (b)). This maximum disappears if we increase the baryon
        or the isospin density and the chiral condensate $c$ monotonically decreases as we
        increase the mass $m$ (see fig~\ref{fig:bgccan} (c)).
	
	Fig.~\ref{fig:bgccan} also shows that  for finite densities, the chiral
        condensate $c$ approaches a fixed non-zero value in the limit $m\to\infty$,
        \ie $T\to 0$. In fig.~\ref{fig:bgccan} (a) this asymptotic value is small due
	to the small baryon density but also non-zero as in fig.~\ref{fig:bgccan}
	(b) and (c). Since the non-zero chiral condensate breaks conformal
	invariance and supersymmetry, these symmetries are always broken in our
	setup even in the limit $T\to 0$, where the AdS black hole background
	becomes conformal and supersymmetric. This symmetry breaking is induced by
	the finite densities and chemical potentials. The finite densities, which
	introduce a new scale, break conformal invariance. The non-zero vev of just
	the time component of the gauge field, which describes the chemical
	potential, breaks Lorentz invariance and therefore supersymmetry.

	Next we present the chemical potentials $\mu^B$ and $\mu^I$ in
	fig.~\ref{fig:bgmucan} for the case $N_f=2$ at different baryon and isospin
	densities. In the limit $m\to\infty$, the baryon chemical potential $\mu^B$
        approaches $M_q$ and the isospin chemical potential $\mu^I$ approaches 0
        (see fig.~\ref{fig:bgmucan} black, green, blue and red line) for $\dt^B>\dt^I\ge 0$
        as shown in equation \eqref{eq:mulargem}. For $\dt^B=\dt^I$ (orange line) the
        baryon and isospin chemical potential both approach $M_q/2$ in the large mass
        limit as shown in equation \eqref{eq:discontidB=dI}. Moreover, fig.~\ref{fig:bgmucan}
        (c) and	(d) show that the baryon chemical potential is independent of the isospin
	density except for the discontinuous step at $\dt^B=\dt^I$ and the isospin
	chemical potential obviously depends on the isospin density.
	Fig.~\ref{fig:bgmucan} (a) and (b) show that the phase transition is also
	visible in the chemical potential. For a baryon density smaller than the
	critical density, the dimensionless quantities $\mu^{B/I}/M_q$ decrease as
	we decrease the mass $m$ and stays constant at a non-vanishing but small
	value after the phase transition. For a baryon density larger than the
	critical density, the dimensionless quantities $\mu^{B/I}/M_q$ diverge in
	the small mass limit $m\to0$.

	\subsection{Numerical results for the background fields\\ at constant baryon
		and isospin chemical potential}
	\label{bggrand} 
	
	In the calculations above we treated the densities $\dt^B$ and $\dt^{I_i}$
	as a independent variables since they are the constants of motion.
	Thermodynamically this means that we consider the canonical ensemble. Since
	we also would like to study the grand canonical ensemble with the chemical
	potentials $\mut^B$ and $\mut^{I_i}$ as independent variables, we must
	analyze the background fields for constant chemical potentials $\mut^B$ and
	$\mut^{I_i}$. To determine the densities $\dt_i$ for a given chemical
	potential $\mut_i$ we must invert equation (\ref{mu}). We use a shooting
	method to obtain these functions $\dt_i(\mut_i,m)$.\footnote{For the exit
	condition of the shooting method we use that the deviation from a given
	chemical potential $\mut^g_i$ to a chemical potential $\mut^c_i(\dt_i)$
	calculated numerically by \eqref{mu} for a chosen $\dt_i$ is at most
	$0.5\%$.}
	
	So far we considered only the case with non-zero densities. We now
	investigate the case with zero densities but still non-zero chemical
	potentials. As shown in \cite{Karch:2007br}, in this case Minkowski
	embeddings must also be included. Since the action only depends on the
	derivative of the gauge field, a Minkowski embedding with constant gauge
	field solves the equation of motion. Therefore, on these Minkowski embeddings the
	chemical potential can take any value. The equation of motion for the
	Minkowski embedding $L$ is given by \cite{Mateos:2007vn}
	\begin{equation}
		\del_r\left[r^3\left(1-\frac{1}{(r^2+L^2)^4}\right)\frac{\del_rL}{\sqrt{1+(\del_rL)^2}}\right]-8\frac{r^3L}{(r^2+L^2)^5}\sqrt{1+(\del_rL)^2}=0\,,\eqlabel{eommink}
	\end{equation}
	where $L$ and $r$ are dimensionless and given by $\rho^2=r^2+L^2$,
        $r=\rho\sin\theta=\rho\sqrt{1-\chi^2}$ and $L=\rho\cos\theta=\rho\chi$. From now
        on we only use these dimensionless coordinates $L$ and $r$. The asymptotic solution
        for the embedding function $L$ close to the boundary is,
	\begin{equation}
		 L=
		 m+\frac{c}{r^2}+\cdots\,.\eqlabel{asymmink} 
	\end{equation}
	The parameters $m$ and $c$ are related to the quark mass $M_q$ and the
	chiral condensate $\langle\bar{\psi}\psi\rangle$ given by equation (\ref{dicmc}).
        Using the initial conditions $L(r=0)=L_0$ and $\del_rL(r=0)=0$, we may solve the
        equation of motion (\ref{eommink}) numerically. The parameters
	$m$ and $c$ may be determined by fitting the numerical solutions to the
	asymptotic form (\ref{asymmink}).
	
\begin{figure}
		\centering
		\psfrag{chi0}[l]{$\stackrel{\chi_0}{L_0}$} \psfrag{m}{$m$}
		\subfigure[]{\includegraphics[width=0.3\textwidth]{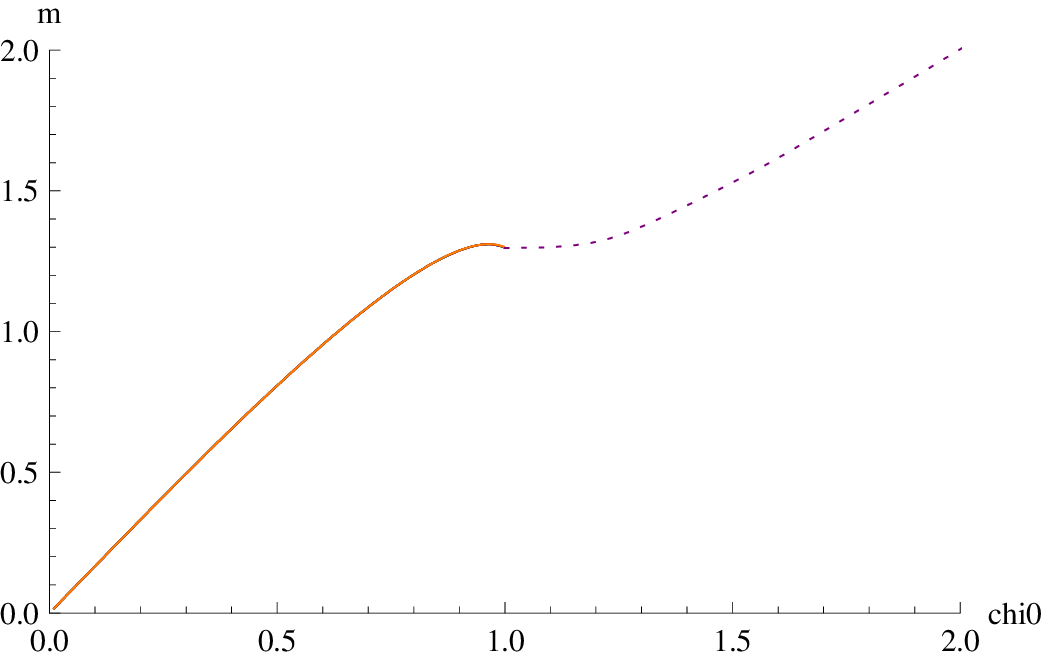}}
		\hspace{8pt}
		\subfigure[]{\includegraphics[width=0.3\textwidth]{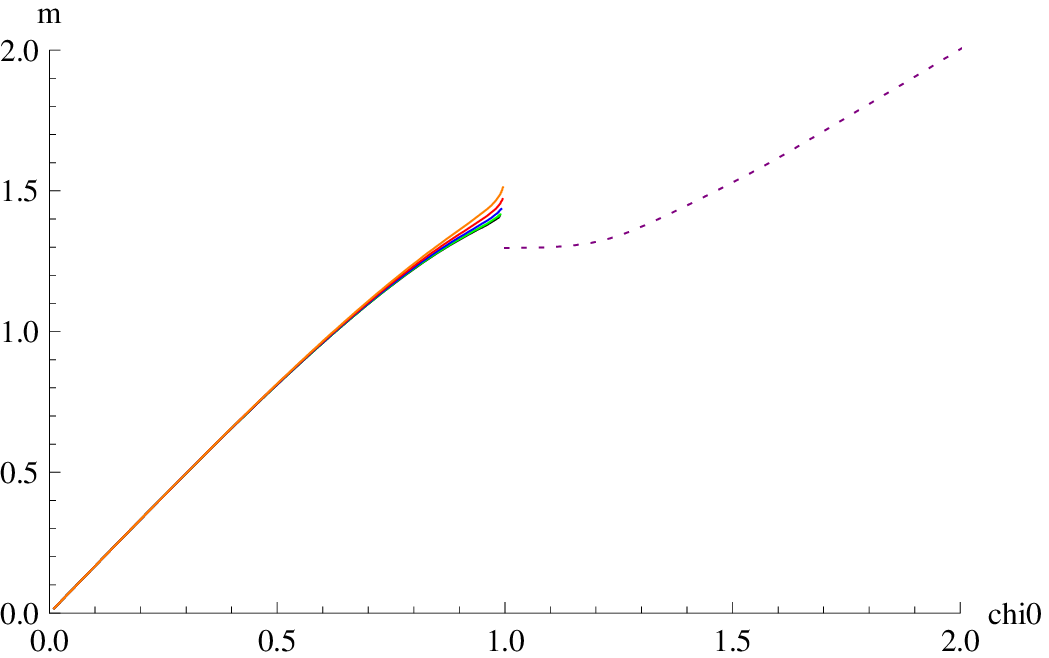}}
		\hspace{8pt}
		\subfigure[]{\includegraphics[width=0.3\textwidth]{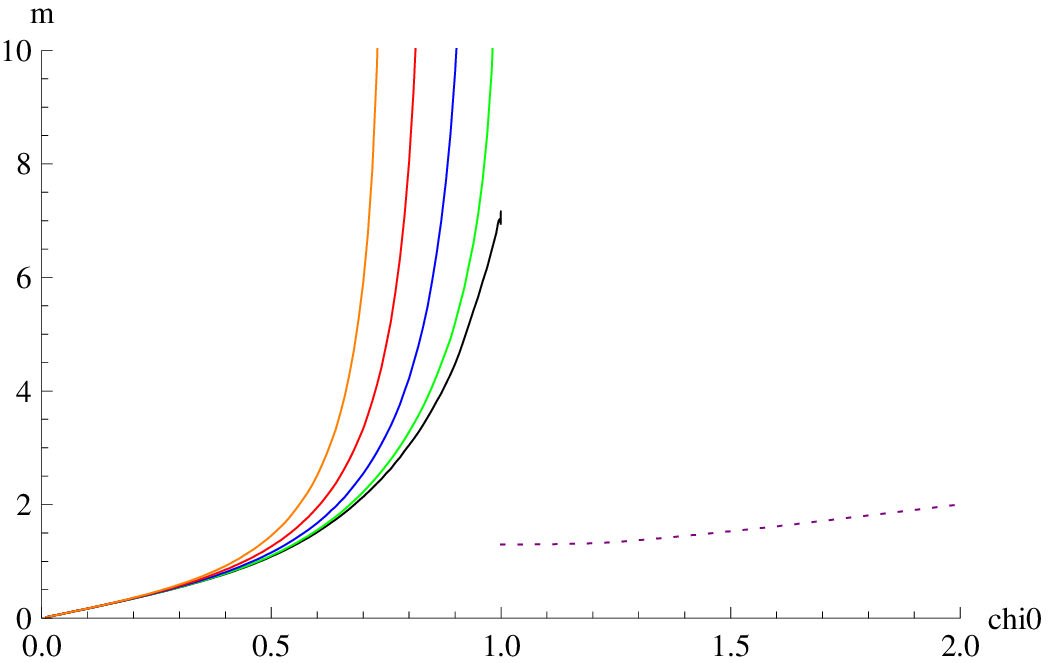}}
		\caption{The dimensionless mass parameter $m$ as defined in equation
			\eqref{dicmc} versus the horizon value of the black hole embeddings
			$\chi_0=\lim_{\rho\to 1}\chi$ between 0 to 1 and the asymptotic
			value of the Minkowski embeddings $L_0=\lim_{r\to 0}L$ between 1 and
			2 at baryon chemical potential $\mu^B/M_q=0.01$ (a),
			$\mu^B/M_q=0.1$ (b) and $\mu^B/M_q=0.8$ (c) for the case $N_f=2$.
			The dotted purple curve corresponds to Minkowski embeddings and the
			five other curves to black hole embeddings with isospin chemical
			potential $\mu^I=0$ (black), $\mu^I=\frac{1}{4}\mu^B$ (green),
			$\mu^I=\frac{1}{2}\mu^B$ (blue), $\mu^I=\frac{3}{4}\mu^B$ (red) and
			$\mu^I=\mu^B$ (orange).}	\label{fig:bgmgrand}
	\end{figure}

\begin{figure}
		\centering
		\psfrag{m}{$m$}
		\psfrag{c}{$c$}
		\subfigure[]{\includegraphics[width=0.3\textwidth]{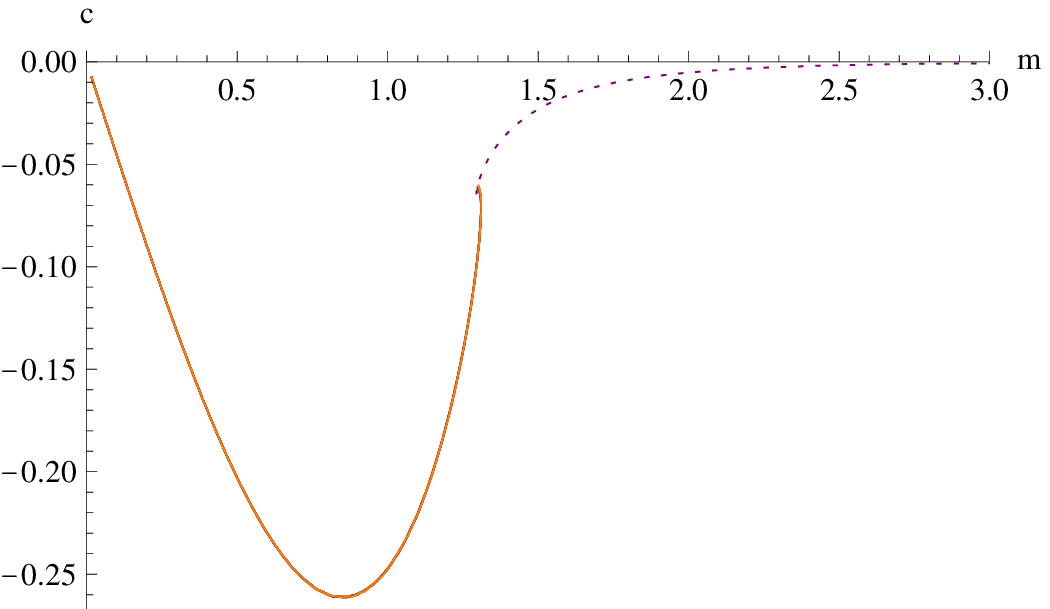}}
		\hspace{8pt}
		\subfigure[]{\includegraphics[width=0.3\textwidth]{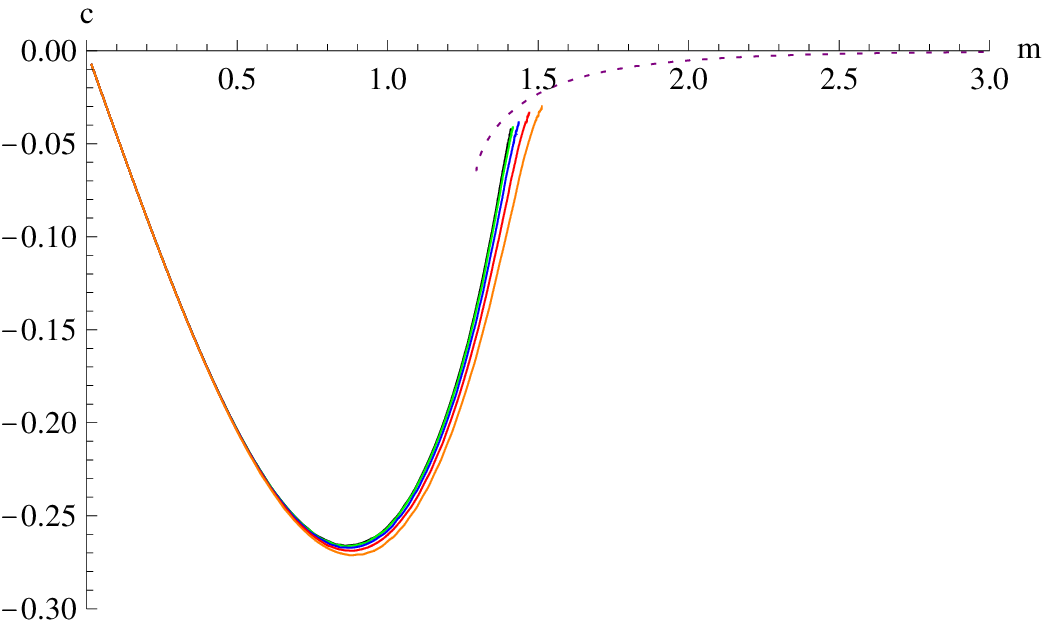}}
		\hspace{8pt}
		\subfigure[]{\includegraphics[width=0.3\textwidth]{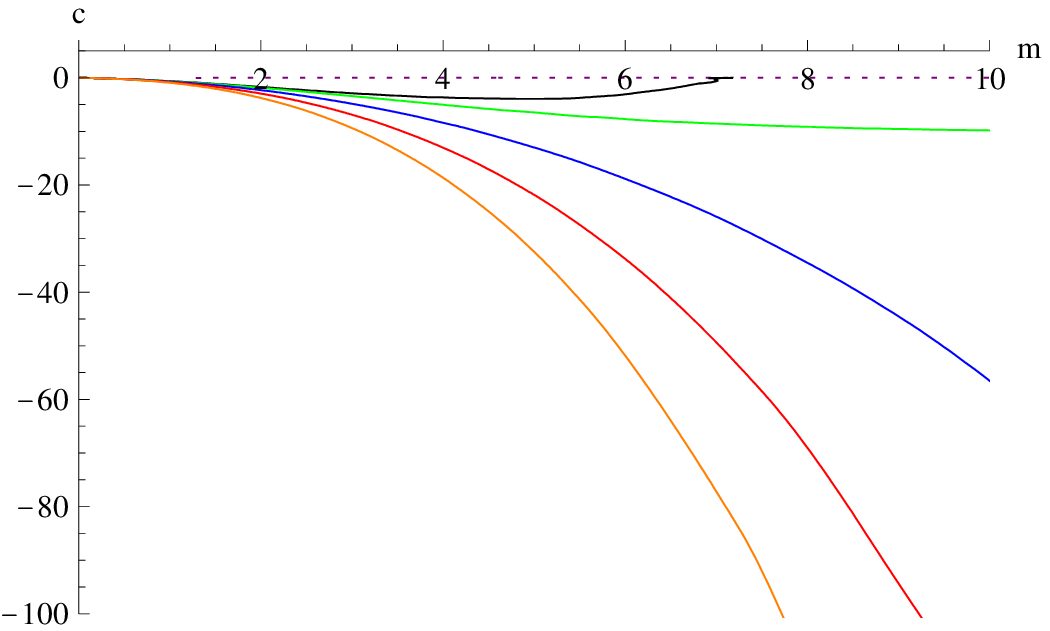}}
		\caption{The dimensionless chiral condensate $c$ versus the mass parameter
			$m$ as defined in equation \eqref{dicmc} at baryon chemical
			potential $\mu^B/M_q=0.01$ (a), $\mu^B/M_q=0.1$ (b) and
			$\mu^B/M_q=0.8$ (c) for the case $N_f=2$. The dotted purple curve
			corresponds to Minkowski embeddings and the five other curves to
			black hole embeddings with isospin chemical potential $\mu^I=0$
			(black), $\mu^I=\frac{1}{4}\mu^B$ (green), $\mu^I=\frac{1}{2}\mu^B$
			(blue), $\mu^I=\frac{3}{4}\mu^B$ (red) and $\mu^I=\mu^B$ (orange).}
		\label{fig:bgcgrand}
	\end{figure}

\begin{figure}
		\centering
		\psfrag{m}{$m$}
		\psfrag{dB}{$\dt^B$}
		\psfrag{dI}{$\dt^I$}
		\subfigure[]{\includegraphics[width=0.3\textwidth]{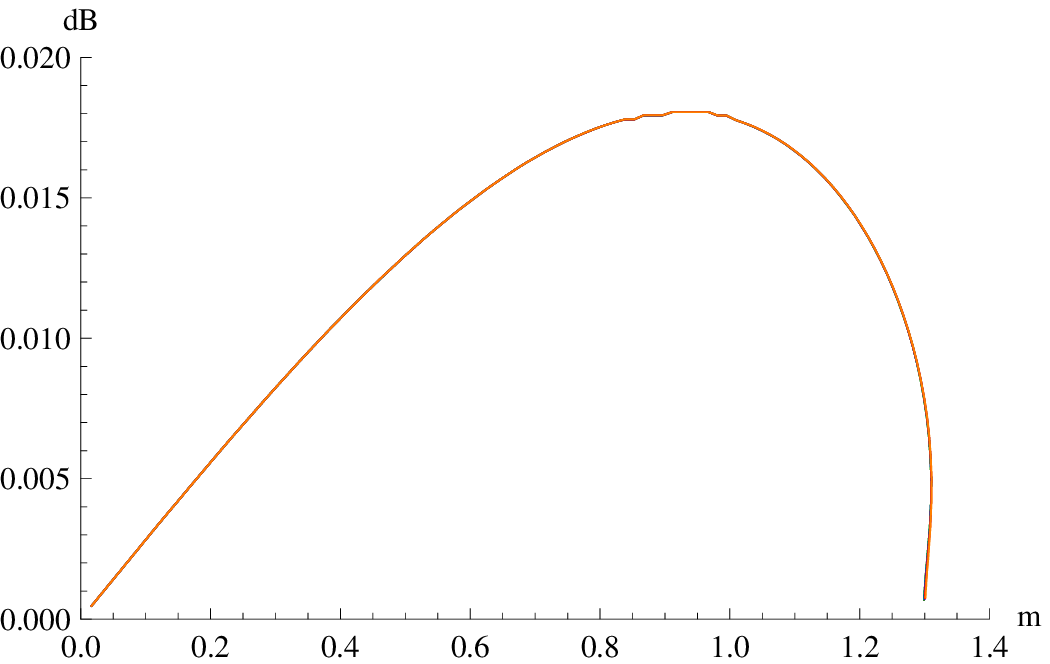}}
		\hspace{8pt}
		\subfigure[]{\includegraphics[width=0.3\textwidth]{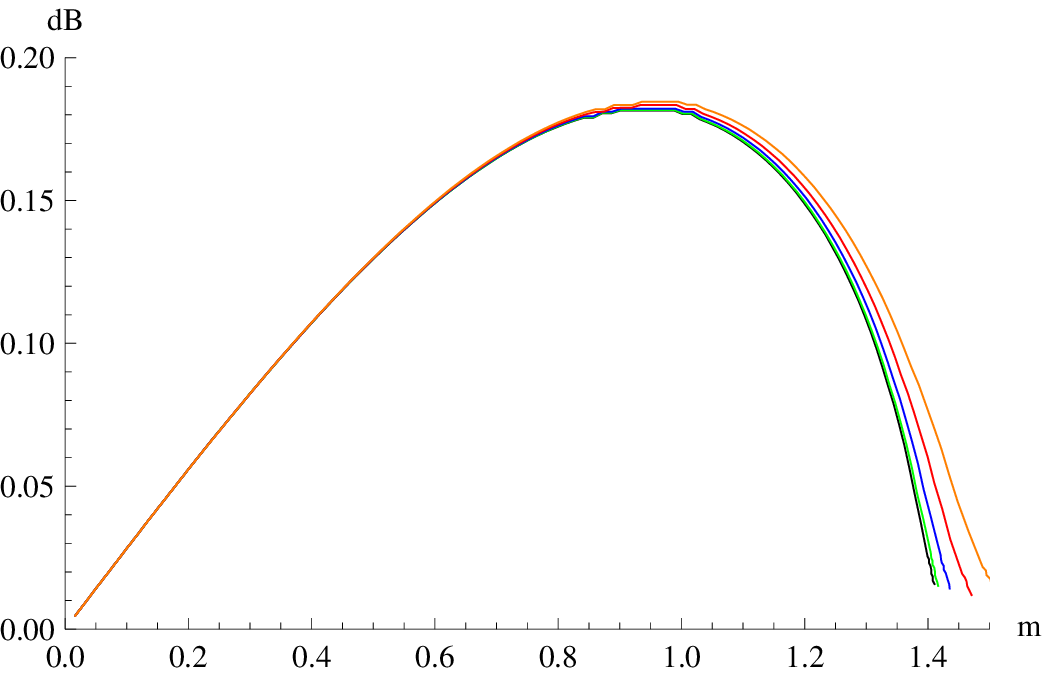}}
		\hspace{8pt}
		\subfigure[]{\includegraphics[width=0.3\textwidth]{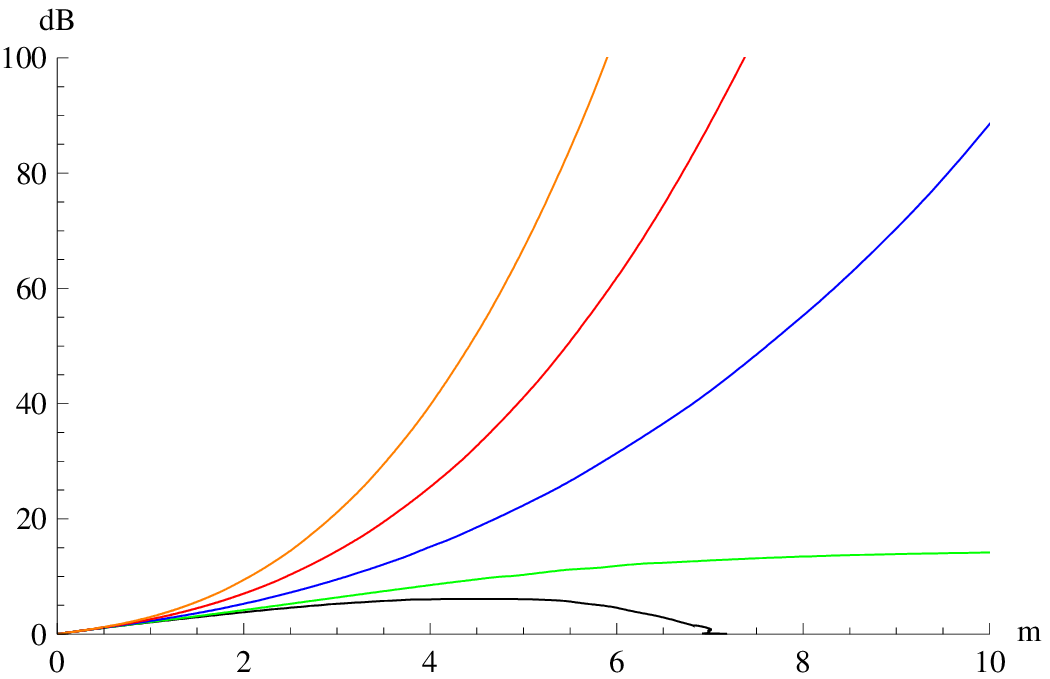}}
		\caption{The dimensionless baryon density $\dt^B$ versus the mass 
			parameter $m$ as defined in equation \eqref{dicmc} at baryon
			chemical potential $\mu^B/M_q=0.01$ (a), $\mu^B/M_q=0.1$ (b) and
			$\mu^B/M_q=0.8$ (c) for the case $N_f=2$. The five different curves
			correspond to black hole embeddings with isospin chemical potential
			$\mu^I=0$ (black), $\mu^I=\frac{1}{4}\mu^B$ (green),
			$\mu^I=\frac{1}{2}\mu^B$ (blue), $\mu^I=\frac{3}{4}\mu^B$ (red) and
			$\mu^I=\mu^B$ (orange). In the Minkowski phase, the baryon density
			is always zero and is therefore not shown in the figures.}
		\label{fig:bgdBgrand}
	\end{figure}

\begin{figure}
		\centering
		\psfrag{m}{$m$}
		\psfrag{dB}{$\dt^B$}
		\psfrag{dI}{$\dt^I$}
		\subfigure[]{\includegraphics[width=0.3\textwidth]{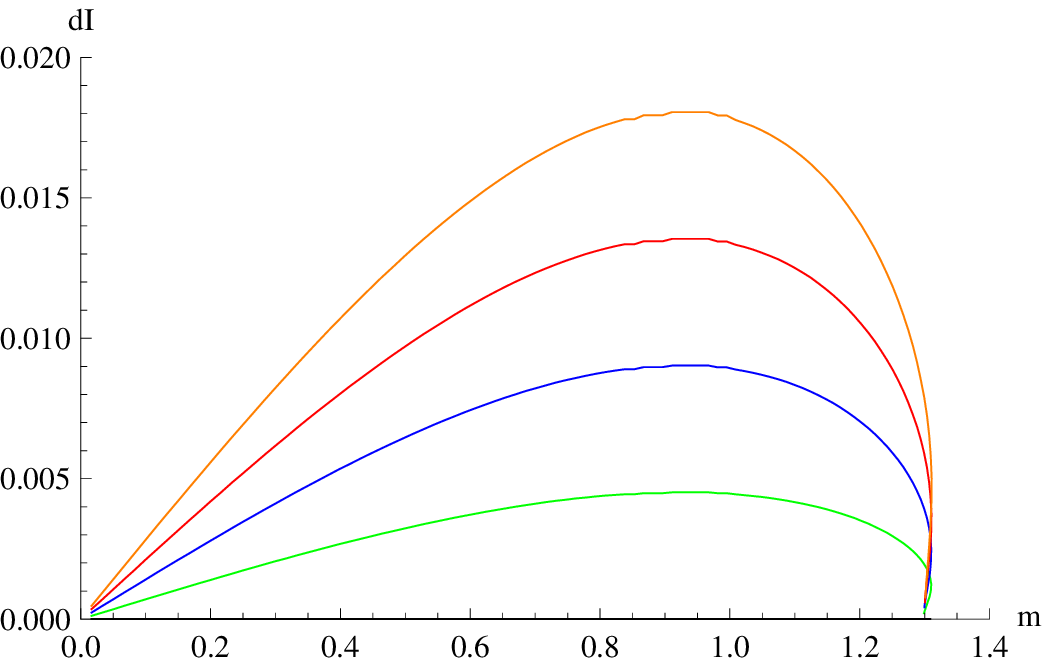}}
		\hspace{8pt}
		\subfigure[]{\includegraphics[width=0.3\textwidth]{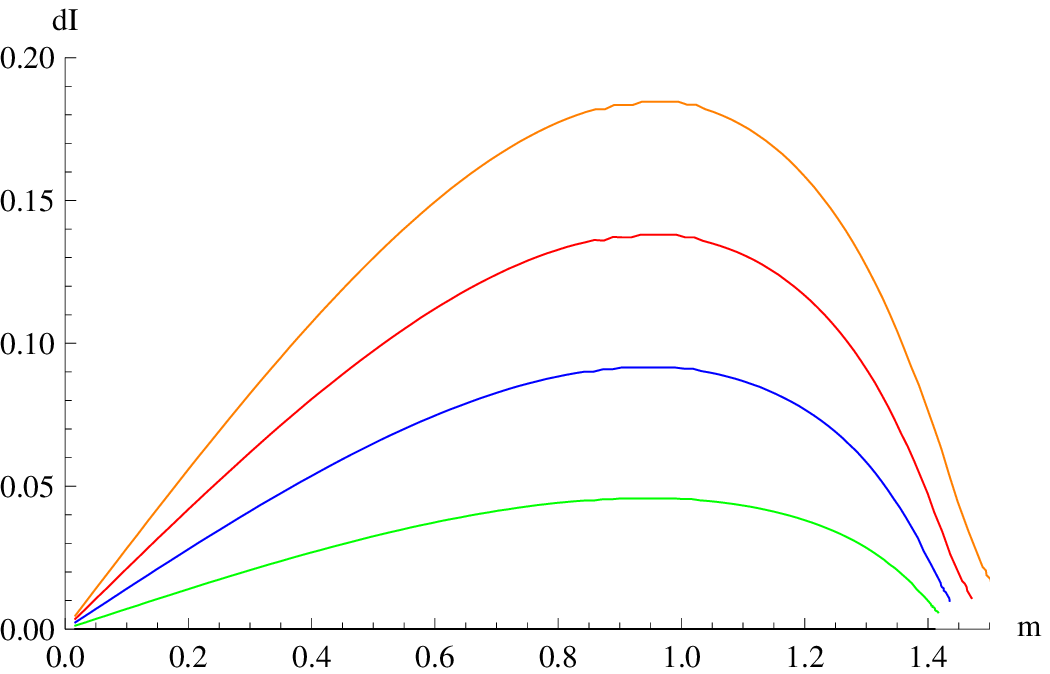}}
		\hspace{8pt}
		\subfigure[]{\includegraphics[width=0.3\textwidth]{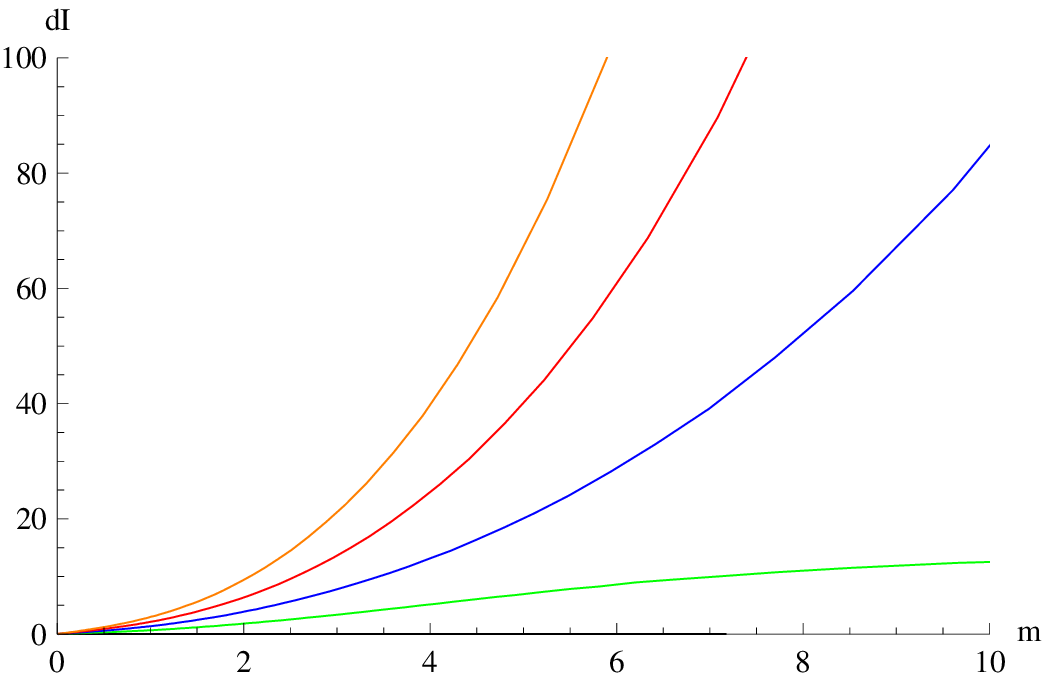}}
		\caption{The dimensionless isospin density $\dt^I$ versus the mass
			parameter $m$ as defined in equation \eqref{dicmc} at baryon
			chemical potential $\mu^B/M_q=0.01$ (a), $\mu^B/M_q=0.1$ (b) and
			$\mu^B/M_q=0.8$ (c) for the case $N_f=2$. The five different curves
			correspond to black hole embeddings with isospin chemical potential
			$\mu^I=0$ (black), $\mu^I=\frac{1}{4}\mu^B$ (green),
			$\mu^I=\frac{1}{2}\mu^B$ (blue), $\mu^I=\frac{3}{4}\mu^B$ (red) and
			$\mu^I=\mu^B$ (orange). In the Minkowski phase, the isospin density
			is always zero and is therefore not shown in the figures. For a zero
			isospin chemical potential $\mu^I=0$, the isospin density $\dt^I$ is
			also zero and therefore coincides with the $m$-axis.}
		\label{fig:bgdIgrand}
	\end{figure}

	In fig.~\ref{fig:bgmgrand}, \ref{fig:bgcgrand}, \ref{fig:bgdBgrand} and
	\ref{fig:bgdIgrand} we show numerical results for the background fields at
	constant chemical potential. Fig.~\ref{fig:bgmgrand} shows the mass
	parameter $m$ versus the horizon value of the black hole embeddings $\chi_0$
	between 0 and 1 and the asymptotic value of the Minkowski embeddings $L_0$
	between 1 and 2 at different baryon and isospin chemical potentials. In
	these figures we see that there is a region where the Minkowski and the
	black hole embeddings generate the same mass parameter $m$. We will show in
	section \ref{phasediagramgrand} that in this overlap there is a first order
	phase transition as in the case of zero chemical potential
	\cite{Babington:2003vm, Mateos:2007vn}. For small chemical potentials (see
	fig.~\ref{fig:bgmgrand} (a)), the mass parameter decreases in the black hole
	phase for $\chi_0\to 1$ as known from \cite{Babington:2003vm,
	Mateos:2007vn}. However, for larger chemical potentials (see
	fig.~\ref{fig:bgmgrand} (b)), the mass parameter $m$ increases monotonically
	as we increase $\chi_0$. In both cases, fig.~\ref{fig:bgmgrand} (a) and (b)
	demonstrate that the mass parameter depends linearly on the the asymptotic
	values $\chi_0$ and $L_0$ in a large region. The only non-linear behavior is
	in the region where both embeddings can be constructed. For chemical
	potentials $(\mu^B+\mu^I)/M_q>1$ (fig.~\ref{fig:bgmgrand} (c)), the mass parameter
   diverges as $\chi_0\to 1$ as in the case of constant densities. This corresponds to
   the fact that in the canonical ensemble, there are black hole embeddings for all mass
   parameters $m$. In the grand canonical ensemble, however, there are also Minkowski embeddings.

	Fig.~\ref{fig:bgcgrand} shows the chiral condensate $c$ versus the mass
	parameter $m$ at different baryon and isospin chemical potentials. For small
	chemical potentials (fig.~\ref{fig:bgcgrand} (a) and (b)), we see a similar
	behavior as in the case of zero chemical potentials \cite{Babington:2003vm,
	Mateos:2007vn}. By increasing the chemical potentials, the spiral behavior
	of the chiral condensate vanishes since the mass parameter increases
	monotonically as we increase $\chi _0$. However, there is still a region
	where the chiral condensate is multivalued since there are black hole and
	Minkowski embeddings which generate the same mass $m$. For large chemical
	potentials $(\mu^B+\mu^I)/M_q>1$ (fig.~\ref{fig:bgcgrand} (c)), the chiral
	condensate diverges as we increase $m$. Since for smaller chemical
	potentials $(\mu^B+\mu^I)/M_q<1$ the chiral condensate in the black hole
	phase approaches the value for Minkowski embeddings at a finite mass $m$,
        the divergent behavior of the chiral condensate for large chemical potentials
	$(\mu^B+\mu^I)/M_q>1$ indicates that there is no phase transition between
        black hole and	Minkowski embeddings anymore.

	Fig.~\ref{fig:bgdBgrand} and \ref{fig:bgdIgrand} show the baryon and isospin
	density versus the mass parameter $m$ calculated by the inversion of
	equation (\ref{mu}) at different baryon and isospin chemical potentials.
	These figures show that for all chemical potentials $\mu^{B/I}$, the baryon
	and the isospin density are zero at zero mass $m=0$. For small chemical
	potential, fig.~\ref{fig:bgdBgrand} (a), (b) and \ref{fig:bgdIgrand} (a),
	(b) show that the densities increase until they reach a maximum at $m\sim
	1$. Beyond the maximum the densities decreases rapidly, but they do not
	reach to zero at the largest $m$ which can be constructed by black hole
	embeddings. Therefore, there is also a discontinuous step in the densities
	as in the chiral condensate at the transition from black hole embeddings to
	Minkowski embeddings. For large chemical potentials $(\mu^B+\mu^I)/M_q>1$
	(fig.~\ref{fig:bgdBgrand} (c) and \ref{fig:bgdIgrand} (c)), the densities
	diverge as we increase $m$, as does the chiral condensate.

\section{Canonical D$7$-brane thermodynamics}
\label{sec:thermoCan}
\subsection{The canonical ensemble:\\ free energy, entropy, energy and speed of sound}
In this section we study the thermodynamic behavior of the quark sector at finite baryon and isospin density which is dual to thermal contributions of the D$7$-branes. Since the partition function $Z$ of the field theory is given according the AdS/CFT dictionary by
\begin{equation}
Z=\ee^{-S_{\text{on-shell}}}\,,
\end{equation}
with the Euclidean on-shell supergravity action $S_{\text{on-shell}}$, the thermodynamical potential, in the canonical ensemble the free energy $F$, is proportional to the Euclidean on-shell action
\begin{equation}
F=-T\ln Z=TS_{\text{on-shell}}\,.
\end{equation}
 To calculate the thermal contributions of the D$7$-branes $F_7$, we have to determine the Euclidean DBI action \eqref{action} on-shell.\par
First  we perform a Wick rotation in the time direction to obtain the Euclidean DBI-action\footnote{The Euclidean time must be periodic with period $\beta=1/T$, such that the geometry is non-singular.}. Next we must renormalize this action by adding the appropriate counterterms $I_{\text{ct}}$ (see \cite{Karch:2005ms} for a review) since it diverges on-shell. Since these counterterms do not depend on the finite densities, we can write them as in \cite{Mateos:2007vn, Kobayashi:2006sb},
\begin{equation}
I_{\text{ct}}=-\frac{\caln_\lambda}{4}\left[\left(\rho_{\text{max}}^2-m^2\right)^2-4mc\right]\,,
\end{equation}
where $\rho_{\text{max}}$ is the UV-cutoff and
\begin{equation}
\caln_\lambda=\frac{T_{D7}V_3\vol(S^3)N_f\vrho_H^4}{4T}=\frac{\lambda N_cN_f V_3T^3}{32}\,,
\end{equation}
whit the Minkowski space volume $V_3$. Then the renormalized Euclidean on-shell action $I_R$ may simply be written as
\begin{equation}
\frac{I_R}{\caln_\lambda}=\frac{1}{N_f} G(m,\mut)-\frac{1}{4}\left[\left(\rho_{\text{min}}^2-m^2\right)^2-4mc\right]\,,\eqlabel{renormaction}
\end{equation}
where $\rho_{\text{min}}$ determines the minimal value of the coordinate $\rho$ on the D$7$-branes, \ie$\rho_{\text{min}}=1$ for black hole embeddings and $\rho_{\text{min}}=L_0$ for Minkowski embeddings and
\begin{equation}
\begin{split}
G(m,\mut)=\int_{\rho_{\text{min}}}^{\infty}\!\dd\rho\;\Bigg(&\rho^3f\ft\left(1-\chi^2\right)\\
&\times\sum_{i=1}^{N_f}\sqrt{1-\chi^2+\rho^2(\del_\rho\chi)^2-2\frac{\ft}{f^2}(1-\chi^2)(\del_\rho \Xt_i)^2}\\
&-N_f\left(\rho^3-\rho m^2\right)\Bigg)\,.
\end{split}\eqlabel{G(m)}
\end{equation}

Since this action depends on the background fields $\chi$ and $X_i$ and therefore on the mass parameter $m$ and the chemical potential $\mu_i$, it is proportional to the grand potential in the grand canonical ensemble. To get the free energy in the canonical ensemble, we must perform a Legendre transformation of the on-shell action $I_R$ as we did in equation (\ref{actiontrans}). The Legendre transformed on-shell action $\tilde{I}_R$ is given by
\begin{equation}
\frac{\tilde{I}_R}{\caln_\lambda}=\frac{1}{N_f}\tilde{G}(m,\dt)-\frac{1}{4}\left[\left(\rho_{\text{min}}^2-m^2\right)^2-4mc\right]\,,
\end{equation}
where
\begin{equation}
\begin{split}
\tilde{G}(m,\dt)=\int_{\rho_{\text{min}}}^\infty\!\dd\rho\;\Bigg(&\rho^3f\ft\left(1-\chi^2\right)\sqrt{1-\chi^2+\rho^2(\del_\rho\chi)^2}\\
&\times\sum_{i=1}^{N_f}\sqrt{1+\frac{8\dt_i^2}{\rho^6\ft^3(1-\chi^2)^3}}-N_f\left(\rho^3-\rho m^2\right)\Bigg)\,.\eqlabel{Gtilde}
\end{split}
\end{equation}
In the following we use this action to calculate the main thermodynamical quantities. The difference to the results obtained in \cite{Kobayashi:2006sb} is induced by the sum over the $N_f$ different densities in equation \eqref{Gtilde}.

\subsubsection{The free energy}
The contribution of the D$7$-branes to the Helmholtz free energy is
\begin{equation}
F_7=T\tilde{I}_R=\frac{\lambda N_cN_f V_3T^4}{32}\calf_7(m,\dt)\eqlabel{freeenergy}
\end{equation}
with the dimensionless quantity
\begin{equation}
\calf_7(m,\dt)=\frac{\tilde{I}_R}{\caln_\lambda}=\frac{1}{N_f}\tilde{G}(m,\dt)-\frac{1}{4}\{(\rho_{\text{min}}^2-m^2)^2-4mc\}\,.\eqlabel{freeenergydim}
\end{equation}
The dimensionless quantity $\calf_7$ determines the dependence of the free energy $F$ of the complete setup on the quark mass $M_q$ and the quark densities $n_q$ at a fixed temperature $T$.\par
Using the example of the free energy we study how a thermodynamic quantity of the complete setup is composed of the quantities of the subsystems. The parts from the D3-branes $F_3$, which is the known quantity for the conformal $\caln=4$ SYM theory, and from D$7$-branes $F_7$ form the free energy of the complete setup
\begin{equation}
\begin{split}
F&=F_3+F_7=-\frac{\pi^2}{8} N_c^2V_3T^4+\frac{\lambda N_cN_f V_3T^4}{32}\calf_7(m,\dt)\\
&=-\frac{\pi^2}{8}N_c^2V_3T^4\left(1-\frac{\lambda N_f}{4\pi^2N_c}\calf_7(m,\dt)\right)\,.
\end{split}\eqlabel{completefreeenergy}
\end{equation}
Here we only get the first order contribution in $N_f/N_c$ since we work in the probe brane limit $N_f\ll N_c$. In the following we define quantities similar to $\calf_7$ for the other thermodynamic quantities and denote them also with a calligraphic letter. Note that these thermal contributions of the D$7$-branes vanish in the limit where supersymmetry and conformal invariance is restored. This limit is given by zero densities and large mass parameters $m\to\infty$.

\subsubsection{The entropy}
As in \cite{Mateos:2007vn, Kobayashi:2006sb}, the contribution to the entropy by the D$7$-branes is given by
\begin{equation}
S_7=-\frac{\del F_7}{\del T}=-\pi R^2\frac{\del F_7}{\del \vrho_H}=\frac{\lambda N_cN_fV_3 T^3}{32}\cals_7(m,\dt)\,,
\end{equation}
with the mass and density dependent part
\begin{equation}
\cals_7(m,\dt)=-4\calf_7(m,\dt)+\frac{12}{N_f}\sum_{i=1}^{N_f}\dt_i\mut_i-2mc\,.\eqlabel{entropydim}
\end{equation}
In comparison to the case without isospin density \cite{Kobayashi:2006sb}, there is sum over the $N_f$ different densities.
\subsubsection{The energy}
Using the thermodynamic relation $E=F+TS$, we calculate the energy
\begin{equation}
E_7=\frac{\lambda N_c N_fV_3 T^4}{32}\cale_7(m,\dt)\eqlabel{energy}\,,
\end{equation}
with the dimensionless quantity
\begin{equation}
\cale_7(m,\dt)=-3\calf_7(m,\dt)+\frac{12}{N_f}\sum_{i=1}^{N_f}\dt_i\mut_i-2mc\,.\eqlabel{energydim}
\end{equation}
Again a sum over the $N_f$ densities appears in the equation above, which is due to the finite isospin densities.
\subsubsection{The speed of sound}\label{sec:speedofsound}
The speed of sound is given by
\begin{equation}
v_s^2=\frac{V_3\del P}{\del E}=-\frac{\del (\Omega_3+\Omega_7)}{\del T}\cdot\frac{\del T}{\del (E_3+E_7)}\,,
\end{equation}
where the pressure $P$ is given by $PV_3=-\Omega=-(\Omega_3+\Omega_7)$. $\Omega$ and $E$ are the grand potential and energy of the complete system. $\Omega_3$ and $E_3$ are the contribution to the grand potential and energy of the D3-branes, respectively. Using the thermodynamic relations $S=-\left(\frac{\del F}{\del T}\right)_{V,N}=-\left(\frac{\del \Omega}{\del T}\right)_{V,\mu}$ and $c_v=\left(\frac{\del E}{\del T}\right)_{V,N}$, we may rewrite the speed of sound in the canonical ensemble
\begin{equation}
v_s^2=\frac{S_3+S_7}{c_{v3}+c_{v7}}\,.
\end{equation}
Using equation (\ref{energydim}) the specific heat $c_{v7}$ becomes
\begin{equation}
c_{v7}=3S_7-\frac{3\lambda N_cN_f V_3T^3}{16}\left(mc-\frac{1}{3}m^2\frac{\del c}{\del m}+\frac{2}{N_f}\sum_{i=1}^{N_f}\dt_i\left(m\frac{\del\mut_i}{\del m}-4\mut_i\right)\right)\,,
\end{equation}
where we used $\del_T=-\frac{m}{T}\del_m$ (see equation (\ref{dicmc})). With the contributions of the D$3$-branes given by the thermodynamics of the conformal $\caln=4$ SYM theory,
\begin{equation}
S_3=\frac{\pi^2}{2}V_3N_c^2T^3\qquad\text{and}\qquad c_{v3}=3S_3\,,
\end{equation}
we evaluate the speed of sound to first order in $N_f/N_c$
\begin{equation}
v_s^2=\frac{1}{3}\left[1+\frac{\lambda N_f}{8\pi^2N_c}\calv_s^2(m,\dt)\right]\,,
\end{equation}
with
\begin{equation}
\calv_s^2(m,\dt)=mc-\frac{1}{3}m^2\frac{\del c}{\del m}+\frac{2}{N_f}\sum_{i=1}^{N_f}\dt_i\left(m\frac{\del\mut_i}{\del m}-4\mut_i\right)\eqlabel{speeddim}\,.
\end{equation}
Notice that the first order correction in $N_f/N_c$ is consistent with the
approximation of probe branes (cf. with equation (\ref{completefreeenergy})).
The last term reflects the presence of the finite baryon and isospin density.
The dependence on the mass parameter $m$ and the chiral condensate $c$ is
given by the result without densities \cite{Mateos:2007vn}. 

\subsubsection{Numerical results for the free energy, entropy, energy and the speed of sound}
\label{sec:canNumRes}  
\begin{figure}
\centering
\psfrag{m}{$m$}
\psfrag{F}{$\calf_7$}
\subfigure[]{\includegraphics[width=0.45\textwidth]{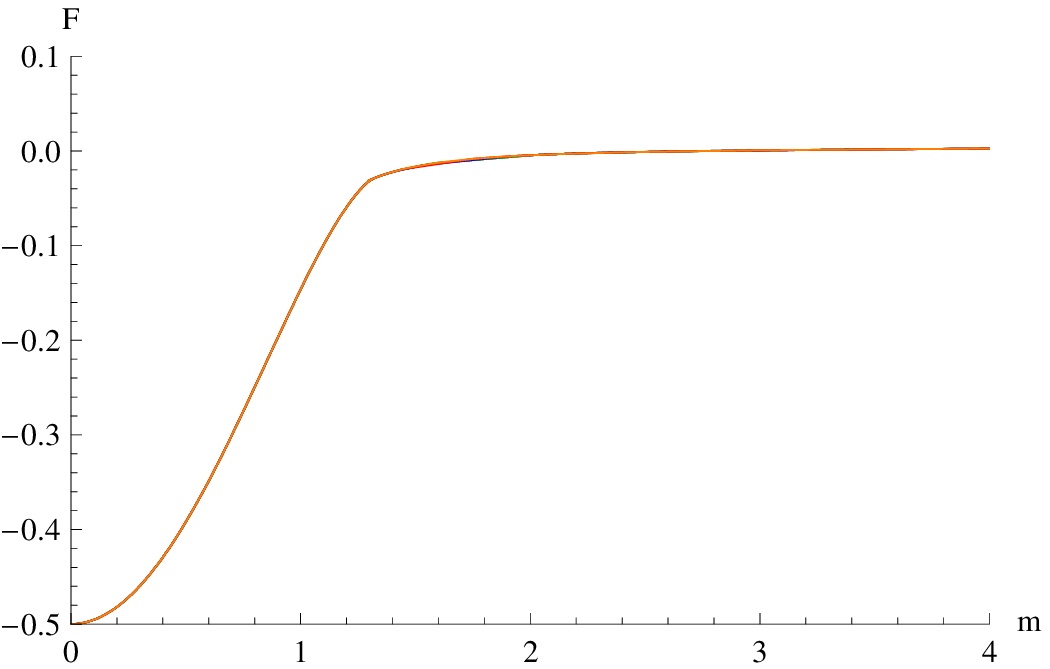}}
\hspace{8pt}
\subfigure[]{\includegraphics[width=0.45\textwidth]{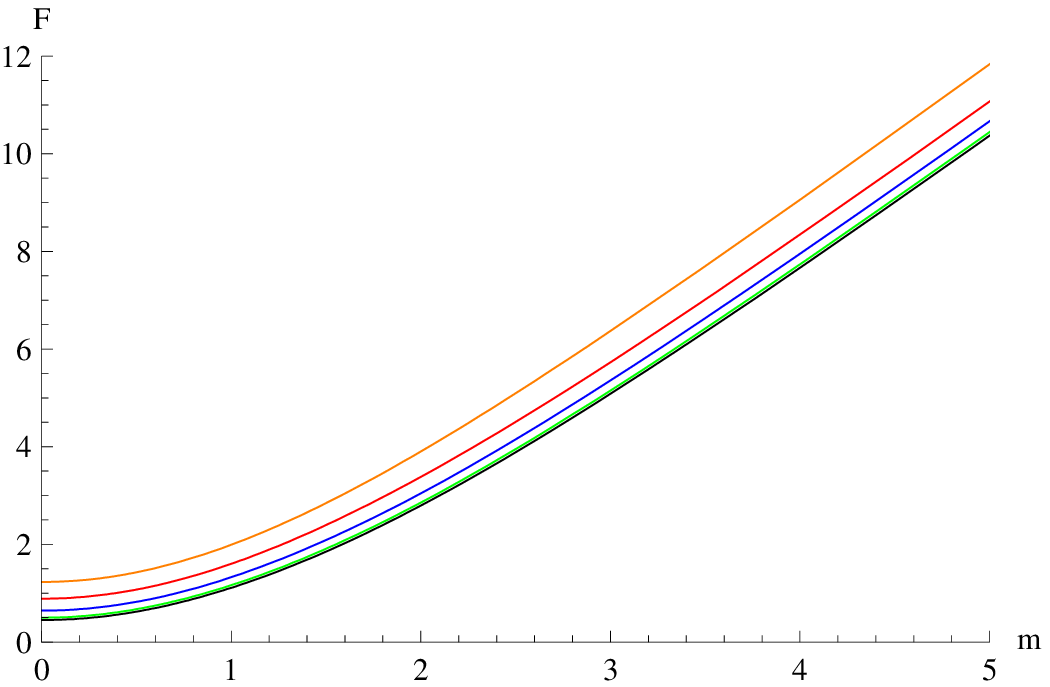}}
\caption{The dimensionless free energy $\calf_7$ versus the mass parameter $m$ as defined in equation \eqref{dicmc} at baryon density $\dt^B=5\cdot 10^{-5}$ (left) and $\dt^B=2$ (right) for the case $N_f=2$. The five different curves in each figure correspond to $\dt^I=0$ (black), $\dt^I=\frac{1}{4}\dt^B$ (green), $\dt^I=\frac{1}{2}\dt^B$ (blue), $\dt^I=\frac{3}{4}\dt^B$ (red) and $\dt^I=\dt^B$ (orange).}\label{fig:Fcan}
\end{figure}

\begin{figure}
\centering
\psfrag{m}{$m$}
\psfrag{S}{$\cals_7$}
\subfigure[]{\includegraphics[width=0.45\textwidth]{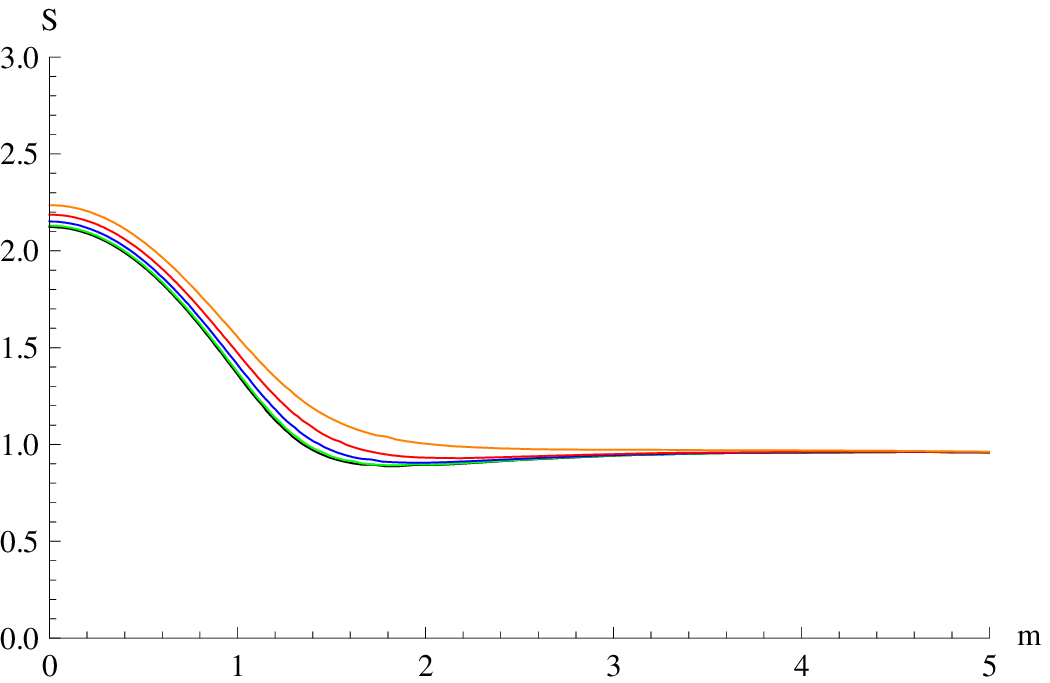}}
\hspace{8pt}
\subfigure[]{\includegraphics[width=0.45\textwidth]{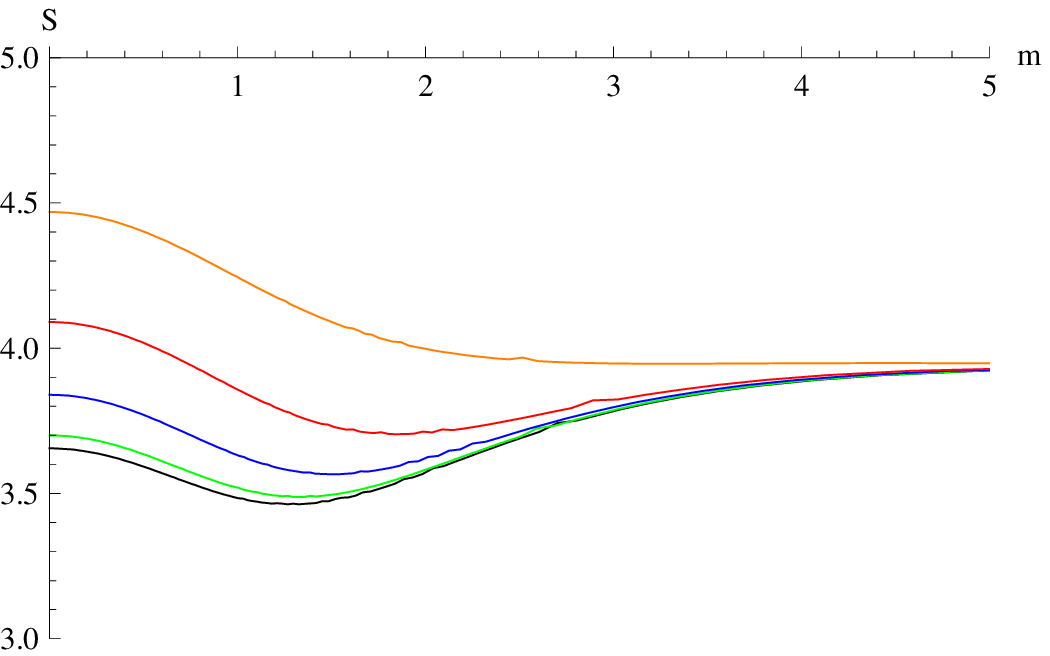}}
\caption{The dimensionless entropy $\cals_7$ versus the mass parameter $m$ as defined in equation \eqref{dicmc} at baryon density $\dt^B=0.5$ (left) and $\dt^B=2$ (right) for the case $N_f=2$. The five different curves in each figure correspond to $\dt^I=0$ (black), $\dt^I=\frac{1}{4}\dt^B$ (green), $\dt^I=\frac{1}{2}\dt^B$ (blue), $\dt^I=\frac{3}{4}\dt^B$ (red) and $\dt^I=\dt^B$ (orange).}\label{fig:Scan}
\end{figure}

\begin{figure}
\centering
\psfrag{m}{$m$}
\psfrag{E}{$\cale_7$}
\subfigure[]{\includegraphics[width=0.45\textwidth]{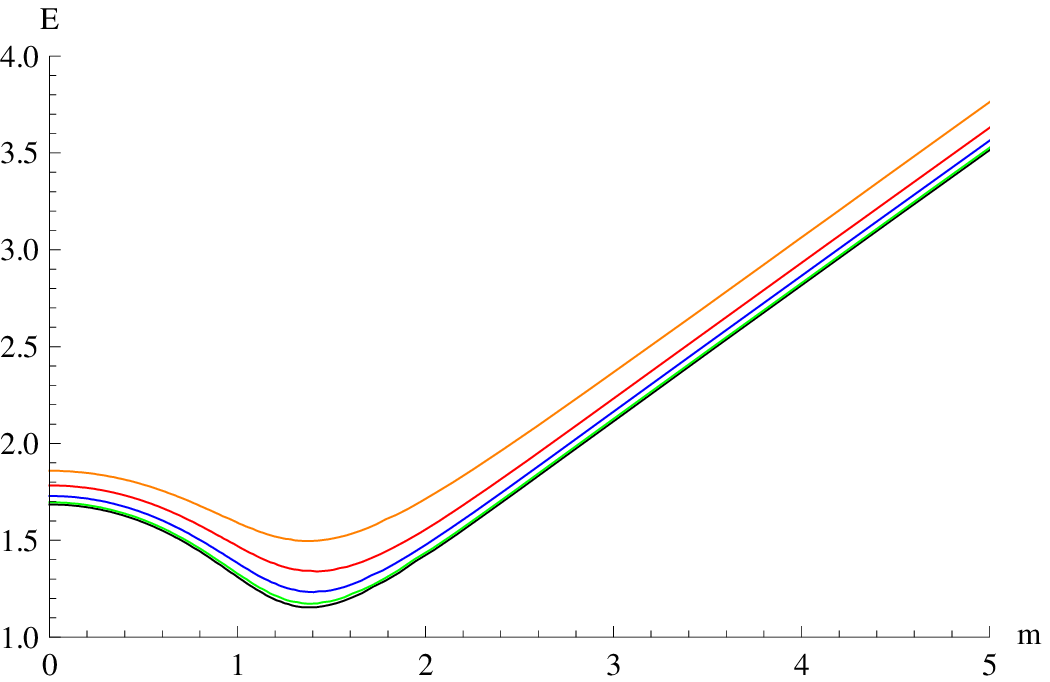}}
\hspace{8pt}
\subfigure[]{\includegraphics[width=0.45\textwidth]{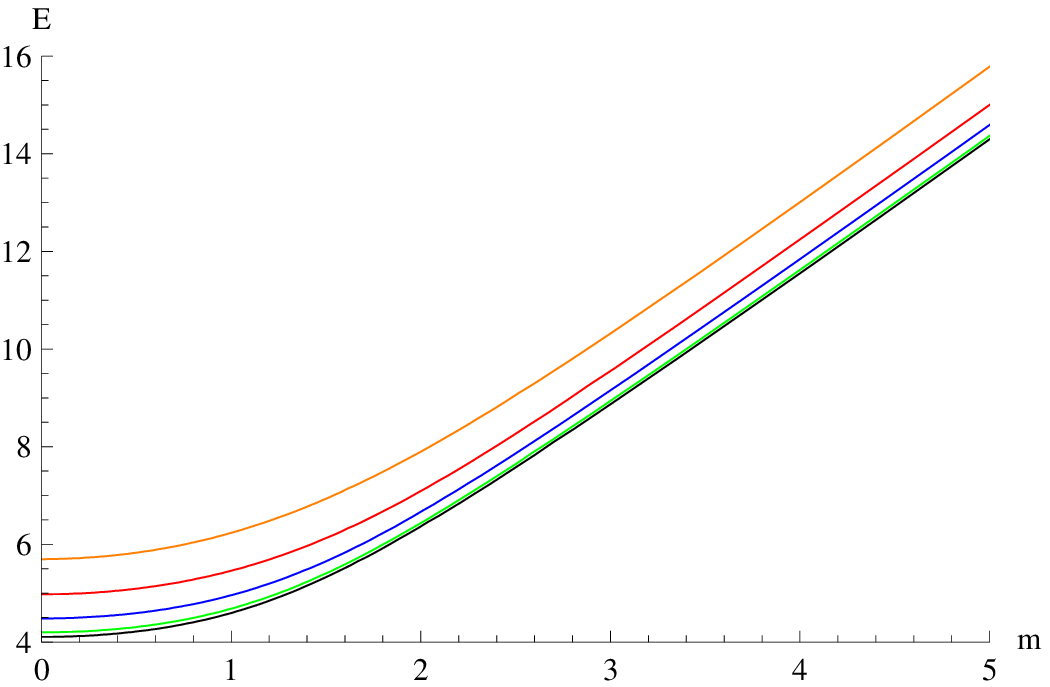}}
\caption{The dimensionless energy $\cale_7$ versus the mass parameter $m$ as defined in equation \eqref{dicmc} at baryon density $\dt^B=0.5$ (left) and $\dt^B=2$ (right) for the case $N_f=2$. The five different curves in each figure correspond to $\dt^I=0$ (black), $\dt^I=\frac{1}{4}\dt^B$ (green), $\dt^I=\frac{1}{2}\dt^B$ (blue), $\dt^I=\frac{3}{4}\dt^B$ (red) and $\dt^I=\dt^B$ (orange).}\label{fig:Ecan}
\end{figure}

\begin{figure}[t]
\centering
\psfrag{dvs2}[b]{$\calv_s^2$}
\psfrag{m}{$m$}
\subfigure[]{\includegraphics[width=0.3\textwidth]{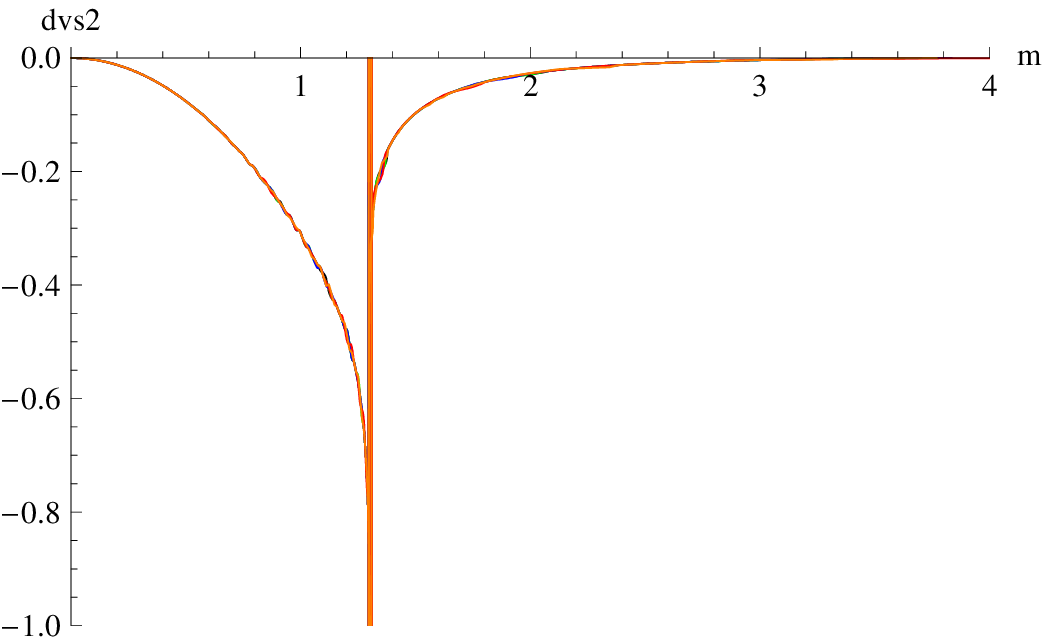}}
\hspace{8pt}
\subfigure[]{\includegraphics[width=0.3\textwidth]{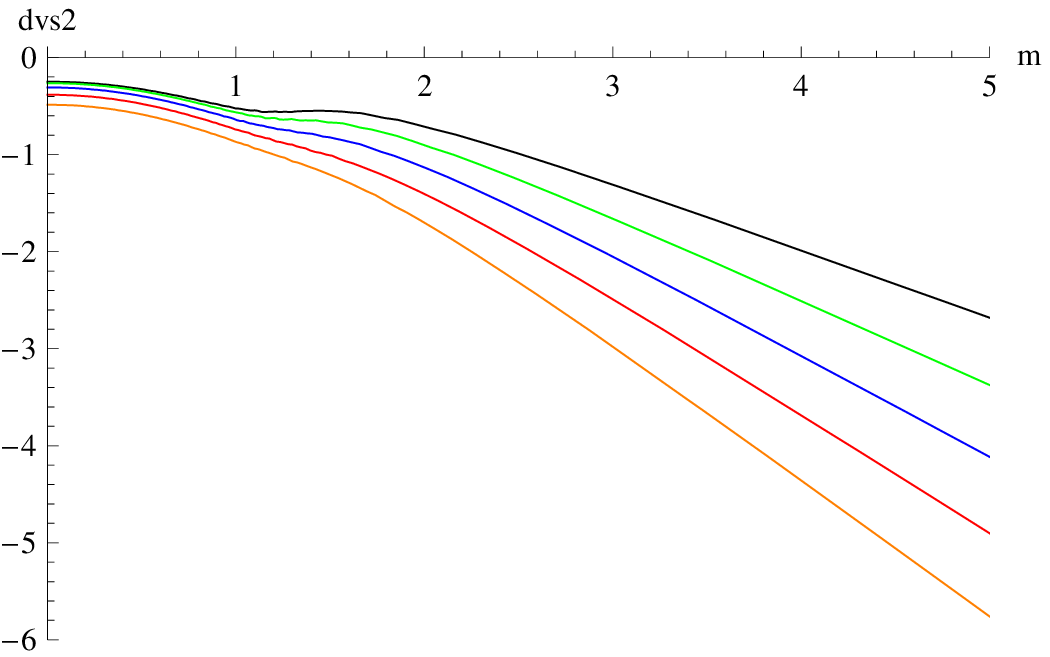}}
\hspace{8pt}
\subfigure[]{\includegraphics[width=0.3\textwidth]{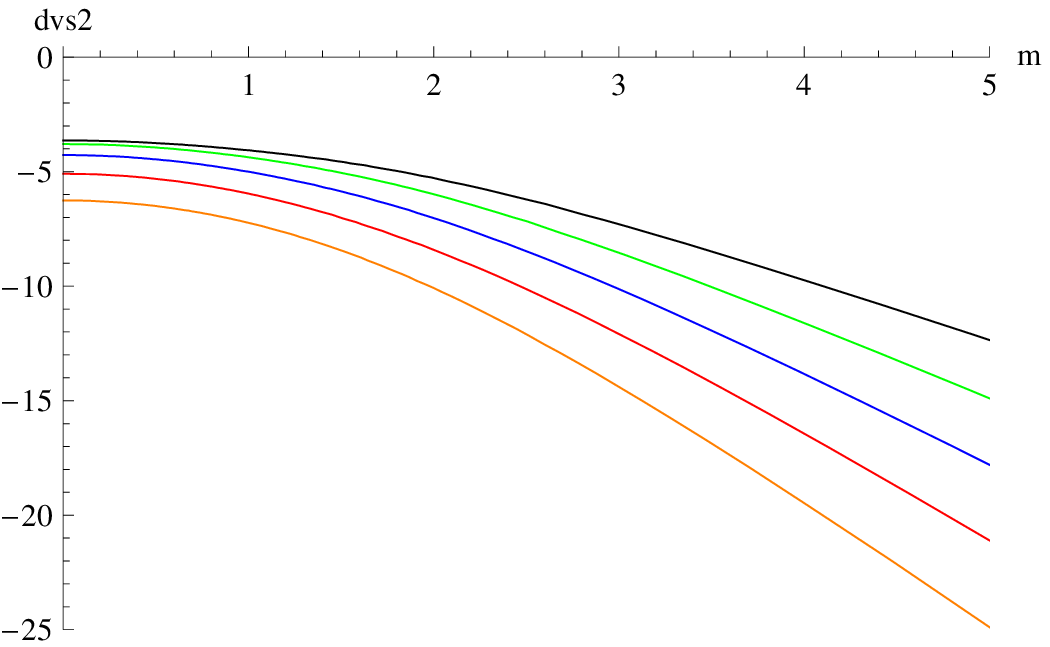}}
\caption{The dimensionless speed of sound relative to the conformal case $\calv_s^2$ versus the mass parameter $m$ as defined in equation \eqref{dicmc} at baryon density $\dt^B=5\cdot 10^{-5}$ (a), $\dt^B=0.5$ (b) and $\dt^B=2$ (c) for the case $N_f=2$. The five different curves in each figure correspond to $\dt^I=0$ (black), $\dt^I=\frac{1}{4}\dt^B$ (green), $\dt^I=\frac{1}{2}\dt^B$ (blue), $\dt^I=\frac{3}{4}\dt^B$ (red) and $\dt^I=\dt^B$ (orange).}\label{fig:Vcan}
\end{figure}

In fig.~\ref{fig:Fcan}, \ref{fig:Scan}, \ref{fig:Ecan} and \ref{fig:Vcan} we present numerical results for the mass and density dependent part of the thermodynamic quantities at different baryon and isospin densities for $N_f=2$.\par 
Fig.~\ref{fig:Fcan} shows the free energy $\calf_7$ versus the mass parameter $m$ at different densities as calculated from equation (\ref{freeenergydim}). For small densities (see fig.~\ref{fig:Fcan} (a)) we see again that the low temperature black hole embeddings have the same free energy as the zero-density Minkowski embeddings. Since the free energy determines the thermodynamics entirely, the same effect is also seen in the other thermodynamic quantities and we therefore do not consider them at small densities. For larger densities (see fig.~\ref{fig:Fcan} (b)) we see a displacement due to the isospin densities, which is constant over a wide range. Moreover, the free energy behaves linearly for $m\gtrsim 2$. The increase of the free energy as $m\to\infty$ demonstrates the deviation from the conformal $\caln=4$ thermodynamics which was discussed below equation \eqref{completefreeenergy}. For small $m$, \ie small quark masses compared to the temperature $M_q\ll T$, we see deviations from the linear behavior due to thermal fluctuations which increase the free energy.\par
In fig.~\ref{fig:Scan} the entropy $\cals_7$ versus the mass parameter $m$ as calculated from equation (\ref{entropydim}) is shown. Again a displacement due to the isospin densities appear. However, the displacement in the entropy becomes smaller as we increase $m$ and the entropy approaches a non-zero value independent of the isospin density as $m\to\infty$. Moreover if the baryon density is above its critical value, a minimum in the entropy appears for zero isospin density. By increasing the isospin density this minimum disappears slowly.\par
 The energy $\cale_7$ shown in fig.~\ref{fig:Ecan} as calculated from equation (\ref{energydim}) shows a very similar behavior as the free energy, discussed above. However, a minimum appears in the energy at $m\approx 1.3$ for a baryon density above the critical density (see fig.~\ref{fig:Ecan} (a)). Increasing the baryon densities (see fig.~\ref{fig:Ecan} (b)), this minimum disappears.\par
Fig.~\ref{fig:Vcan} shows the speed of sound $\calv_s^2$ calculated from equation (\ref{speeddim}). Again a displacement due to the isospin densities appear. In the speed of sound this displacement grow as we increase $m$. For a baryon density above the critical density and zero isospin density, the speed of sound is constant at $m\approx 1.3$ as shown in fig.~\ref{fig:Vcan} (b). Increasing the baryon or isospin density the speed of sound monotonously decreases as $m$ increases (fig.~\ref{fig:Vcan} (b) and (c)).\par

\subsubsection{Interpretation: Quark mass selection and temperature change}\label{quarkmassdet}
Here we discuss the effect of a non-zero isospin density on the quark mass~$M_q$.   
We argue that under certain conditions explained below, the thermodynamic quantities given in
the previous section~\ref{sec:canNumRes} select preferred values~$M_q^*$ for the mass~$M_q$ of 
the quarks inside the quark gluon plasma. In a slightly different setup we find that our system
prefers a distinct temperature~$T=T^*$.

In our model the quark mass~$M_q$ is a free parameter. 
We fix the temperature~$T$ and charge densities~$n^i$
and keep the volume~$V$ of our system constant.  
The canonical ensemble describes this situation with the  
free energy being the relevant thermodynamic potential.  
We argue how the quark mass~$M_q$ as a free parameter has 
to be selected in order to satisfy the second law of thermodynamics.
In fact we will find a competing behavior between entropy and energy requirements. 
The second law of thermodynamics implies that in thermal equilibrium, the entropy is maximized
at fixed thermodynamic potential. This is equivalent to minimzing the thermodynamic
potential at a fixed entropy {\cite{Landau:1980sm}}. In our case this potential is the free energy.  
However, competing situations between entropy and energy requirements are well known
 from frustrated systems, \eg polymers or anti-ferromagnets
on a triangle lattice, as well as glass states, \eg spin glass~\cite{spinGlass}. 
In these systems the temperature is usually the 
thermodynamic variable which adjusts at a distinct value where the criteria are optimally satisfied. 
In our setup we have
the quark mass~$M_q$ as a free parameter, which in analogy to the case discussed above adjusts 
 at a specific value~$M_q^*$.
In order to find this optimal value~$M_q^*$, we examine figures~\ref{fig:Fcan} and~\ref{fig:Scan}.  
Recall from equations~\eqref{dicmc} and~\eqref{eq:ndt} that the mass parameter~$m$ is related to the 
temperature and quark mass by~$m=2 M_q/(\sqrt{\lambda}T)$ and~$n^i\propto \tilde d^i$.
We observe that in the region $m\gtrsim 4$ for all densities, 
the free energy decreases at a constant entropy as
the quark mass $M_q$ is increased. At smaller values of the mass parameter $m\lesssim 4$,   
the entropy changes as we vary the quark mass $M_q$. 
The second law of thermodynamics forbids the system to be in equilibrium. 
From these considerations and from looking at figures~\ref{fig:Fcan} and~\ref{fig:Scan} we conclude
that the quark mass has to take a value of~$0\le M_q^* \le \sqrt{\lambda} T m/2$,    
such that~$0\le m\le 4$. Since in this regime the entropy and free energy
both vary when the densities are changed, we expect the value of~$M_q$ to 
depend on the densities~$n^i$,  
i.e.~$M_q=M_q^*(n^i)$.   
Therefore, our system with the quark mass as a free parameter shows similar properties as the frustrated systems
discussed above.

It is amusing to apply the results of this analysis to the quark gluon plasma formed after the big bang. 
This would imply that our universe could have evolved in such a way that the quark mass 
was chosen such that it  
optimizes the free energy and entropy requirements. Note that during this mass selection
process we have to stay in a regime 
where the metric does not change considerably in order for this  
system to be approximately closed~\cite{Landau:1980sm}.

Now we consider a different scenario, by which a distinct temperature~$T^*$ is favored. 
We work in the microcanonic ensemble by fixing the entropy instead of temperature.
An example of such a situation  
is the collision of heavy ions in vacuum.     
Since the natural temperature variable for our system is
the temperature~$T$ measured in units of the quark mass~$M_q$, we let~$m$ change freely. 
In this case the quark mass~$M_q$ is fixed and the volume of plasma evolves  
to a preferred value of temperature~$T$, in a reversible thermodynamic process.  
The system seeks to minimize its internal
energy~$\cale_7$ while its entropy~$\cals_7$ is fixed by definition. In this setup the plasma 
relaxes isentropically to a favored temperature~$T=T^*$.

\subsection{Phase transition/phase diagram in the canonical ensemble}\label{phasediagramcan}

\begin{figure}[t]
\psfrag{m}{$m$}
\psfrag{F}{$\calf_7$}
\psfrag{dvs2}{$\calv_s^2$}
\psfrag{E}{$\cale_7$}
\psfrag{S}{$\cals_7$}
\psfrag{c}{$c$}
\psfrag{muB}[b]{$\mu^B/M_q$}
\psfrag{muI}[b]{$\mu^I/M_q$}
\centering
\subfigure[]{\includegraphics[width=0.3\textwidth]{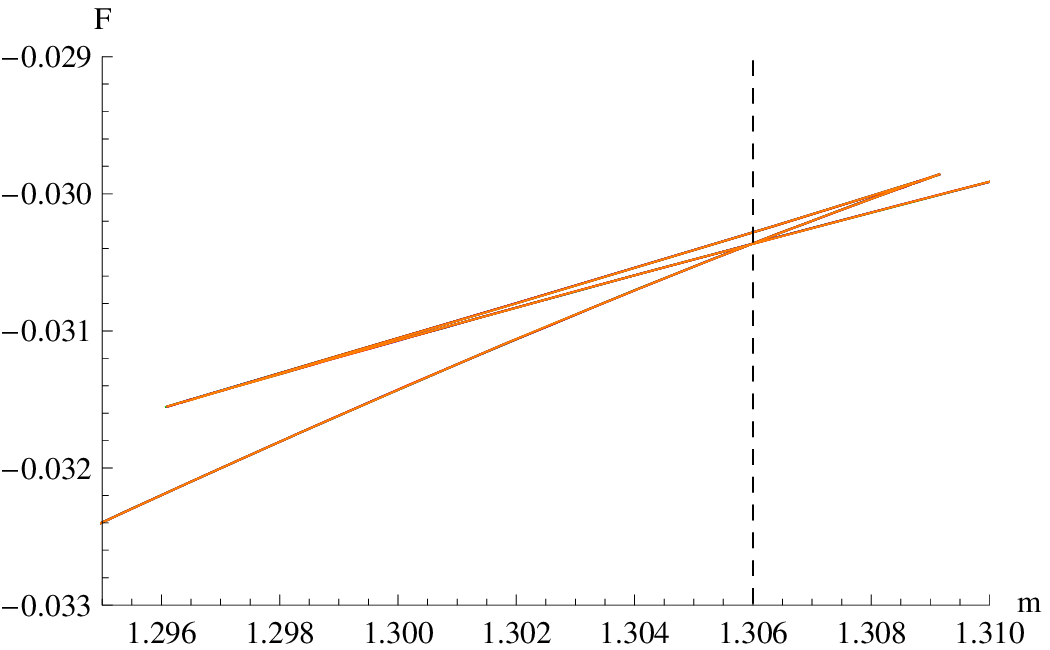}}
\hspace{8pt}
\subfigure[]{\includegraphics[width=0.3\textwidth]{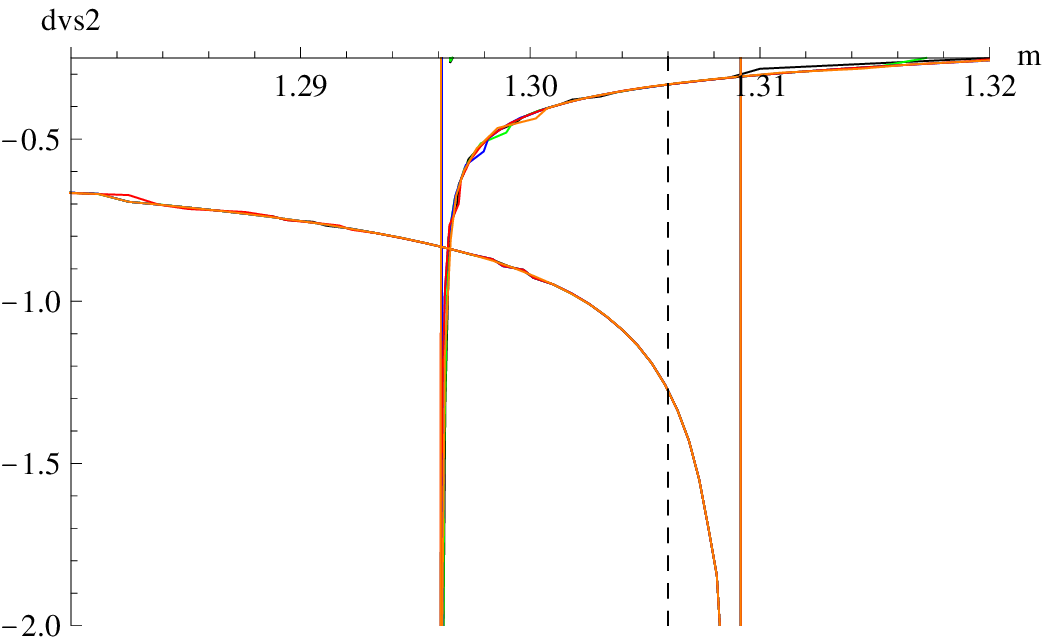}}
\hspace{8pt}
\subfigure[]{\includegraphics[width=0.3\textwidth]{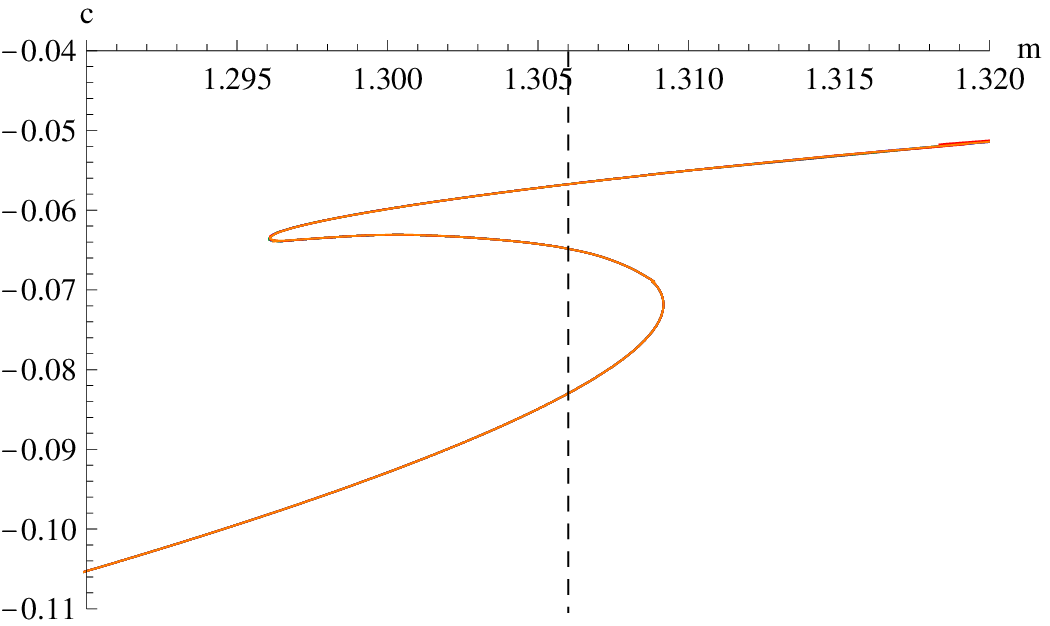}}
\subfigure[]{\includegraphics[width=0.3\textwidth]{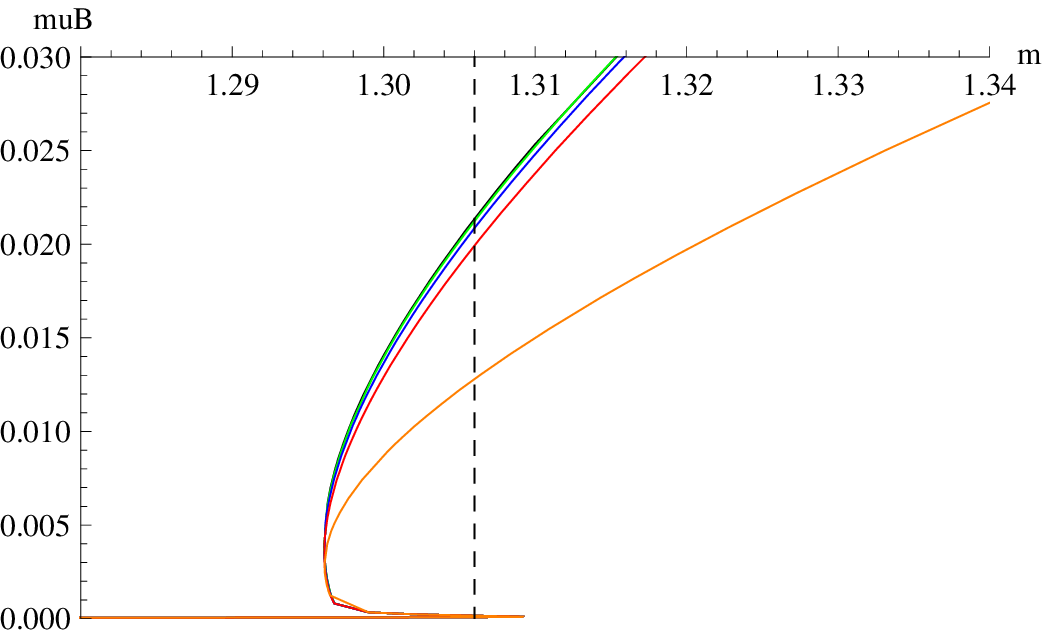}}
\hspace{24pt}
\subfigure[]{\includegraphics[width=0.3\textwidth]{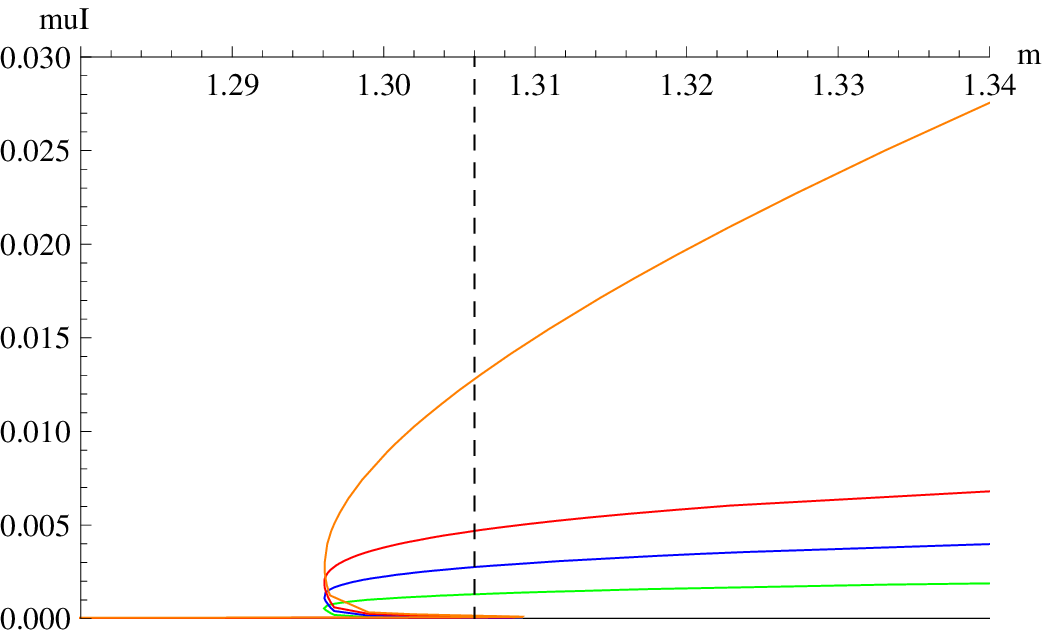}}
\caption{(a) The dimensionless free energy $\calf_7$, (b) the speed of sound $\calv_s^2$, (c) the chiral condensate $c$, (d) the baryon chemical potential $\mu^B/M_q$ and (e) the isospin chemical potential $\mu^I/M_q$ versus the mass parameter $m$ as defined in equation \eqref{dicmc} near the phase transition at the baryon density $\dt^B=5\cdot 10^{-5}$ for $N_f=2$. The resolution of the figures is not high enough to dissolve the curves at different isospin density $\dt^I=0,\frac{1}{4}\dt^B,\frac{1}{2}\dt^B,\frac{3}{4}\dt^B,\dt^B$. The dashed line at $m=1.306$ marks the phase transition.}\label{fig:phasetranscan}
\end{figure}

\begin{figure}[t]
\psfrag{dtB}[t]{$\dt^B$}
\psfrag{dtI}[lb]{$\dt^I$}
\psfrag{dB}[t]{$\dt^B$}
\psfrag{dI}[r][][1][-90]{$\dt^I$}
\psfrag{m}[r]{$m$}
\centering
\subfigure[]{\includegraphics[width=0.45\textwidth]{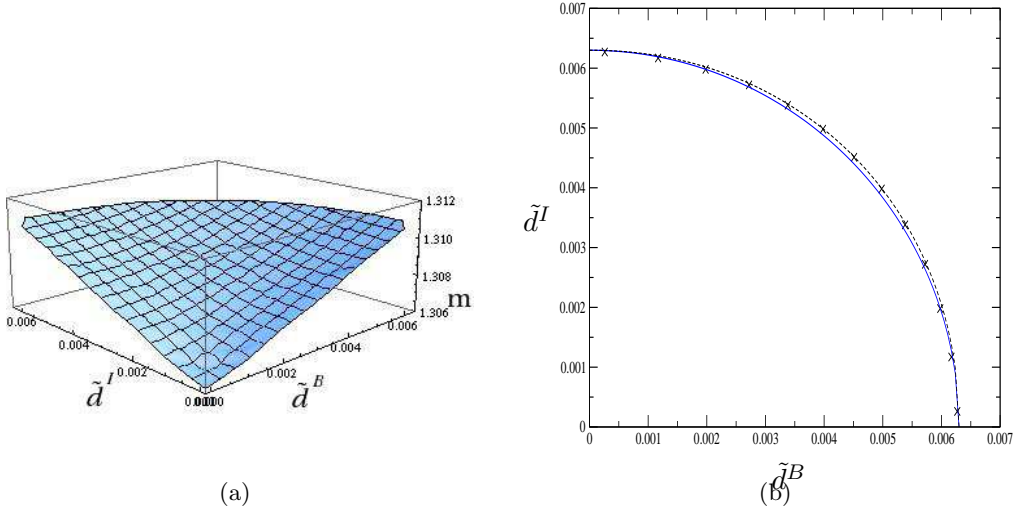}}
\hspace{20pt}
\subfigure[]{\includegraphics[width=0.45\textwidth,height=0.45\textwidth]{./figs/critpoints.eps}}
\caption{(a) The phase diagram in the canonical ensemble for $N_f=2$: The cusp marks the phase transition for $\dt^B=\dt^I=0$ and the upper rim the critical points. Notice that the phase diagram is approximately $O(2)$ invariant in agreement with the symmetries discussed in section \ref{sym}. (b) Position of the critical points: Comparison of the position of the critical points (cross) with the $O(2)$ invariant circle (blue) and a superellipse $(\dt^I)^a=(\dt^B_{\text{crit}})^a-(\dt^B)^a$ with $a=2.08$ (dashed). The difference of the parameter a from 2 determines the deviation from the $O(2)$ symmetry.}\label{fig:phasediagram}
\end{figure}

In the pure baryonic case \cite{Kobayashi:2006sb} it was shown that there is a
first order phase transition between two black hole embeddings for densities
below the critical density $\dt^B_{\text{crit}}=0.0063$.\footnote{Note that
  the normalization we use in this paper \eqref{dt} differ to the
  normalization used in \cite{Kobayashi:2006sb} by $N_f$.} In this section we
analyze how the isospin density influences this phase transition. In
fig.~\ref{fig:phasetranscan} we give an example of the behavior of the free
energy, the speed of sound, the chiral condensate, the baryon and the isospin
chemical potential $\mu^{B/I}/M_q$ close to the phase transition. The free
energy, the speed of sound and the chiral condensate behave as in the pure
baryonic case \cite{Kobayashi:2006sb}. The isospin chemical potential
$\mu^I/M_q$ takes different values depending on the isospin density after the
phase transition (see fig.~\ref{fig:phasetranscan} (e)). The baryon chemical
potential (fig.~\ref{fig:phasetranscan} (d)) instead is almost independent of
the isospin density. However, the behavior of both chemical potentials changes
dramatically if the baryon and isospin densities are equal in agreement with
equation \eqref{eq:discontidB=dI}.\par 
Fig.~\ref{fig:phasetranscan} shows that the phase transition is first order since the free energy is continuous and the speed of sound, the chiral condensate and the chemical potentials show a multivalued behavior and therefore a discontinuous step at the phase transition. The phase transition is marked by the crossing point in the free energy of the two branches coming in from small $m$ and large $m$ (see fig.~\ref{fig:phasetranscan} (a)).\par
In fig.~\ref{fig:phasediagram} (a) we map out the phase diagram by varying the densities and determining the crossing point in the free energy. As outlined in section \ref{sym}, the phase diagram is approximately $O(2)$ invariant. We find a line of critical points, which mark the critical mass parameter $m_{\text{crit}}$ and critical densities $\dt^B_{\text{crit}}$, $\dt^I_{\text{crit}}$. For densities around the critical points, we expect the largest deviation to the $O(2)$ symmetry in the phase diagram (see equation \eqref{approxO(2)}). To get a quantitative description of this deviation, we plot the critical points and compare them to the $O(2)$ invariant circle and the superellipse $(\dt^I)^a=(\dt^B_{\text{crit}})^a-(\dt^B)^a$ with $a=2.08$ in fig.~\ref{fig:phasediagram} (b). The deviation $a-2=0.08$ is still small at the critical points. Thus, the complete phase boundary in the canonical ensemble is $O(2)$ invariant in a good approximation.\par
More advanced thermodynamical investigations of the pure baryonic case \cite{Kobayashi:2006sb,Mateos:2007vc} show that there is a region were the black hole embeddings are not a stable ground state for this theory. The points where the phase transition between the two black hole embeddings occurs is at the boundary of this unstable region. In \cite{Mateos:2007vc} the authors expect that in this region an inhomogeneous mixture of black hole and Minkowski embeddings is the stable ground state. Therefore the phase transition is not between two black hole embeddings, but between a black hole embedding and this inhomogeneous mixture. The existence of a mixed phase is typical for a first order phase transition and can \eg be observed in boiling water. The water does not instantly turn into gas but forms droplets consisting of a mixture of water and water vapor.

\subsection{Canonical thermodynamics in the large quark mass limit}\label{canlargem}

In the large mass limit $m\to\infty$, \ie $M_q\gg T$, the dimensionless
thermodynamic quantities and the chiral condensate vanish if the baryon and
isospin densities are zero because supersymmetry and conformal symmetry are
restored and therefore the thermodynamics are given by the results of the stack
of D3-branes dual to the conformal $\caln=4$ SYM theory \cite{Mateos:2007vn}.
Introducing finite baryon and isospin densities break these symmetries even
though the AdS black hole background becomes symmetric and we get deviations in
the thermodynamic quantities from the conformal results (see \eg
fig.~\ref{fig:Fcan}). In the following we investigate the dependence of these
deviations on the densities $\dt^{I_i}$ further.

\subsubsection{Numerical results}

\begin{figure}[t]
\centering
\psfrag{c}{$c$}
\psfrag{F}[l]{$\calf_7$}
\psfrag{muBm}{$\frac{\mu^B}{M_q}$}
\psfrag{muIm}{$\frac{\mu^I}{M_q}$}
\psfrag{dB}[t]{$\dt^B$}
\psfrag{dI}[b]{$\dt^I$}
\subfigure[]{\includegraphics[width=0.4\textwidth]{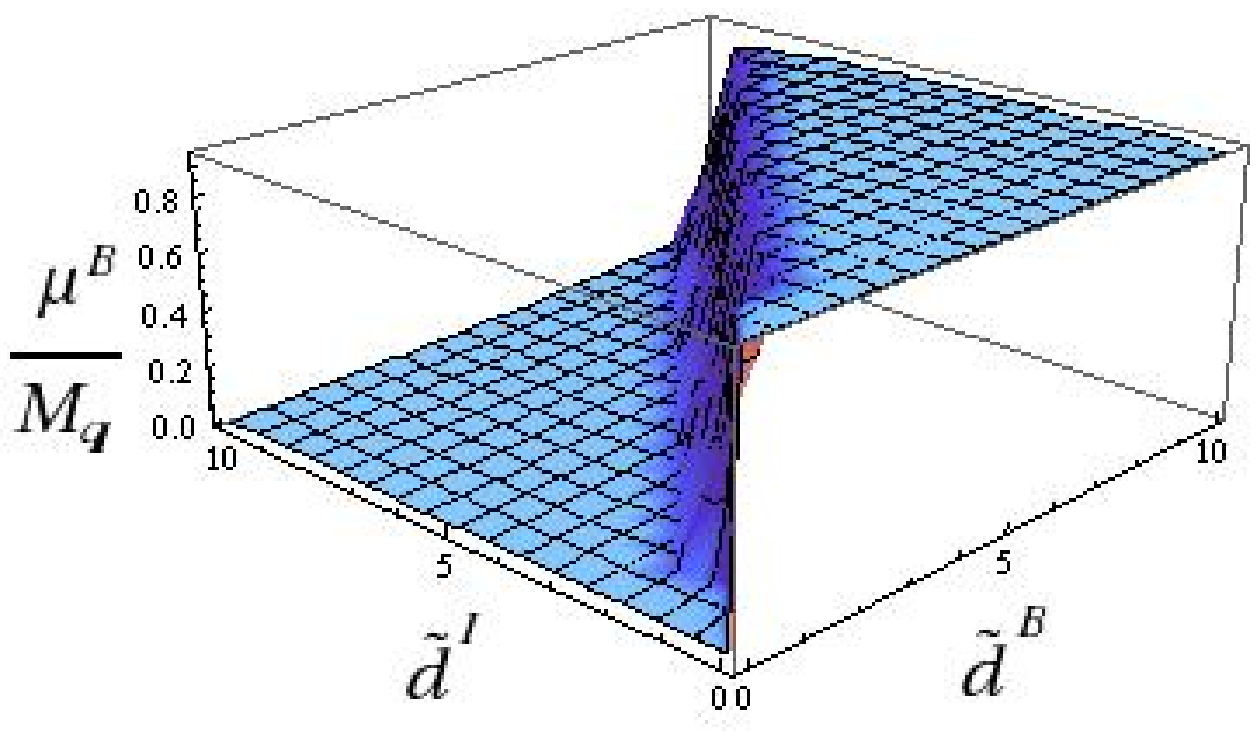}}
\hspace{16pt}
\subfigure[]{\includegraphics[width=0.4\textwidth]{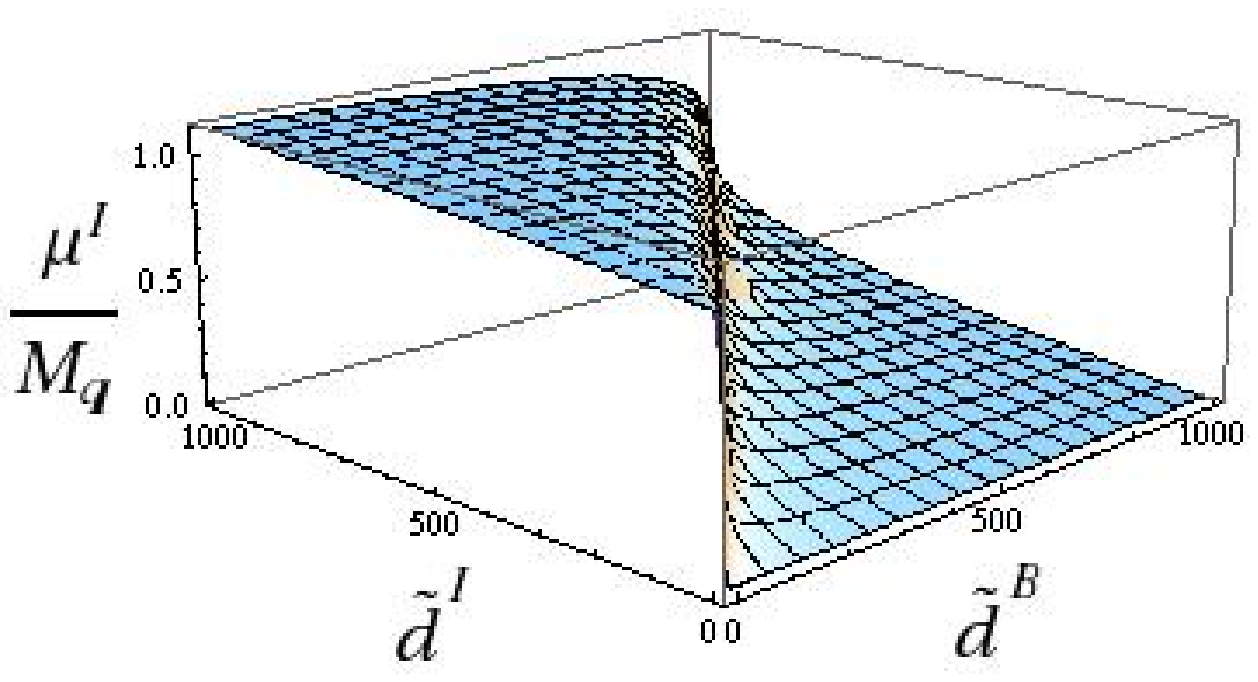}}
\subfigure[]{\includegraphics[width=0.4\textwidth]{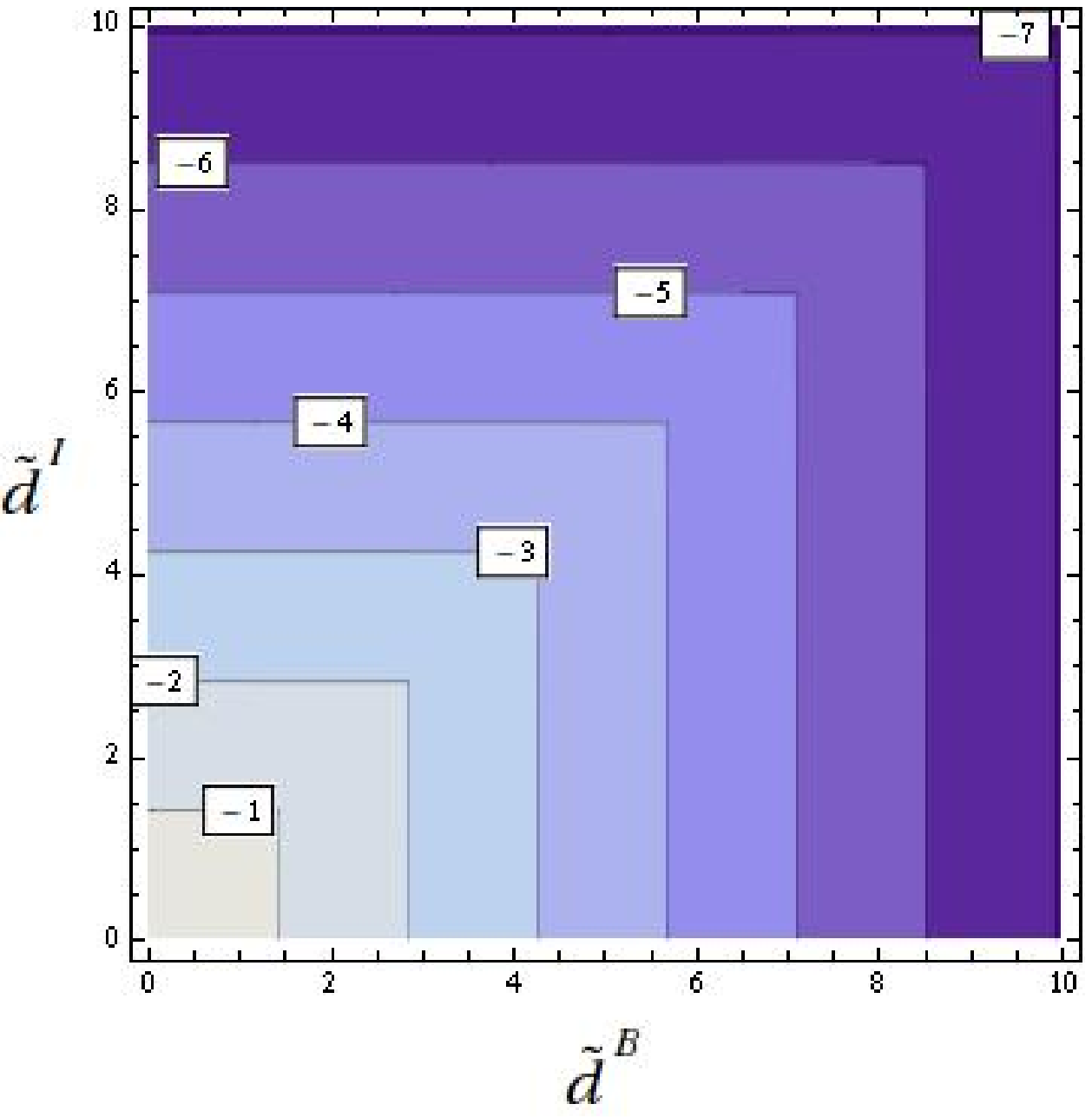}}
\hspace{16pt}
\subfigure[]{\includegraphics[width=0.4\textwidth]{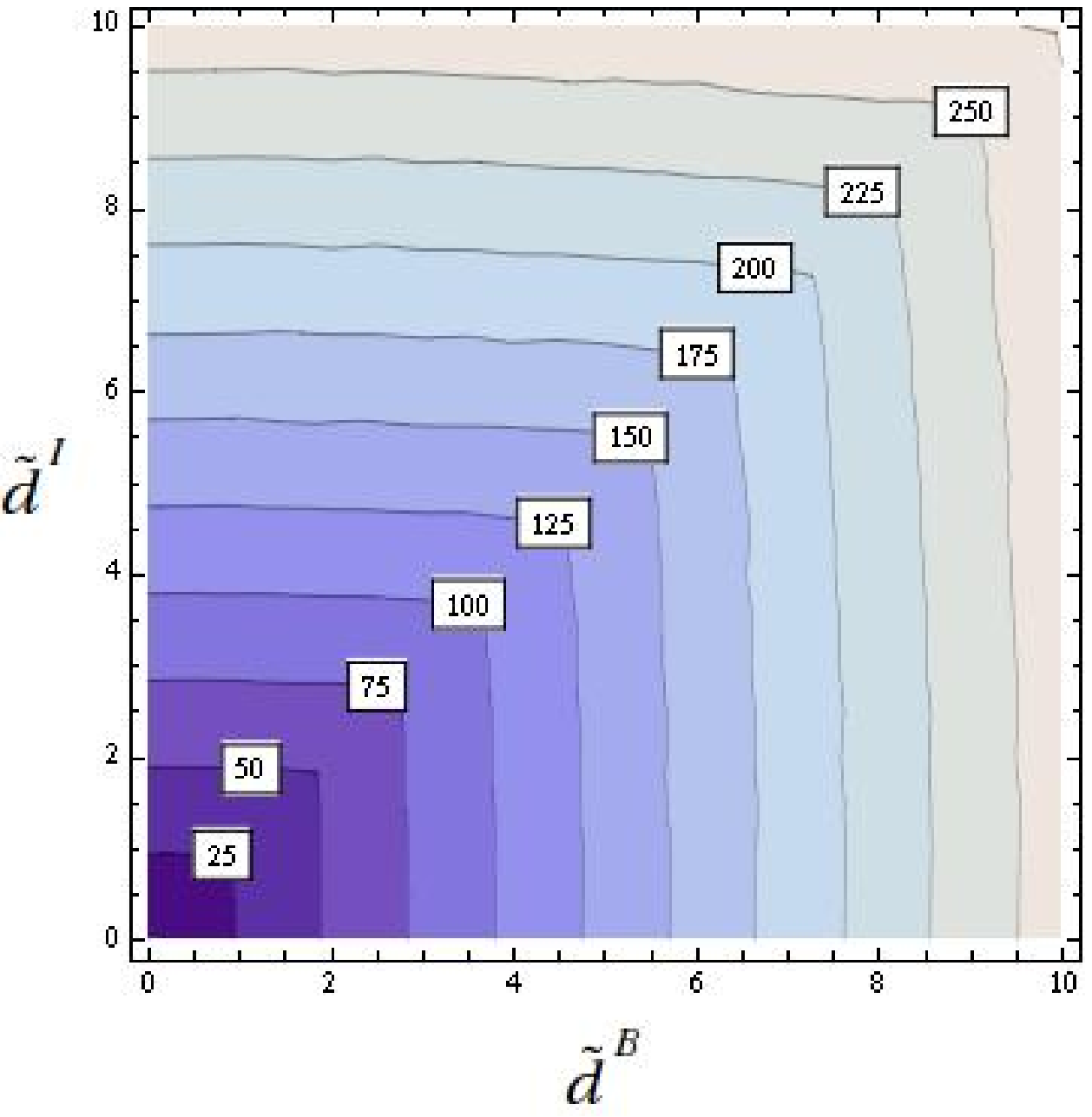}}
\caption{(a) The baryon chemical potential $\mu^B/M_q$, (b) the isospin chemical potential $\mu^I/M_q$, (c) the chiral condensate $c$ and (d) the free energy $\calf_7$ versus the baryon and isospin density at $m=20$, where the mass parameter $m$ is defined in equation \eqref{dicmc}, in the case $N_f=2$. (c) and (d) are contour plots of the chiral condensate and the free energy, respectively.}\label{fig:largemcan}
\end{figure}

\begin{figure}
\centering
\scriptsize
\psfrag{muB}{$\mu^B/M_q=0$}
\psfrag{muBn}{$\mu^B/M_q=1$}
\psfrag{muI}{$\mu^I/M_q=0$}
\psfrag{muIn}{$\mu^I/M_q=1$}
\psfrag{muBnn}{$\mu^B/M_q=-1$}
\psfrag{muInn}{$\mu^I/M_q=-1$}
\psfrag{dB}{\normalsize $\dt^B$}
\psfrag{dI}{\normalsize $\dt^I$}
\includegraphics[height=0.5\textwidth]{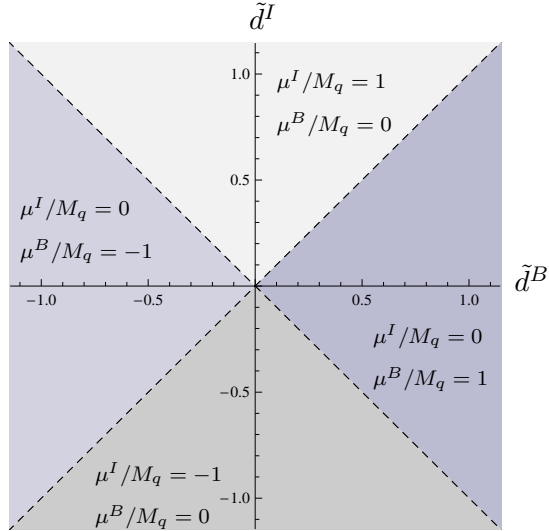}
\caption{The phase diagram for $m\to\infty$, \ie $M_q\gg T$: The dashed lines marks the phase transitions where the chemical potentials develop a discontinuous gap. Notice that the dashed lines are the fixed points of the symmetries $\dt_1\leftrightarrow -\dt_1$, $\dt_2\leftrightarrow -\dt_2$.}\label{phasediagramlargemcan}
\end{figure}

First we discuss the numerical results for the thermodynamic quantities, the chemical potentials and the chiral condensate at a fixed mass $m=20$ in dependence of the baryon and isospin density for $N_f=2$, presented in fig.~\ref{fig:largemcan}. We restrict our analysis to non-negative baryon and isospin densities $\dt^B,\dt^I\ge 0$. The results may be continued into the other three quadrants by the accidental symmetries, discussed in section \ref{sym}.The discontinuous step in the chemical potentials along the line $\dt^I=\dt^B\ge 0$ described in \eqref{eq:mulargem} and \eqref{eq:discontidB=dI} are shown in fig.~\ref{fig:largemcan} (a) and (b). We can easily characterize this behavior in the large mass limit $m\to\infty$ by
\begin{equation}
\frac{\mu^B}{M_q}\to H(\dt^B-\dt^I)\qquad\text{and}\qquad\frac{\mu^I}{M_q}\to H(\dt^I-\dt^B)\,,
\end{equation}
where $H$ is the Heaviside function. This discontinuity suggests that there is a first order phase transition between the two regions $0\le \dt^I<\dt^B$ and $0\le \dt^B<\dt^I$. At the same line the chiral condensate and the thermodynamic quantities develop a kink (see fig.~\ref{fig:largemcan}), which also indicates that there is a phase boundary along the line $\dt^B=\dt^I\ge 0$. Using the accidental symmetries discussed in section \ref{sym}, we give an overview of the four different phases and the values of the chemical potentials in the corresponding region in fig.~\ref{phasediagramlargemcan}. The same four different phases are also seen in QCD-like theories studied in \cite{Splittorff:2000mm}. Since in \cite{Splittorff:2000mm} the authors study the grand canonical ensemble, the phase transition is marked by a discontinuous step in the baryon and isospin density and phase boundaries are given by $|\mu^B|=|\mu^I|$. Along these phase boundaries the chiral condensate develops a similar kink as we observe here.\par
From the contour plot in fig.~\ref{fig:largemcan} (c), we see that the chiral condensate is independent of the isospin density for large masses in the region $0\le \dt^I<\dt^B$ since the contour lines are parallel to the lines of constant baryon density in the region $0\le\dt^I<\dt^B$. To investigate the region $0\le \dt^B<\dt^I$, we can simply interchange the baryon and isospin density by the accidental symmetries discussed in section \ref{sym}. Therefore, the chiral condensate only depends on the isospin density in the region $0\le \dt^B<\dt^I$. Fig.~\ref{fig:bgccan} shows that the chiral condensate is also independent of the mass parameter $m$ since it approaches a constant value as $m\to\infty$. Approximately, the dependence of the chiral condensate is given by
\begin{equation}\eqlabel{largemcanc}
c\to
\begin{cases}
-\frac{\dt^B}{\sqrt{2}}\qquad\text{for}\qquad 0\le\dt^I<\dt^B\\[2mm]
-\frac{\dt^I}{\sqrt{2}}\qquad\text{for}\qquad 0\le\dt^B<\dt^I
\end{cases}
\qquad\text{as}\qquad m\to\infty\,.
\end{equation}
Next we analyze the thermodynamic quantities and start with the free energy. The contour plot of the free energy in fig.~\ref{fig:largemcan} (d) directly shows a dependence on the isospin and the baryon density in both regions $0\le\dt^I<\dt^B$ and $0\le \dt^B<\dt^I$. In the region $0\le \dt^I<\dt^B$, the dependence appears to be linear on the baryon density but nonlinear on the isospin density. Fig.~\ref{fig:Fcan} shows also a linear dependence of the free energy on the mass parameter $m$ in the large mass limit. In order collect this behavior in a formula, we make a suitable ansatz and fit to the numerical data
\begin{equation}
\calf_7\sim
\begin{cases}
\sqrt{2}m\dt^B-2\dt^B+0.6\frac{(\dt^I)^{2.25}}{(\dt^B)^{0.9}}\quad\text{for}\quad 0\le\dt^I<\dt^B\\[2mm]
\sqrt{2}m\dt^I-2\dt^I+0.6\frac{(\dt^B)^{2.25}}{(\dt^I)^{0.9}}\quad\text{for}\quad 0\le \dt^B<\dt^I
\end{cases}
\quad\text{as}\quad m\to\infty\,.\eqlabel{largemcanfreeenergy}
\end{equation}
Similarly we write for the energy
\begin{equation}
\cale_7\sim
\begin{cases}
\sqrt{2}m\dt^B+0.6\frac{(\dt^I)^{2.25}}{(\dt^B)^{0.9}}\qquad\text{for}\quad 0\le\dt^I<\dt^B\\
\sqrt{2}m\dt^I+0.6\frac{(\dt^B)^{2.25}}{(\dt^I)^{0.9}}\qquad\text{for}\quad 0\le \dt^B<\dt^I
\end{cases}
\qquad\text{as}\qquad m\to\infty\,.\eqlabel{largemcanenergy}
\end{equation}
The entropy behaves similarly as the chiral condensate and we therefore may approximate it with
\begin{equation}
\cals_7\to
\begin{cases}
2\dt^B\qquad\text{for}\qquad 0\le\dt^I<\dt^B\\
2\dt^I\qquad\text{for}\qquad 0\le\dt^B<\dt^I
\end{cases}
\qquad\text{as}\qquad m\to\infty\,.\eqlabel{largemcanentropy}
\end{equation}
In the following we interpret the large mass limit of the chiral condensate, the free energy, the energy and the entropy found in \eqref{largemcanfreeenergy}-\eqref{largemcanentropy}.

\subsubsection{Interpretation of the numerical results}
For the interpretation of the results found in \eqref{largemcanfreeenergy}-\eqref{largemcanentropy}, the difference of the two regions $0\le\dt^I<\dt^B$ and $0\le \dt^B<\dt^I$ as displayed in fig.~\ref{fig:largemcan} is essential. First we show that in both regions the larger density is a measure for the degrees of freedom. For this purpose, we map our theory to a QCD-like theory. In our setup the open D$3$-D$7$ strings model quarks\footnote{Strictly speaking the D$3$-D$7$ strings model a SUSY multiplet consisting of quarks and squarks.} of a QCD-like theory\cite{Karch:2002sh}. For $N_f=2$ we consider two different quarks, \eg {\it u} and {\it d} quarks. We assign a baryon charge $q^B$ and isospin charge $q^I$ to these quarks,
\begin{equation}
u=|q^B,q^I\rangle=|1,1\rangle\,,\qquad d=|1,-1\rangle\,.
\end{equation}
The effect of the symmetries \eqref{symNf=2} on these charges is given by
\begin{equation}
\begin{aligned}
d_1\leftrightarrow&\; d_2:\\
& u=|1,1\rangle\mapsto |1,-1\rangle=d && \text{and} && d=|1,-1\rangle\mapsto |1,1\rangle=u\\
d_1\leftrightarrow& -d_1:\\
& u=|1,1\rangle\mapsto |-1,-1\rangle=\bar u\quad && \text{and}\quad && d=|1,-1\rangle\mapsto |1,-1\rangle=d\\
d_2\leftrightarrow& -d_2:\\
& u=|1,1\rangle\mapsto |1,1\rangle=u && \text{and} && d=|1,-1\rangle\mapsto |-1,1\rangle=\bar d\,,
\end{aligned}
\end{equation}
where we defined the anti-quarks $\bar u$ and $\bar d$ consistently with the ideas given in section \ref{sym} and the convention used in QCD-like theories\footnote{Recall that the pion $\pi^+=u\bar d=|0,2\rangle$ ($\pi^-=d\bar u=|0,-2\rangle$) has positive (negative) isospin charge in QCD. Notice that the normalization of our charges is not the same as in QCD.}.  Using the charges of the quarks we may assign a baryon charge to the plasma by 
\begin{equation}
q^B_{\text{plasma}}=n_u+n_d-n_{\bar u}-n_{\bar d}\,,
\end{equation}
and the isospin charge by
\begin{equation}
q^I_{\text{plasma}}=n_u+n_{\bar d}-n_{\bar u}-n_{d}\,,
\end{equation}
where $n_i$ counts the different quarks in the plasma. A plasma with $0\le q^I_{\text{plasma}}<q^B_{\text{plasma}}$, \ie $0\le\dt^I<\dt^B$, may be constructed using $u$ and $d$ quarks only. Experimentally such a plasma can be produced by heavy ion collisions. However, for a plasma with $0\le q^B_{\text{plasma}}<q^I_{\text{plasma}}$, \ie $0\le \dt^B<\dt^I$,  we must consider the free quarks $u$ and anti-quarks $\bar d$. Therefore for positive densities, the larger one of the two quantities $q^B_{\text{plasma}}$ and $q^I_{\text{plasma}}$ counts the degrees of freedom of the plasma consisting of the free quarks and anti-quarks. This explains the difference in the two regions displayed in fig.~\ref{fig:largemcan}.\par
First we consider the entropy $\cals_7$ given in equation \eqref{largemcanentropy}. The entropy measures the logarithm of the number of states the plasma could be in. The degrees of freedom increase these number of states by their phase space volume. Since the larger density determines the degrees of freedom, it is clear that the entropy increases linearly with the larger density.\par
Next we consider the free energy and the energy found in \eqref{largemcanfreeenergy} and \eqref{largemcanenergy}. Since the degrees of freedom contribute to the free energy (energy) by their rest mass, a term appears which is proportional to $\dt^B m$ and $\dt^I m$. In the free energy (energy) there is also a nonlinear dependence on the smaller density, $(\dt^I)^{2.25}/(\dt^B)^{0.9}$ and $(\dt^B)^{2.25}/(\dt^I)^{0.9}$, respectively. We expect that this contribution is due to a charge of the plasma which is induced by the non-Abelian charges of the quarks and measured by the smaller density. This charge is similar to an electrostatic charge in the sense that the free energy (energy) increases as we increase the amount of the charge of the plasma. The deviation in the exponent of $(\dt^{I/B})^{2.25}$ from the naively expected Coulomb behavior $(\dt^{I/B})^2$ is probably due to non-perturbative effects of vacuum polarisations since the gauge coupling is large\footnote{In QED, where the coupling constant is small, the perturbative corrections due to vacuum  polarisations are sometimes called Uehling effect \cite{PhysRev.48.55}.}. The fact that there is a term depending on the larger density $(\dt^B)^{-0.9}$ may be due to a screening of the plasma charge by the quarks and anti-quarks.

\subsubsection{Expected behavior of the system perturbed by interactions\\ which break baryon and isospin conservation}

In this section we study the possible behavior of our system if the baryon and
isospin density are \emph{not} conserved. Interactions which break baryon
(isospin) conservation are \eg included in the MSSM (electroweak theory). We are
not able to include such interactions explicitly in our setup. However, we may
analyze the modification in the plasma induced by variation of the density using
the thermodynamic results found in \eqref{largemcanfreeenergy} to
\eqref{largemcanentropy}. The variation in the plasma must be such that the free
energy becomes minimal at a fixed entropy (see section~\ref{quarkmassdet}). We
discuss the reaction of our system separately for the four different regions in
the phase diagram shown in fig.~\ref{phasediagramlargemcan}.

Experimentally the region $\dt^B>|\dt^I|$ may be observed by heavy ion
collisions. In this region we learn from equation (\ref{largemcanentropy}) that
the entropy is independent of the isospin density but proportional to the baryon
density. Therefore, at constant entropy the baryon density must be constant,
too. Only the isospin density may change. For constant baryon density the
minimum of the free energy is located at zero isospin density, see
fig.~\ref{fig:largemcan}~(d). Thus, we expect our system, perturbed by weak
interactions, to decrease the amount of isospin density to zero. This effect is
well known in nuclear physics. The nuclei stable with respect to $\beta$-decay
consists of the same number of protons and neutrons\footnote{For heavy nuclei
there is a deviation due to the electric charge of the protons.}. As is well
known the $\beta$-decay is induced by electroweak interactions, which we
consider here to change the isospin density.

The region $\dt^B<-|\dt^I|$ is similar to the region discussed above since they
are connected by the symmetry $\dt^B\leftrightarrow -\dt^B$ which interchanges
particles with anti-particles. Thus, this region determines the physics in a
situation where anti-particles dominate. In this region we also expect that the
system approaches a state of zero isospin density. For the example of nuclear
physics this means that nuclei are also stable under $\beta$ decay if they
consist of the same number of anti-protons and anti-neutrons.

In the other two regions the amount of isospin charges is bigger than the amount
of baryon charges $|\dt^B|<|\dt^I|$. To relate these two regions with the
regions discussed above we make use of the symmetry $\dt^B\leftrightarrow\dt^I$
which interchanges the baryon and isospin densities. Thus, we expect that for
these regions the system moves to a state of zero baryon density at a fixed
isospin density. Such a state can be achieved by pairs of particle and
anti-particle of the same isospin since these pairs only contribute to the
isospin density but not to the baryon density. An example for such pairs are
mesons as \eg $\pi^+$ and $\pi^-$.

\subsection{Canonical thermodynamics in the small quark mass limit}\label{cansmallm}
\begin{figure}
\centering
\psfrag{Flowm}{$\calf_7^{\text{thermal}}$}
\psfrag{m}{$m$}
\subfigure[]{\includegraphics[width=0.45\textwidth]{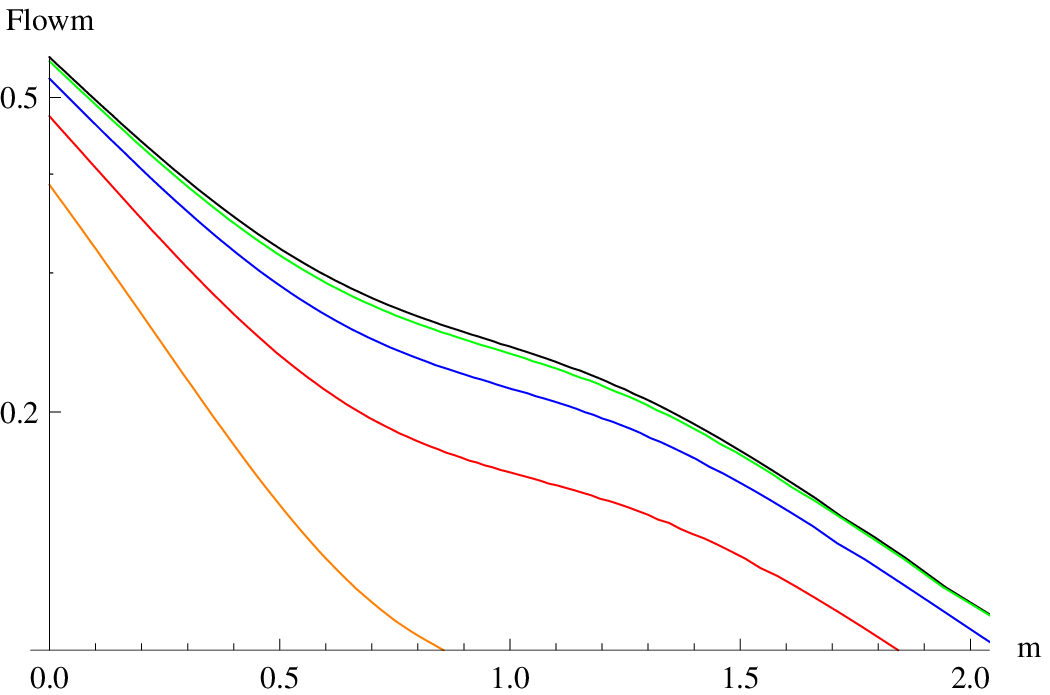}}
\hspace{8pt}
\subfigure[]{\includegraphics[width=0.45\textwidth]{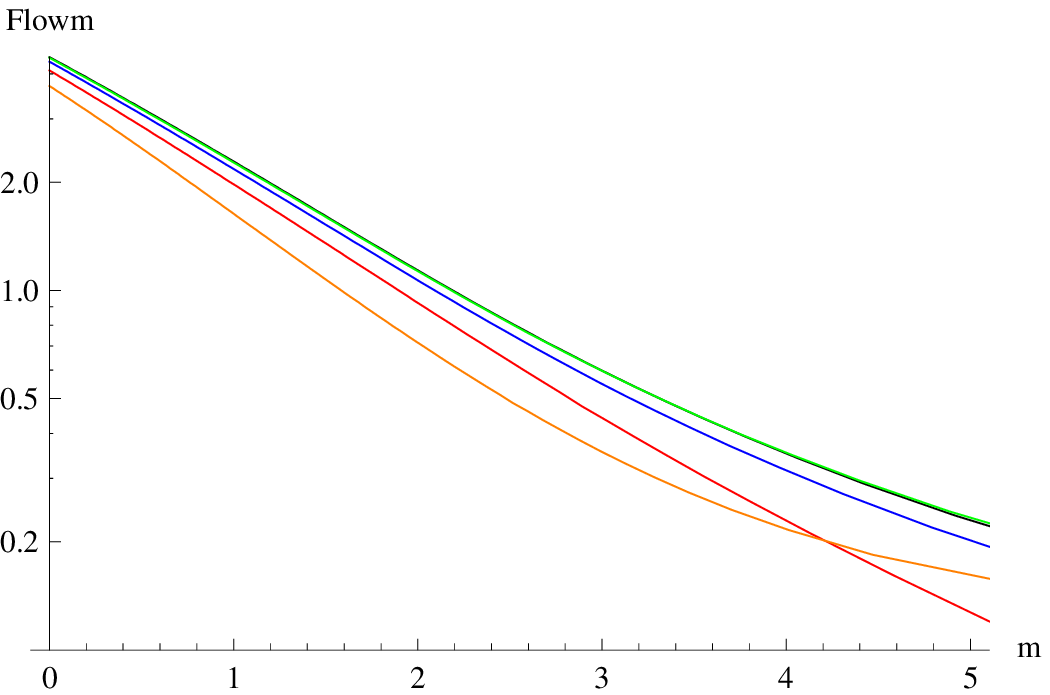}}
\caption{The deviation in the free energy due to thermal fluctuations $\calf_7^{\text{thermal}}$ versus the mass parameter $m$ as defined in equation \eqref{dicmc} at a baryon $\dt^B=0.5$ (left), $\dt^B=2$ (right) and different isospin density $\dt^I$ for $N_f=2$ in a logarithmic plot. The different colors correspond to different isospin densities $\dt^I=0$ (black), $\dt^I=\frac{1}{4}\dt^B$ (green), $\dt^I=\frac{1}{2}\dt^B$ (blue), $\dt^I=\frac{3}{4}\dt^B$ (red) and $\dt^I=\dt^B$ (orange).}\label{fig:smallmcan}
\end{figure}

In the small mass limit $m\to 0$, \ie $M_q\ll T$, the thermal energy is sufficient to excite the plasma, \eg by producing pairs of particles and anti-particles. The probability of these thermal excitations is in the large temperature limit, \ie small mass limit, approximately described by the Boltzmann factor
\begin{equation}\eqlabel{boltzmann}
P=\ee^{-\beta(E-\sum_{i}\mu^{I_i}n_q^{I_i})}\,,
\end{equation}
where $E$ is the energy of the plasma. In our setup this expression may be approximated by
\begin{equation}
P=\ee^{-Am+B\sum_{i}\dt^{I_i}\mut^{I_i}(m)}\,,
\end{equation}
where $A,B$ are some constants. The dependence of the chemical potentials $\mut^{I_i}$ on the mass parameter $m$ induces a slow variation of
\begin{equation}
K(m)=B\sum_{i}\dt^{I_i}\mut^{I_i}(m)\,,
\end{equation}
since the dimensionless chemical potentials $\mut^{I_i}$ are almost constant for small masses $m$ \cite[fig.~8(a)]{Kobayashi:2006sb}. Thus, the Boltzmann factor may be written as
\begin{equation}
P=\ee^{-Am+K(m)}\,,\eqlabel{boltzmannend}
\end{equation}
where $K(m)$ is slowly varying. We expect that the deviation in the thermodynamic quantities due to thermal fluctuations, \ie the deviation from the large mass limit studied in section \ref{canlargem}, behaves as the Boltzmann factor described in equation (\ref{boltzmannend}). In the example of the free energy the thermal fluctuations $\calf_7^{\text{thermal}}$ are given by the dimensionless free energy calculated in equation \eqref{freeenergydim} minus the large mass limit of the free energy given in equation \eqref{largemcanfreeenergy},
\begin{equation}
\calf_7^{\text{thermal}}=\calf_7-\begin{cases}
\sqrt{2}m\dt^B-2\dt^B+0.6\frac{(\dt^I)^{2.25}}{(\dt^B)^{0.9}}\qquad\text{for}\qquad \dt^B>\dt^I\\[2mm]
\sqrt{2}m\dt^I-2\dt^I+0.6\frac{(\dt^B)^{2.25}}{(\dt^I)^{0.9}}\qquad\text{for}\qquad \dt^B<\dt^I
\end{cases}\,,
\end{equation}
 A selection of our numerical results at different baryon and isospin densities is given in the logarithmic plot in fig.~\ref{fig:smallmcan}, where we see the expected result at small masses $m$. We also observe deviations in the large mass limit, \ie $T\ll M_q$, since the Boltzmann factor given in equation \eqref{boltzmann} does not give the full quantum statistic needed to describe the system.

\section{Grand canonical D$7$-brane thermodynamics}
\label{grandcanonical}
\subsection{The grand canonical ensemble:\\ Grand potential, entropy, energy and speed of sound}
In this chapter we investigate the properties of the grand canonical ensemble. For this purpose, we have to use the renormalized Euclidean action (\ref{renormaction}). It is proportional to the grand potential 
\begin{equation}
\Omega_7=TI_R=\frac{\lambda N_cN_f V_3 T^4}{32}\calw_7(m,\mut)\,,\eqlabel{grandpot}
\end{equation}
with the dimensionless quantity
\begin{equation}
\calw_7(m,\mut)=\frac{1}{N_f}G(m,\mut)-\frac{1}{4}\left[\left(\rho_{\text{min}}^2-m^2\right)^2-4mc\right]\,.\eqlabel{grandpotdim}
\end{equation}
Since we must also consider Minkowski embeddings in the grand canonical ensemble, we rewrite $G(m,\mut)$ from equation (\ref{G(m)}) in the coordinates $L$ and $r$ suitable for Minkowski embeddings and for constant gauge fields
\begin{equation}
G(m)=N_f\int_0^\infty \,\dd r\left[f\ft r^3\sqrt{1+(\del_rL)^2}+(r+L\del_rL)\left[m^2-\left(r^2+L^2\right)\right]\right]\,.
\end{equation}
For Minkowski embeddings, the grand potential does not depend on the chemical potential and therefore coincides with the free energy at zero chemical potential \cite{Mateos:2007vn}. Since the free energy determines the thermodynamics entirely, we may apply the results of \cite{Mateos:2007vn}.\par
We calculate the thermodynamic quantities in the black hole phase numerically
by evaluating the canonical results at a fixed chemical potentials. The
transformation to fixed chemical potential can again be done, as in section
\ref{bggrand}, by the functions $\dt_i(\mut_i,m)$, which is the inverse of
(\ref{mu}).  Only the speed of sound has to be recalculated as it is no
thermodynamic state function. The difference to the result (\ref{speeddim}) in
the canonical ensemble is that the derivative now acts on the density instead
of the chemical potentials, 
\begin{equation}
\calv_s^2(m,\mut)=mc-\frac{1}{3}m^2\frac{\del c}{\del m}+\frac{2}{N_f}\sum_{i=1}^{N_f}\mut_i\left(m\frac{\del\dt_i}{\del m}-4\dt_i\right)\eqlabel{speedgrand}
\end{equation}
since the specific heat in the grand canonical ensemble is determined by $c_V=\left(\frac{\del E}{\del T}\right)_{V,\mu}$ (cf.~derivation in section \ref{sec:speedofsound}).

\subsubsection{Numerical results for the grand potential, entropy, energy and speed of sound}

\begin{figure}[t]
\centering
\psfrag{m}{$m$}
\psfrag{Omega}{$\calw_7$}
\subfigure[]{\includegraphics[width=0.3\textwidth]{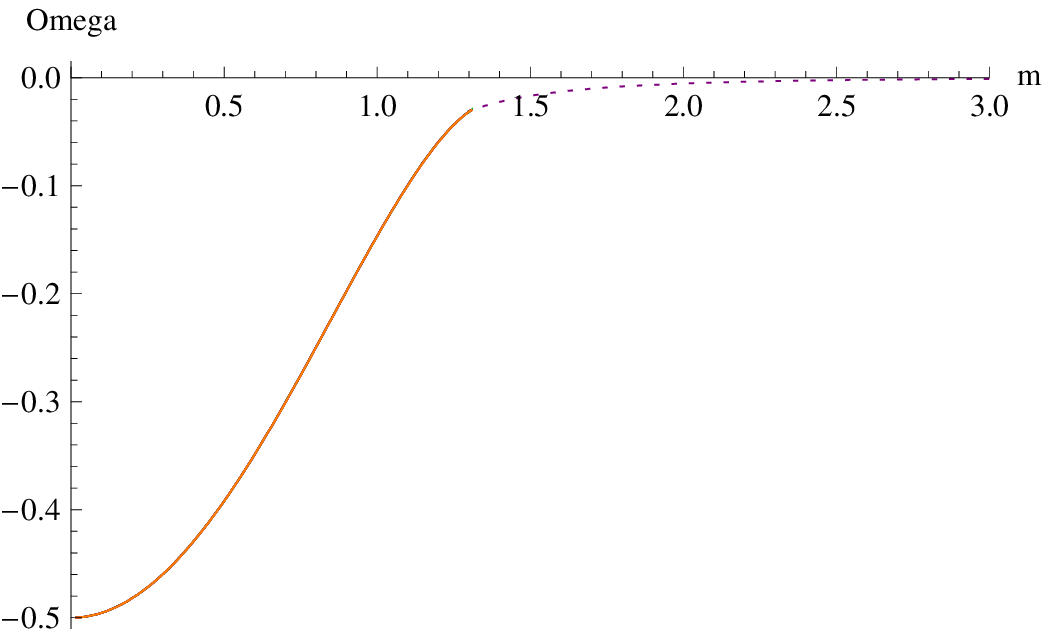}}
\hspace{8pt}
\subfigure[]{\includegraphics[width=0.3\textwidth]{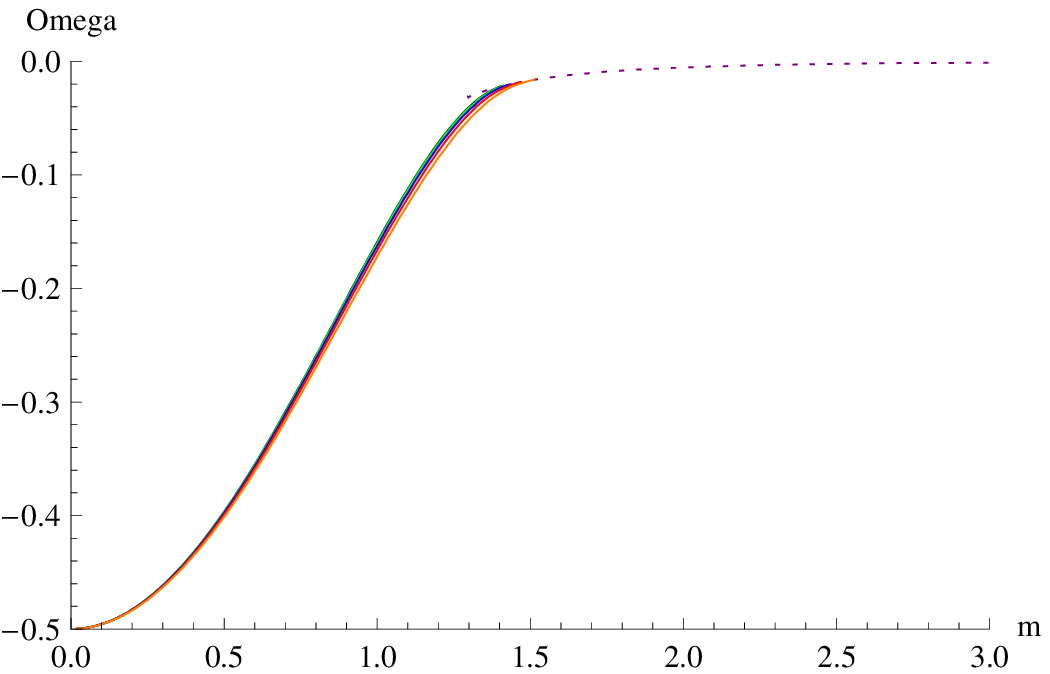}}
\hspace{8pt}
\subfigure[]{\includegraphics[width=0.3\textwidth]{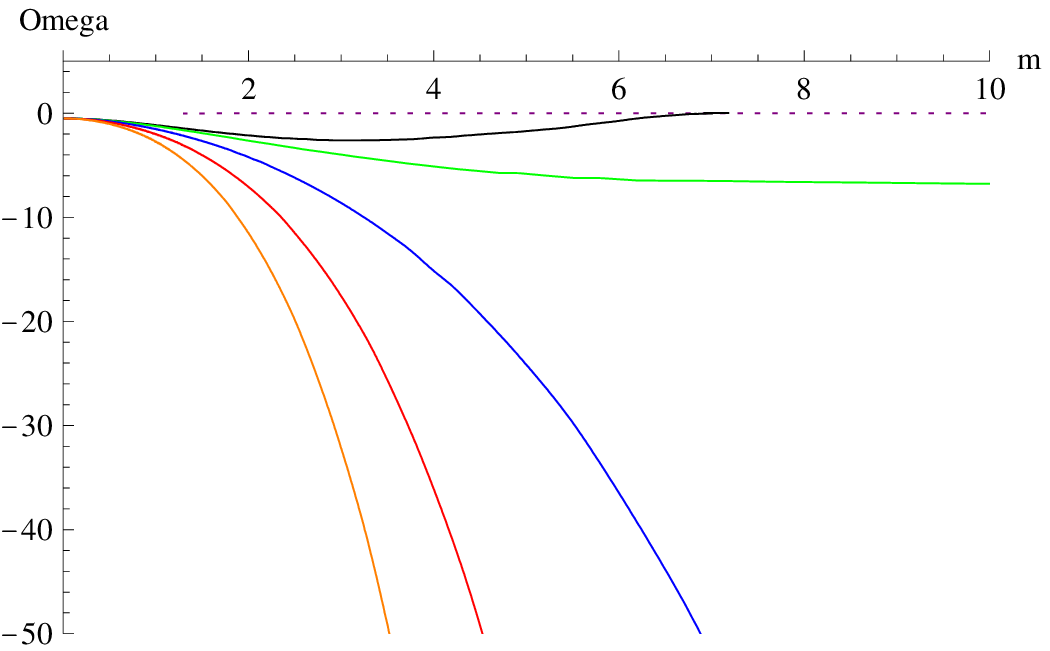}}
\caption{The dimensionless grand potential $\calw_7$ versus the mass parameter $m$  as defined in equation \eqref{dicmc} at baryon chemical potential $\mu^B/M_q=0.01$ (a), $\mu^B/M_q=0.1$ (b) and $\mu^B/M_q=0.8$ (c) for the case $N_f=2$. The dotted purple curve corresponds to Minkowski embeddings and the five other curves to black hole embeddings with isospin chemical potential $\mu^I=0$ (black), $\mu^I=\frac{1}{4}\mu^B$ (green), $\mu^I=\frac{1}{2}\mu^B$ (blue), $\mu^I=\frac{3}{4}\mu^B$ (red) and $\mu^I=\mu^B$ (orange).}\label{fig:Omegagrand}
\end{figure}

\begin{figure}[t]
\centering
\psfrag{m}{$m$}
\psfrag{S}{$\cals_7$}
\subfigure[]{\includegraphics[width=0.45\textwidth]{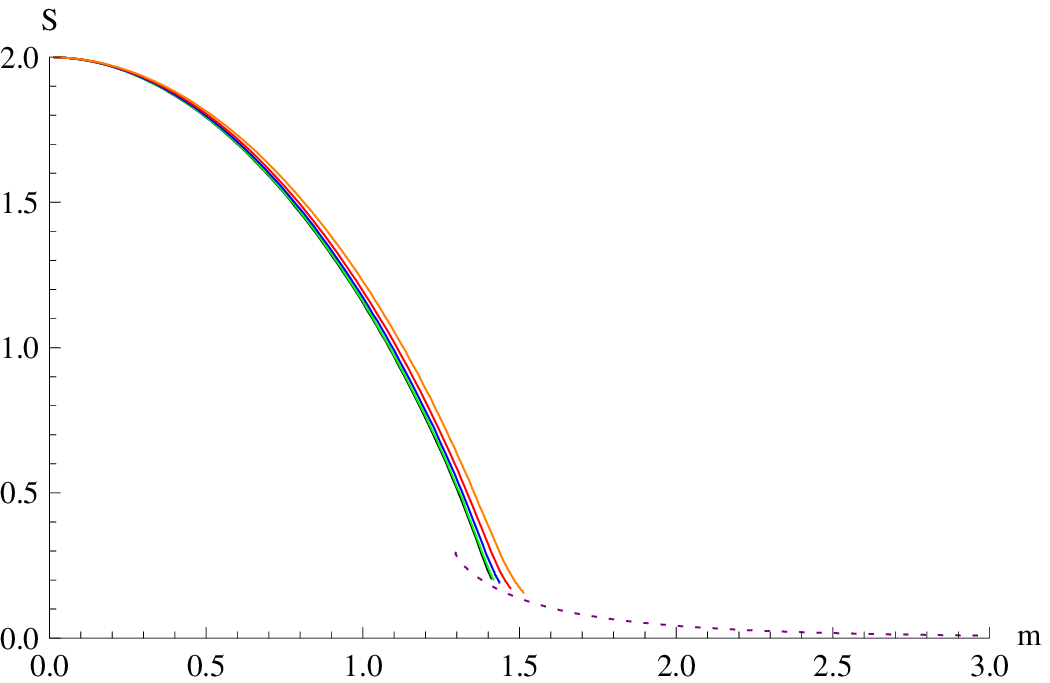}}
\hspace{8pt}
\subfigure[]{\includegraphics[width=0.45\textwidth]{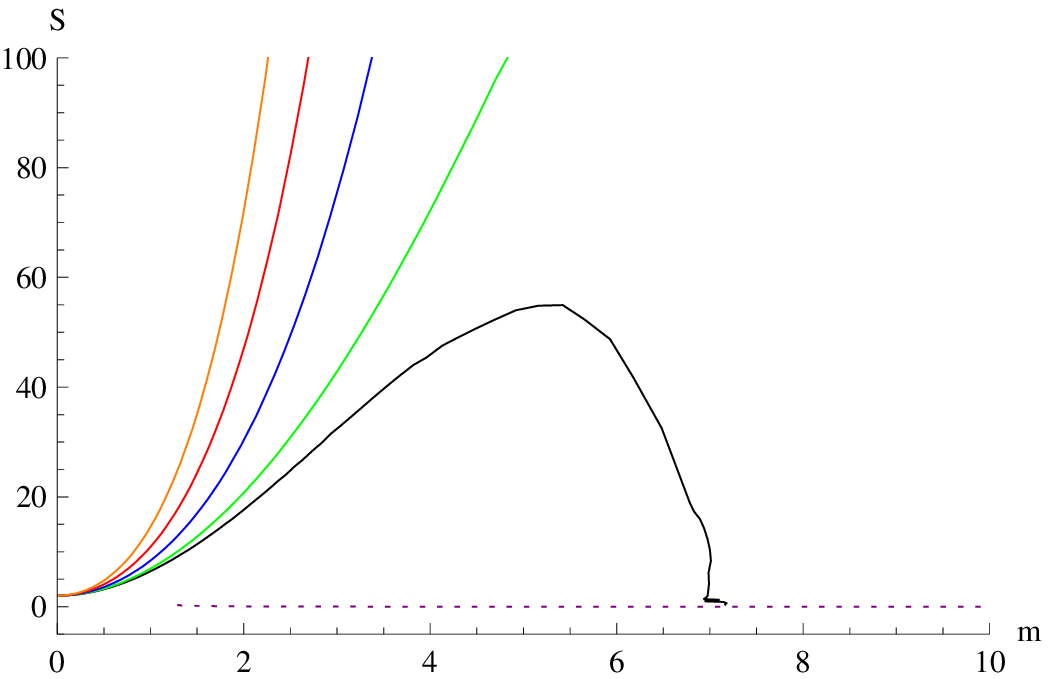}}
\caption{The dimensionless entropy $\cals_7$ versus the mass parameter $m$ as defined in equation \eqref{dicmc} at baryon chemical potential $\mu^B/M_q=0.1$ (left) and  $\mu^B/M_q=0.8$ (right) for the case $N_f=2$. The dotted purple curve corresponds to Minkowski embeddings and the five other curves to black hole embeddings with isospin chemical potential $\mu^I=0$ (black), $\mu^I=\frac{1}{4}\mu^B$ (green), $\mu^I=\frac{1}{2}\mu^B$ (blue), $\mu^I=\frac{3}{4}\mu^B$ (red) and $\mu^I=\mu^B$ (orange).}\label{fig:Sgrand}
\end{figure}

\begin{figure}[t]
\centering
\psfrag{m}{$m$}
\psfrag{dvs2}[b]{$\calv_s^2$}
\subfigure[]{\includegraphics[width=0.3\textwidth]{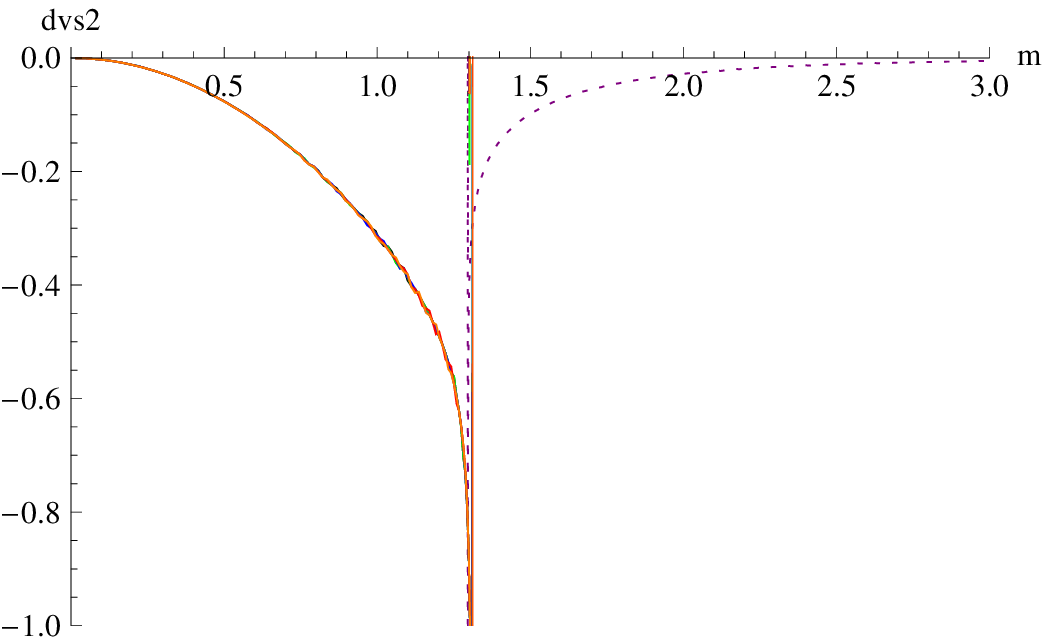}}
\hspace{8pt}
\subfigure[]{\includegraphics[width=0.3\textwidth]{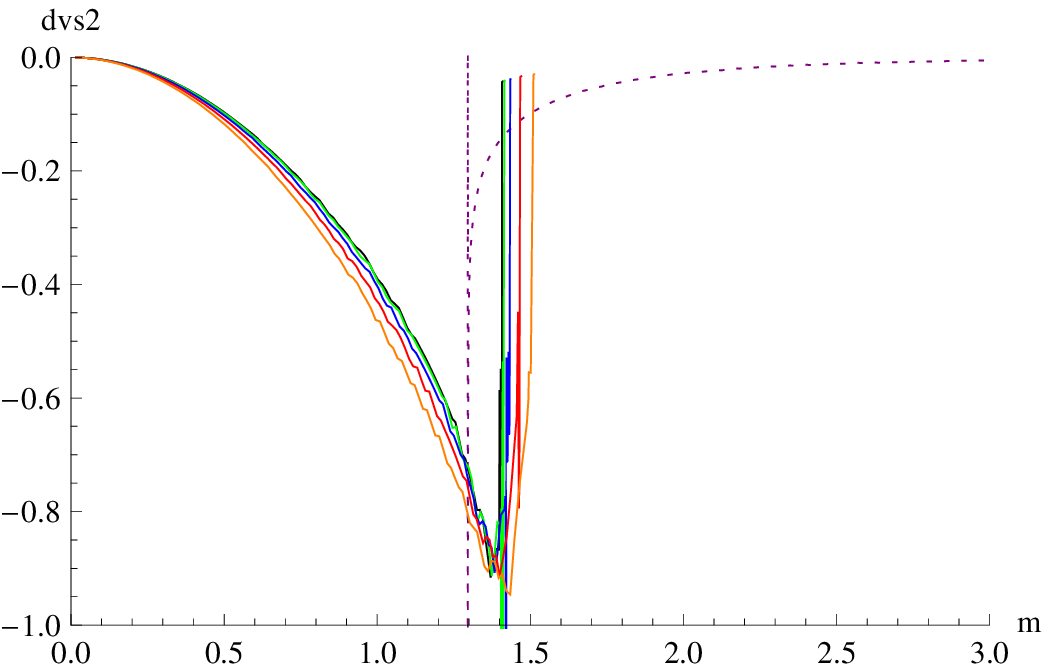}}
\hspace{8pt}
\subfigure[]{\includegraphics[width=0.3\textwidth]{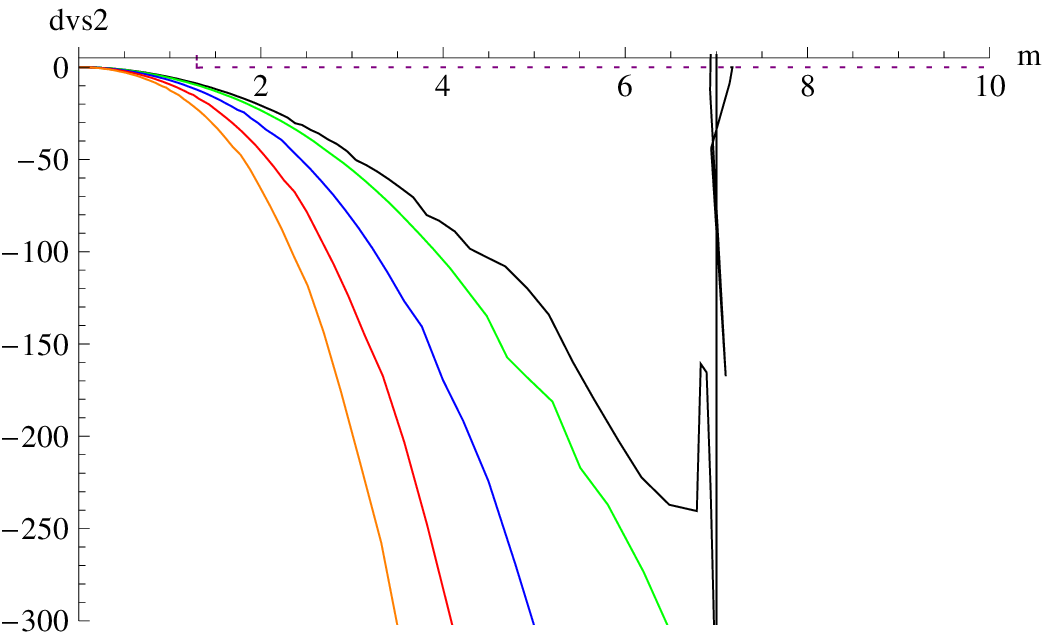}}
\caption{The dimensionless speed of sound relative to the conformal case $\calv_s^2$ versus the mass parameter $m$ as defined in equation \eqref{dicmc} at baryon chemical potential $\mu^B/M_q=0.01$ (a), $\mu^B/M_q=0.1$ (b) and $\mu^B/M_q=0.8$ (c) for the case $N_f=2$. The dotted purple curve corresponds to Minkowski embeddings and the five other curves to black hole embeddings with isospin chemical potential $\mu^I=0$ (black), $\mu^I=\frac{1}{4}\mu^B$ (green), $\mu^I=\frac{1}{2}\mu^B$ (blue), $\mu^I=\frac{3}{4}\mu^B$ (red) and $\mu^I=\mu^B$ (orange).}\label{fig:Vgrand}
\end{figure}

In fig.~\ref{fig:Omegagrand}, \ref{fig:Sgrand} and \ref{fig:Vgrand} we present numerical results for the thermodynamic quantities in the grand canonical ensemble for $N_f=2$ at different baryon and isospin chemical potentials.\par
Fig.~\ref{fig:Omegagrand} shows the dimensionless grand potential $\calw_7$ versus the mass parameter $m$ as calculated by equation (\ref{grandpotdim}). we compare the grand potential given in fig.~\ref{fig:Omegagrand} with the result where no finite charges are considered \cite{Mateos:2007vn}. The deviation between these two cases is induced by the finite densities given in fig.~\ref{fig:bgdBgrand} and \ref{fig:bgdIgrand}. Since the densities $\dt^{I_i}$ are all zero for $m=0$, there is no deviation. By increasing the mass $m$, the densities and the deviation grow. At $m\approx 1$ the densities are maximal for small chemical potentials and therefore,  the deviation becomes smaller as we increase $m$ further and we get a phase transition similar to the case without chemical potentials (see fig.~\ref{fig:Omegagrand} (a)). For larger chemical potentials but still $(\mu^B+\mu^I)/M_q<1$ (fig.~\ref{fig:Omegagrand} (b) and black curve in fig.~\ref{fig:Omegagrand} (c)), the deviation to the case without chemical potential is much larger and the position of the phase transition moves to larger mass $m$. For $(\mu^B+\mu^I)/M_q=1$ (green curve in fig.~\ref{fig:Omegagrand} (c)), the grand potential approaches a constant value in the limit $m\to\infty$. This value is smaller than the value in the Minkowski phase, such that there is no phase transition between black hole and Minkowski embeddings possible. For even larger chemical potentials $(\mu^B+\mu^I)/M_q>1$, the grand potential diverges to $-\infty$ as $m\to\infty$. We expect the black hole embeddings to be thermodynamically favored for all masses $m$ if $(\mu^B+\mu^I)/M_q> 1$.\par
Fig.~\ref{fig:Sgrand} shows the entropy $\cals_7$ versus the mass parameter $m$. For small chemical potentials, the entropy behaves as the zero chemical potential result from \cite{Mateos:2007vn}. By increasing the chemical potentials the spiral behavior disappears and the entropy decreases monotonically (see fig.~\ref{fig:Sgrand} (a)). For large chemical potentials presented in fig.~\ref{fig:Sgrand} (b) the entropy changes its behavior dramatically. It increases to a maximum and decreases again towards the values obtained for the Minkowski embeddings as we increase the mass $m$. For even larger chemical potentials the entropy diverges to $\infty$ as $m\to\infty$. The energy behaves like the entropy and we therefore do not present it separately.\par
Fig.~\ref{fig:Vgrand} shows the speed of sound $\calv_s^2$ as calculated from equation (\ref{speedgrand}). In the speed of sound there is no dramatic change aside from the transition at $(\mu^B+\mu^I)/M_q=1$ as we increase the chemical potential. By increasing the mass $m$ it always decreases until the phase transition and increases again in the Minkowski phase. The magnitude of the speed of sound close to the phase transition increases as we increase the chemical potentials.

\subsection{Phase transition/phase diagram in the grand canonical ensemble}\label{phasediagramgrand}

\begin{figure}
\centering
\psfrag{m}{$m$}
\psfrag{Omega}[b]{$\calw_7$}
\psfrag{dvs2}[b]{$\calv_s^2$}
\psfrag{c}[b]{$c$}
\subfigure[]{\includegraphics[width=0.3\textwidth]{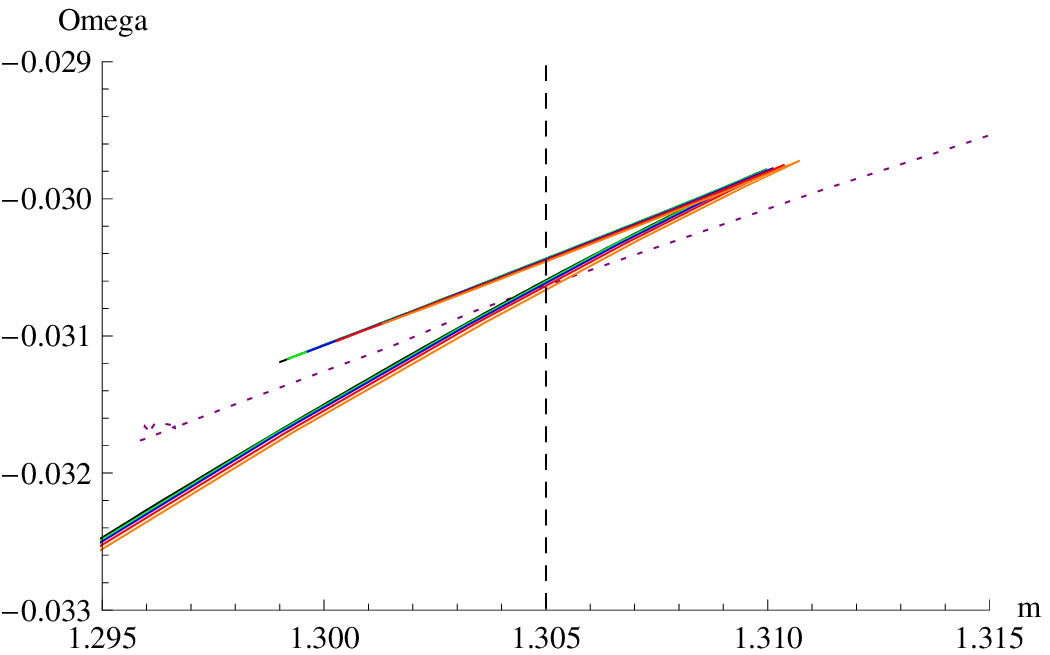}}
\hspace{8pt}
\subfigure[]{\includegraphics[width=0.3\textwidth]{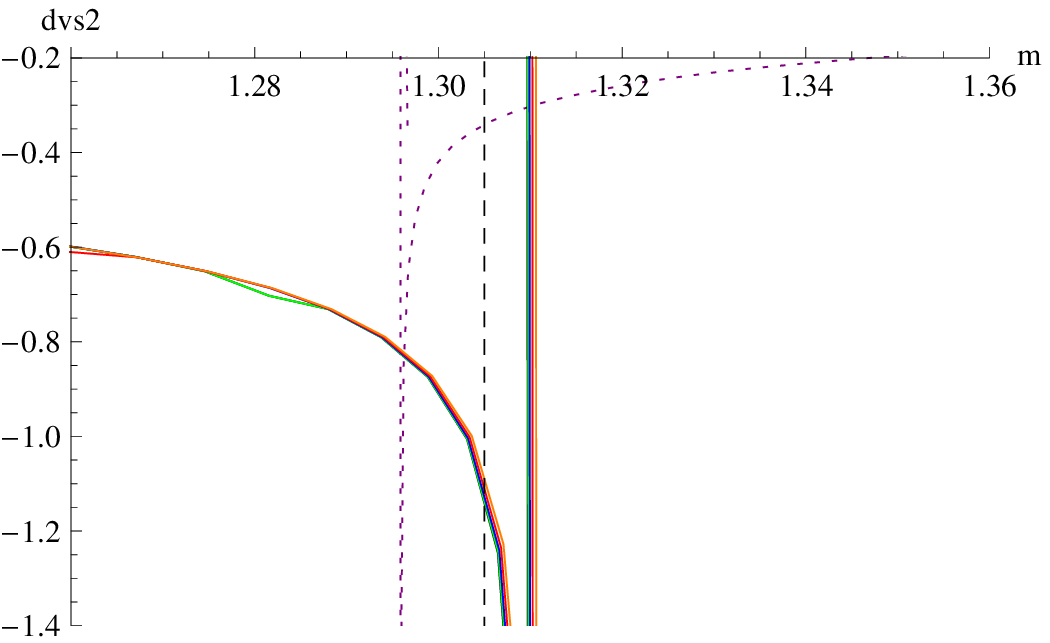}}
\hspace{8pt}
\subfigure[]{\includegraphics[width=0.3\textwidth]{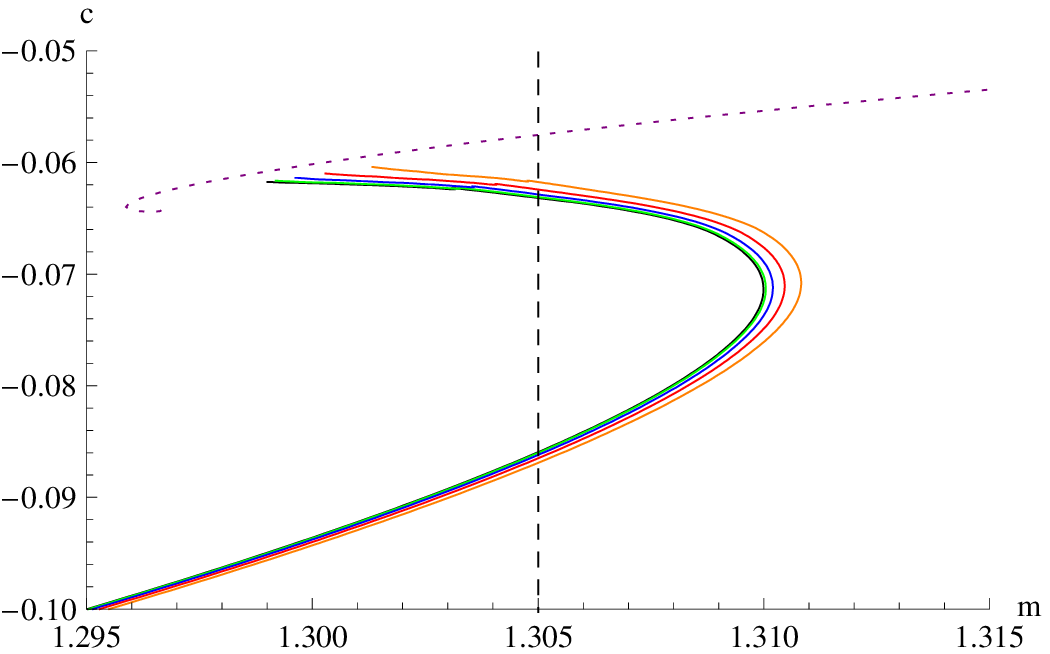}}
\caption{(a) The dimensionless grand potential $\calw_7$, (b) speed of sound $\calv_s^2$ and (d) the chiral condensate $c$ versus the mass parameter $m$ as defined in equation \eqref{dicmc} near the phase transition at $\mu^B/M_q=0.01$ for $N_f=2$. The different colors of the curves corresponds to the different isospin chemical potentials $\mu^I=0$ (black), $\mu^I=\frac{1}{4}\mu^B$ (green), $\mu^I=\frac{1}{2}\mu^B$ (blue), $\mu^I=\frac{3}{4}\mu^B$ (red) and $\mu^I=\mu^B$ (orange). The purple curve corresponds to Minkowski embeddings.}\label{fig:phasetransgrand}
\end{figure}

\begin{figure}
\centering
\psfrag{mu/M}[r][][1][-90]{$\frac{\mu}{M_q}$}
\psfrag{T/M}[t]{$m^{-1}$}
\psfrag{A}{A}
\psfrag{B}{B}
\psfrag{C}{C}
\psfrag{m}[l]{$m^{-1}$}
\subfigure[]{\includegraphics[width=0.45\textwidth]{./figs/sketchgrand.eps}}
\hspace{8pt}
\subfigure[]{\includegraphics[width=0.45\textwidth]{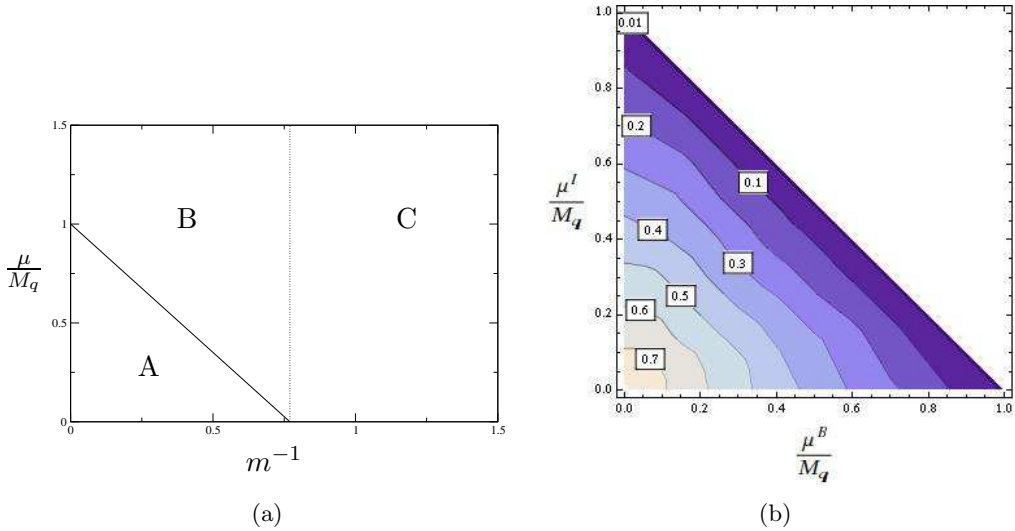}}
\caption{(a) Sketch of the possible embeddings in the grand canonical ensemble for one brane: In (A) only Minkowski embeddings, in (B) Minkowski and black hole embeddings and in (C) only black hole embeddings can exist. In (B) only one or a mixture of the two embeddings can be thermodynamically preferred.\newline
(b) Contour plot of the phase diagram in the grand canonical ensemble for $N_f=2$: Contours label the value $m^{-1}$ of the phase boundary. For small chemical potential the phase diagram is approximately $O(2)$ invariant in agreement with the symmetry discussed in section \ref{sym}. For larger chemical potentials the rotation symmetry breaks down and there is only an $\Z4$ symmetry left. For $(\mu^B+\mu^I)/M_q>1$, there is no transition between black hole and Minkowski embeddings. Below the hypersurface the Minkowski embeddings are thermodynamically favored. Above the hypersurface black hole embeddings are preferred.}\label{fig:phasediagramgrand}
\end{figure}

We review the regions in the phase diagram for the pure baryonic case, where Minkowski and black hole embeddings can be constructed. From \cite{Babington:2003vm, Mateos:2007vn} it is known that Minkowski embeddings can only exist for $m\gtrsim 1.3$. In \cite{Mateos:2007vc} it is shown that there is a region (large $m$ and small $\mu/M_q$), where black hole embeddings cannot exist (see also our numerical results in fig.~\ref{fig:bgmucan} (a) and (b)). Nevertheless, there is a region where Minkowski and black hole embeddings both can be constructed, but only one or a mixture of both is thermodynamically favored. In fig.~\ref{fig:phasediagramgrand} (a) we sketch these regions.\par
In the following we determine the position of the phase transition numerically. For this purpose, we look for a crossing point of the grand canonical potential calculated using black hole and Minkowski embeddings. Fig.~\ref{fig:phasetransgrand} shows the free energy, the speed of sound and the chiral condensate close to this transition. The entropy and the energy show a similar behavior as the chiral condensate.\par
By varying the chemical potentials, we map out the phase diagram for the grand
canonical ensemble shown in Fig.~\ref{fig:phasediagramgrand} (b). For lower
values of $m^{-1}$, the Minkowski embeddings are the thermodynamically favored
embeddings and for larger ones above the hypersurface the black hole
embeddings are preferred. The contours at small chemical potentials show that
the phase diagram is approximately $O(2)$ invariant, in agreement with the
discussion in section \ref{sym}. This rotation symmetry breaks down for larger
chemical potentials and the discrete $\Z4$ symmetry is left. For
$(\mu^B+\mu^I)/M_q>1$ the black hole embeddings are the thermodynamically
preferred embeddings for all masses $m$. Therefore, there is no phase
transition between black hole and Minkowski embeddings for chemical potentials
$(\mu^B+\mu^I)/M_q>1$. This is the natural extension of the pure baryonic
result \cite{Mateos:2007vc, Karch:2007br} considering the sum of the baryon
and isospin chemical potential $\mu^B+\mu^I$ as the relvant quantity.\par 
Our numerical results suggest that the phase transition between black hole and Minkowski embeddings is first order for finite chemical potentials as in the case of zero chemical potentials \cite{Babington:2003vm,Mateos:2007vn}. However, in \cite{Karch:2007br} the authors show analytically that the phase transition in the pure baryonic case is second order at zero temperature. We expect that the same is true if we include the isospin charges. Therefore, the order of the phase transition must change from first to second order at one point of the phase boundary. Due to numerical errors we cannot strictly distinguish between first and second order phase transitions\footnote{It is numerically very challenging to construct all black hole embeddings close to the phase transition.}. However, we argue in the following that this point should be located at zero temperature. In general at a critical point, where a phase boundary ends, a first order phase transition becomes second order \cite{0034-4885-50-7-001}. We do not observe that the phase transition disappears. Thus, a critical point can only be located a the end of the phase boundary, \ie at zero temperature or zero chemical potentials. Since the phase transition at zero chemical potentials is clearly first order\cite{Babington:2003vm,Mateos:2007vn}, the critical point can only be located at zero temperature. Therefore, we expect a second order phase transition at zero temperature. To strengthen this argument one has to expand the analytic studies in \cite{Karch:2007br} to the next order in the temperature $T$ and determine analytically the order of the phase transition for a small but finite temperature $T$.

\subsection{Grand canonical thermodynamics in the large quark mass limit}\label{largemgrand}

\begin{figure}
\centering
\subfigure[]{\includegraphics[width=0.45\textwidth]{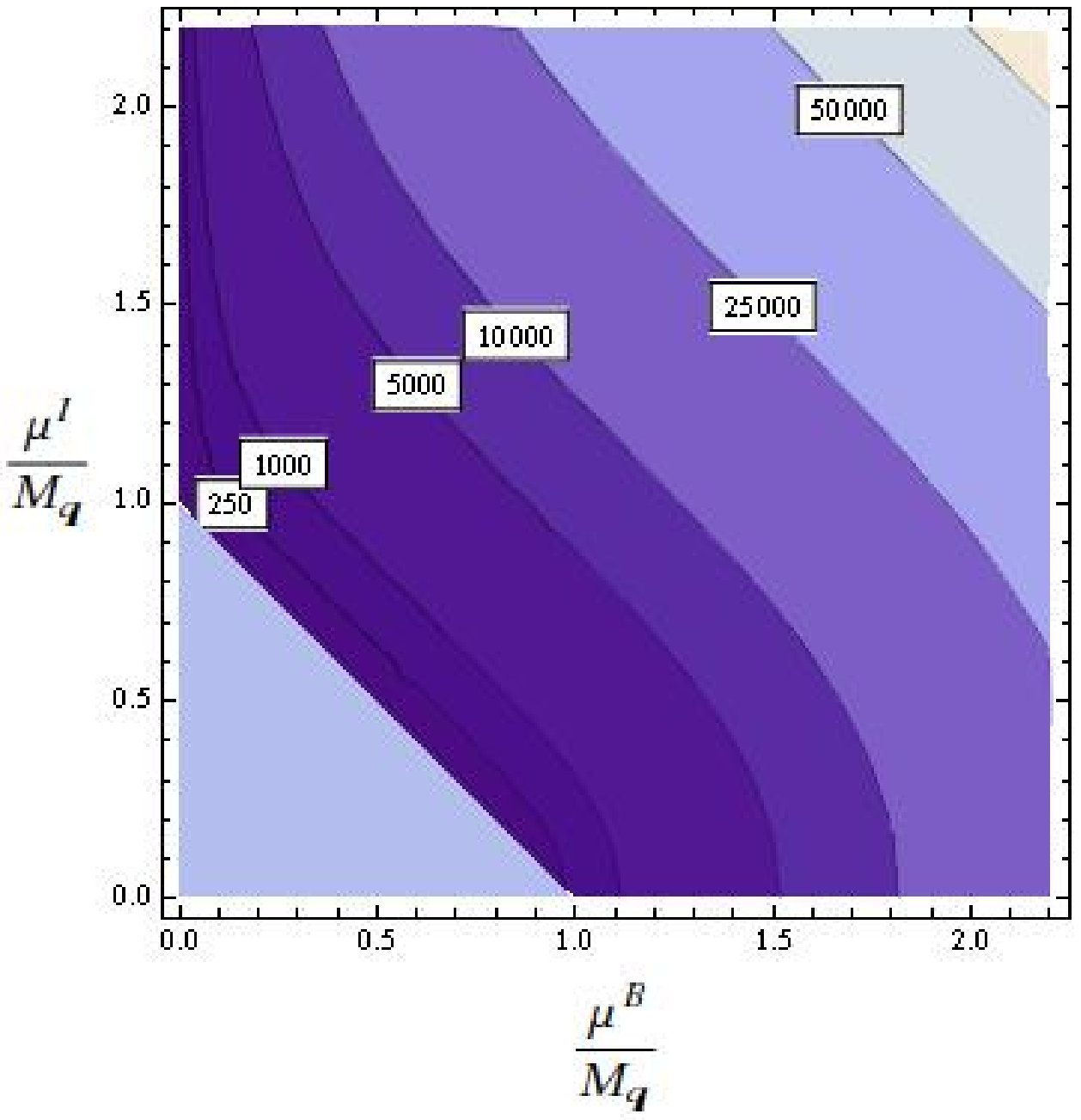}}
\hspace{8pt}
\subfigure[]{\includegraphics[width=0.45\textwidth]{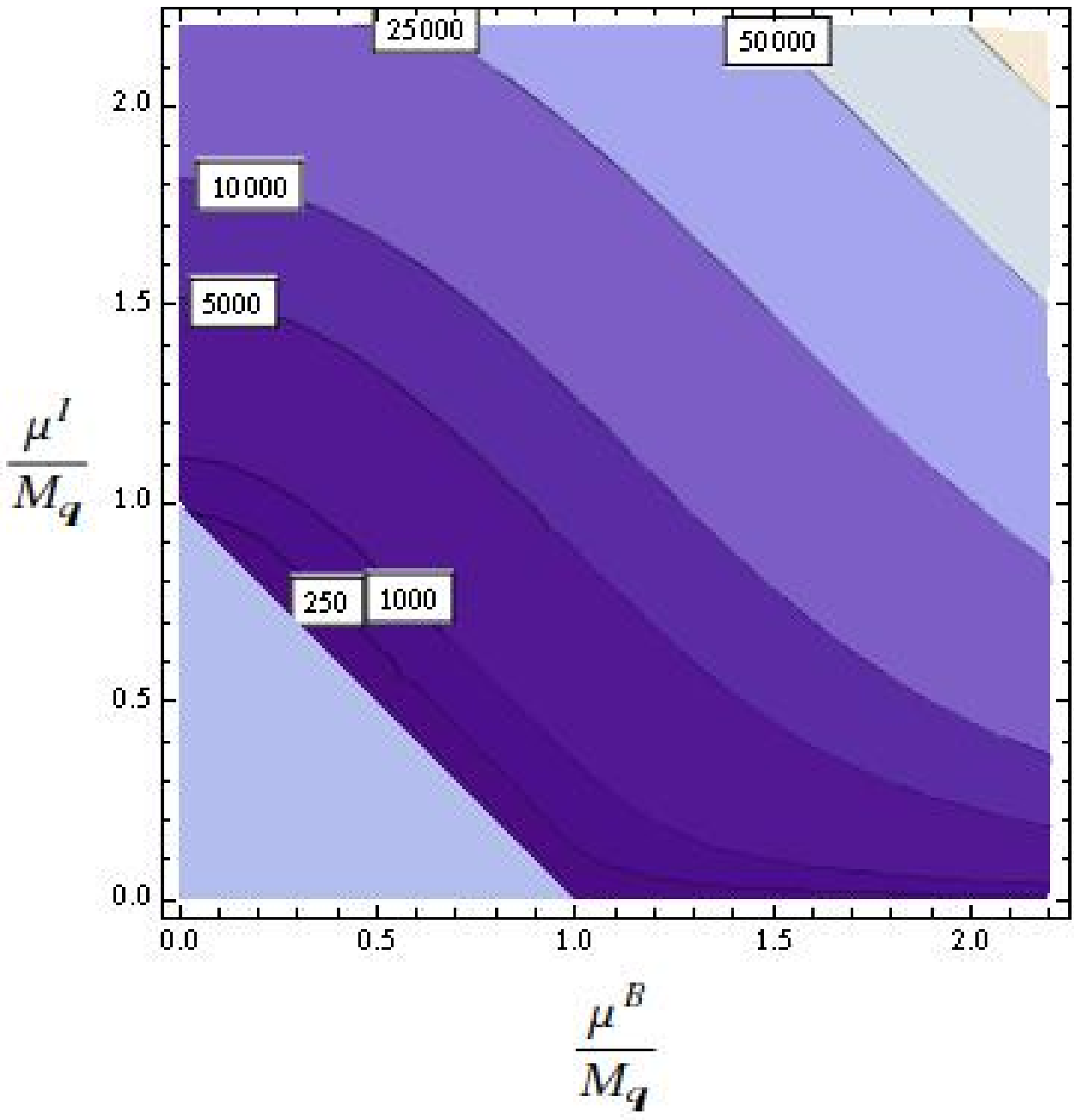}}
\caption{(a) The baryon density $\dt^B$, (b) the isospin density $\dt^I$  versus the baryon and isospin chemical potential $\mu^B/M_q$, $\mu^I/M_q$ at a fixed mass parameter $m=20$, the mass parameter $m$ is defined in equation \eqref{dicmc}.}\label{fig:largemdgrand}
\end{figure}

\begin{figure}
\centering
\subfigure[]{\includegraphics[width=0.45\textwidth]{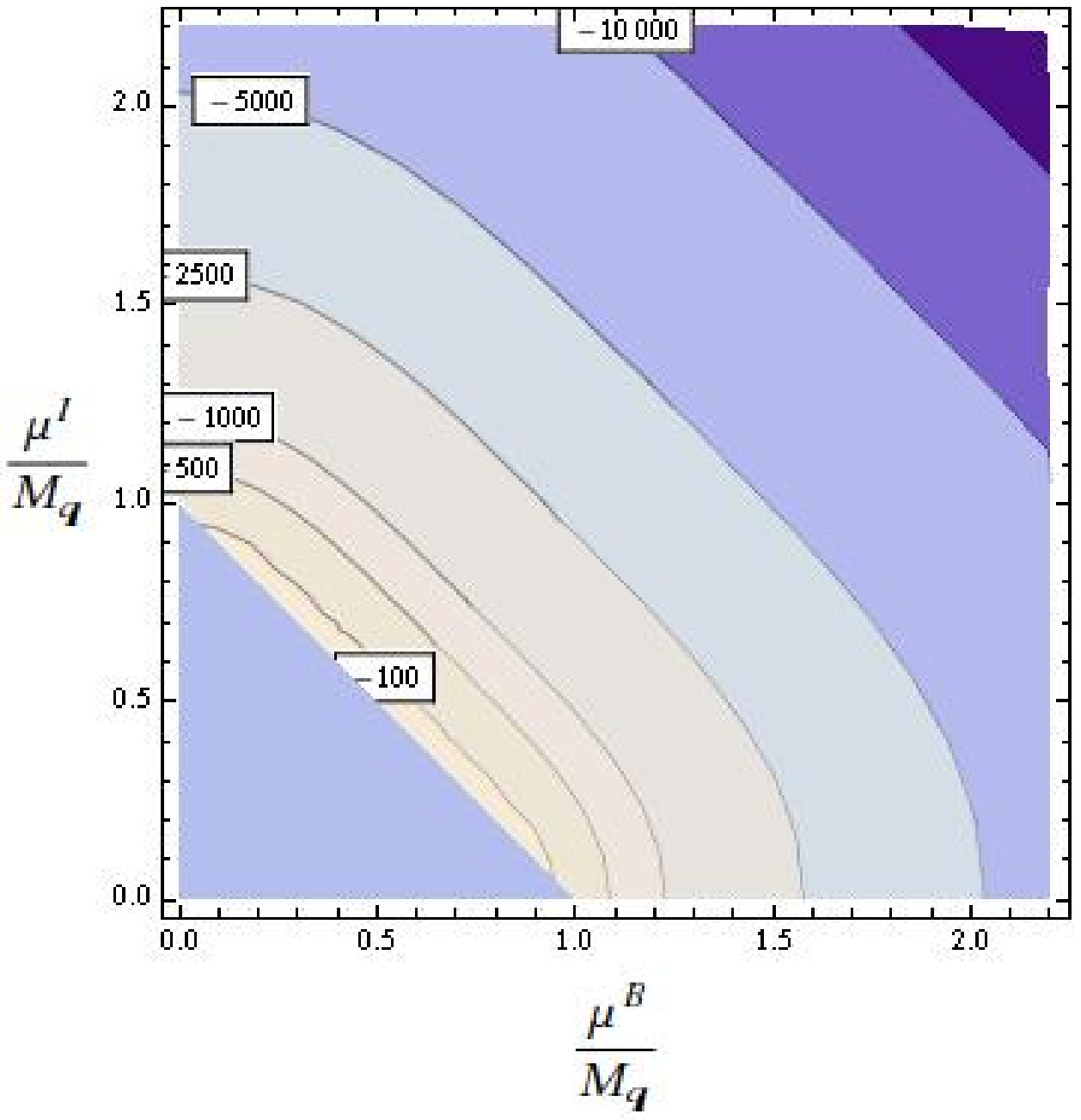}}
\hspace{8pt}
\subfigure[]{\includegraphics[width=0.45\textwidth]{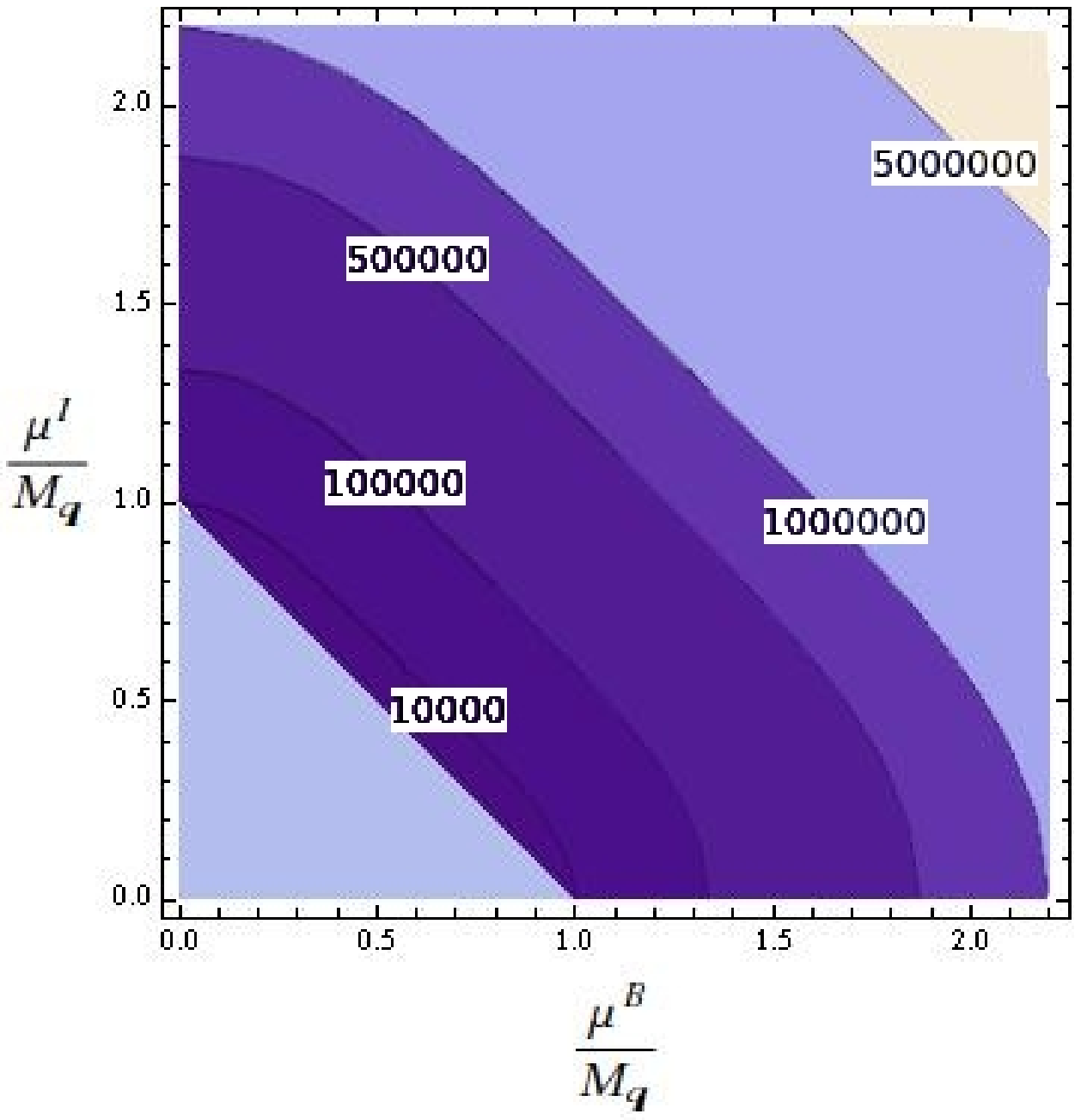}}
\caption{(a) The chiral condensate $c$ and (d) the energy $\cale_7$ versus the baryon and isospin chemical potential $\mu^B/M_q$, $\mu^I/M_q$ at a fixed mass parameter $m=20$, where the mass parameter $m$ is defined in equation \eqref{dicmc}.}\label{fig:largemcEgrand}
\end{figure}

In the large quark mass limit $m\to\infty$, \ie $M_q\gg T$, the Minkowski embeddings are thermodynamically favored for small chemical potentials $(\mu^B+\mu^I)/M_q\lesssim1$. Since in the Minkowski embedding the chiral condensate and the thermodynamic quantities are independent on the chemical potentials, they all vanish for $m\to\infty$ as in the case of zero chemical potentials \cite{Mateos:2007vn}. The asymptotic behaviour is given by
\begin{equation}
\calw_7\sim -\frac{1}{12m^4}\,,\qquad \cals_7\sim\frac{2}{3m^4}\,,\qquad \cale_7\sim\frac{7}{12m^4}\,,\qquad c\sim-\frac{1}{6m^5}\,.
\end{equation}
However for chemical potentials greater than the critical values, black hole embeddings with finite baryon and isospin densities are preferred. From the studies of the canonical ensemble, we know that the finite densities break supersymmetry and conformal invariance and therefore the thermodynamic quantities and the chiral condensate do not vanish in the large mass limit any more.\par
In this section we study the system in the grand canonical ensemble and compare it with the canonical ensemble, which we studied in section \ref{canlargem}. In fig.~\ref{fig:largemdgrand} and \ref{fig:largemcEgrand} we present the baryon and isospin density, the chiral condensate and energy at a fixed mass $m=20$ versus the baryon and isospin chemical potential. The grand potential and the entropy behave like the energy or the chiral condensate. The phase transition between Minkowski and black hole embeddings is along the line $(\mu^B+\mu^I)/M_q\approx 1$. Since the quantities do not depend on the chemical potentials in the Minkowski embeddings, we see a plateau in the region $(\mu^B+\mu^I)/M_q\lesssim 1$ where all the presented quantities in fig.~\ref{fig:largemdgrand} and \ref{fig:largemcEgrand} are approximately zero. After the phase transition the baryon and isospin density increase rapidly ($\dt^B, \dt^I\approx 8\cdot 10^4$ at $\mu^B/M_q=\mu^I/M_q=2.5$ see fig.~\ref{fig:largemdgrand}). Since we learn from the investigation of the canonical ensemble that the thermodynamic quantities and the chiral condensate depend at least linearly on the densities, the quantities also increase or decrease very fast after the phase transition (see fig.~\ref{fig:largemcEgrand}).\par
In the contour plots of the baryon and isospin density (see fig.~\ref{fig:largemdgrand}) we see a turning point in the contour lines at $\mu^B=\mu^I$ for small chemical potentials. Since there is a phase transition between the regions $0\le \dt^I<\dt^B$ and $0\le \dt^B<\dt^I$ in the canonical ensemble, we also expect a phase transition between the regions $0\le \mu^I<\mu^B$ and $0\le \mu^B<\mu^I$ in the grand canonical ensemble. In the canonical ensemble this phase transition is clearly marked by a discontinuous step in the chemical potentials, and by a kink in the thermodynamic quantities and the chiral condensate (see fig.~\ref{fig:largemcan}). In the grand canonical ensemble this phase transition is not visible in the contour plots of the baryon and isospin density. The thermodynamic quantities and the chiral condensate also appear to behave smoothly at this point (see fig.~\ref{fig:largemcEgrand}). Close to the line $\mu^B=\mu^I$, the thermodynamic quantities depend only on the sum of baryon and isospin chemical potential, $\mu^B+\mu^I$. It appears that there is an inconsistency in the transformation between the grand canonical and canonical ensemble since we only observe a phase transition in the canonical ensemble.\par
In the following we resolve the inconsistency and show that a new phase must be included in the grand canonical phase diagram (see fig.~\ref{fig:phasediagramgrand}). For this purpose, we determine the different regions in the grand canonical phase diagram fig.~\ref{fig:phasediagramgrand}, which the canonical and the grand canonical ensemble control. In the canonical ensemble, the possible values for the chemical potentials $\mu^{B/I}$ are $0$, $M_q/2$ and $M_q$ (see fig.~\ref{phasediagramlargemcan} and equations \eqref{eq:mulargem}, \eqref{eq:discontidB=dI}) in the limit $m\to\infty$. Different states in the canonical ensemble determined by $(\dt^B,\dt^I,m\to\infty)$ are mapped to a single state in the grand canonical ensemble, \eg $(\mu^B=M_q,\mu^I=0,m\to\infty)$. Therefore, the phase transition observed in the canonical ensemble (see fig.~\ref{phasediagramlargemcan}) in the limit $m\to\infty$ cannot be seen in the grand canonical ensemble since the two dimensional density plane is mapped to three different points in the grand canonical ensemble $(\mu^B=M_q,\mu^I=0)$, $(\mu^B=M_q/2,\mu^I=M_q/2)$ and $(\mu^B=0,\mu^I=M_q)$. These three points are located along the phase boundary of the Minkowski phase $(\mu^B+\mu^I)/M_q\approx 1$ in fig.~\ref{fig:largemdgrand} and \ref{fig:largemcEgrand}. However, we can analyze the plots in the grand canonical ensemble in this section for finite mass. We also observe in the canonical ensemble that for finite mass and large densities the phase transition in the density plane disappears slowly, such that there is no inconsistency between the canonical and grand canonical ensemble in the region of finite mass $m$.\par
Since we cannot map bijectively the canonical to grand canonical ensemble in the region of large mass  $m$ and chemical potentials $\mu^{B/I}\approx M_q$, the considered state of our system cannot be globally stable, \eg\cite[sec. III]{Cvetic:1999ne}. Since the energy develops huge values ($\cale_7\approx 10^6$ at $\mu^B=\mu^I=2M_q$ fig.~\ref{fig:largemcEgrand}), this also suggest that the considered state is globally unstable.\par
 In QCD, it is expected that mesons condense at small temperature and large isospin chemical potential, 
which corresponds to our unstable region. 
See 
\cite{Voskresensky:1997ub,Lenaghan:2001sd,Sannino:2002wp} 
for examples of vector condensation  
and~\cite{Splittorff:2000mm,Son:2000by,Son:2000xc}, as well as~\cite{Ebert:2008tp,He:2005nk} for examples of pion condensation.    
We expect similarly to QCD a new stable phase, which is determined by the condensation of mesons. These mesons are dual to fluctuations about the background. In section \ref{sec:mesons} we will discuss these fluctuations and indeed find the expected new phase.

\section{Hydrodynamics}
\label{sec:hydrodynamics}

\begin{figure}
\centering
\psfrag{dB}{$\dt^B$}
\psfrag{dI}{$\dt^I$}
\psfrag{m}{$m$}
\psfrag{DT}[b]{$\frac{DT}{2\pi}$}
\subfigure[]{\includegraphics[width=0.45\textwidth]{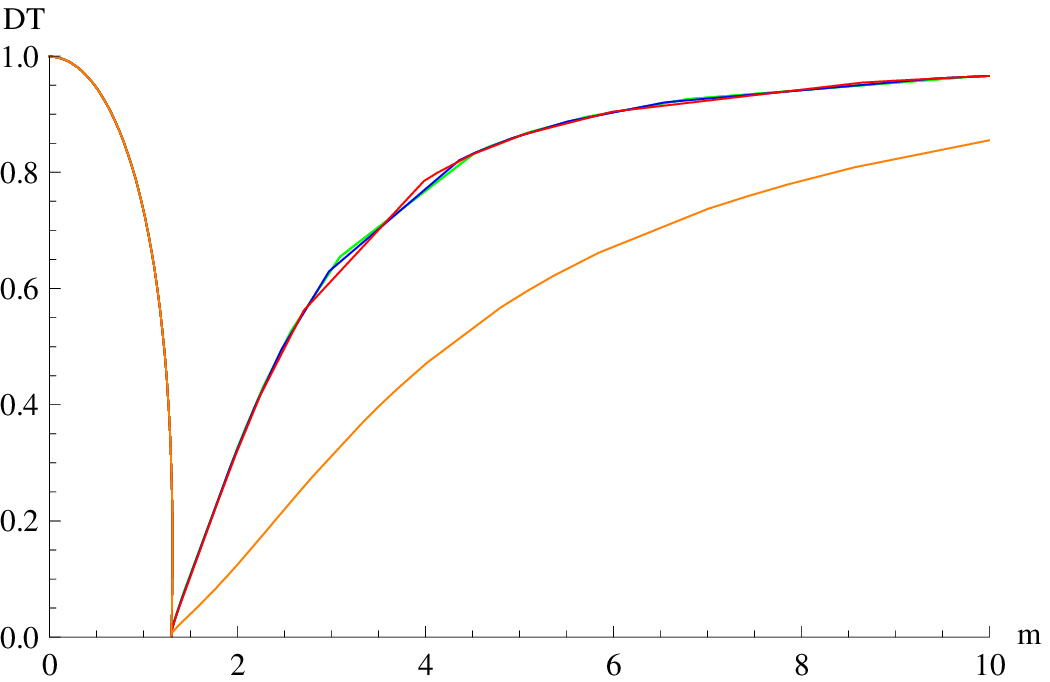}}
\hspace{8pt}
\subfigure[]{\includegraphics[width=0.45\textwidth]{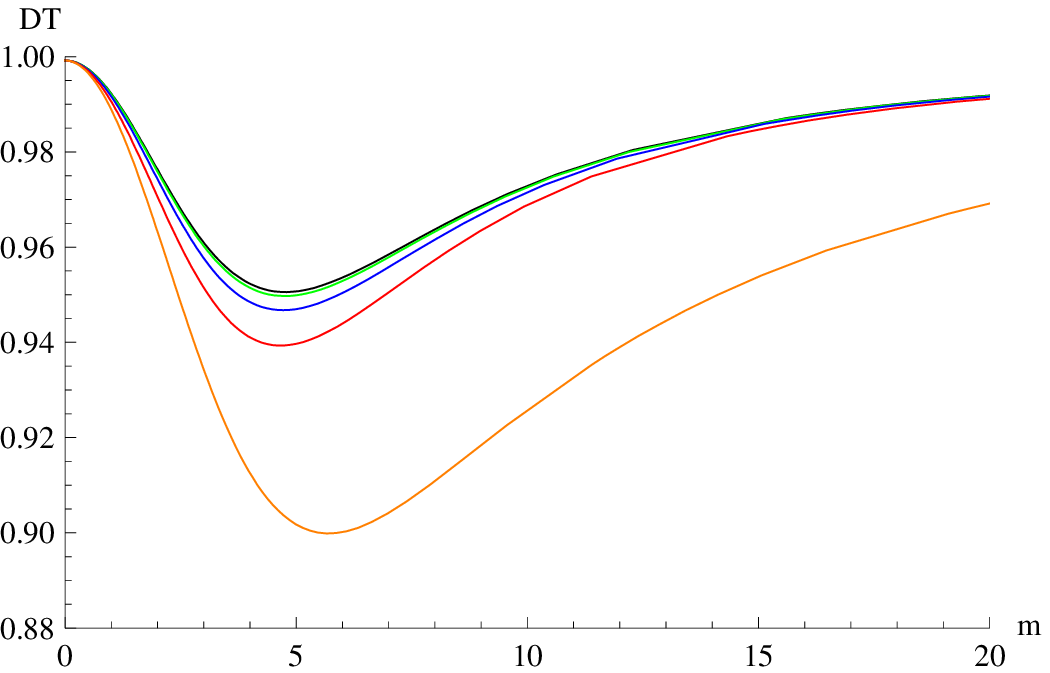}}
\subfigure[]{\includegraphics[width=0.45\textwidth]{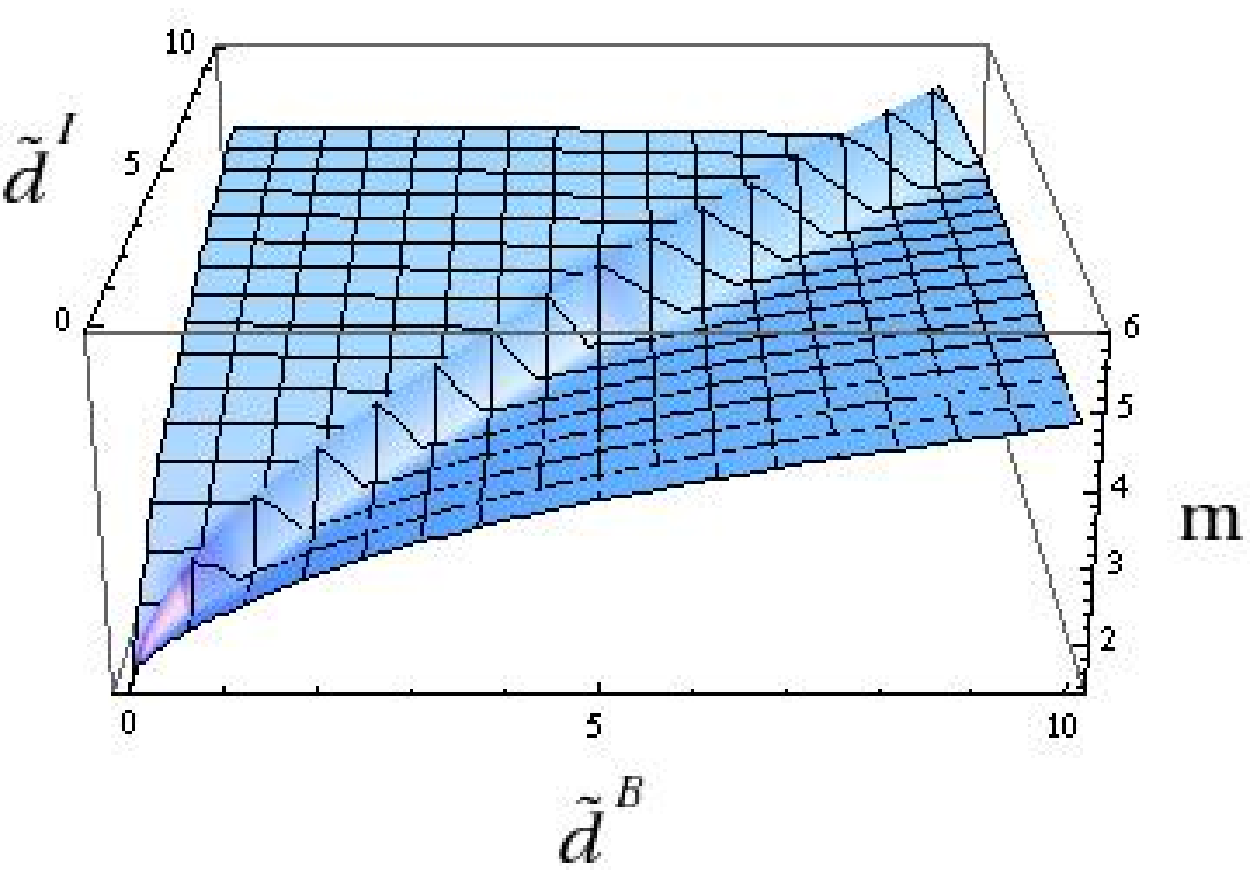}}
\hspace{20pt}
\subfigure[]{\includegraphics[width=0.45\textwidth]{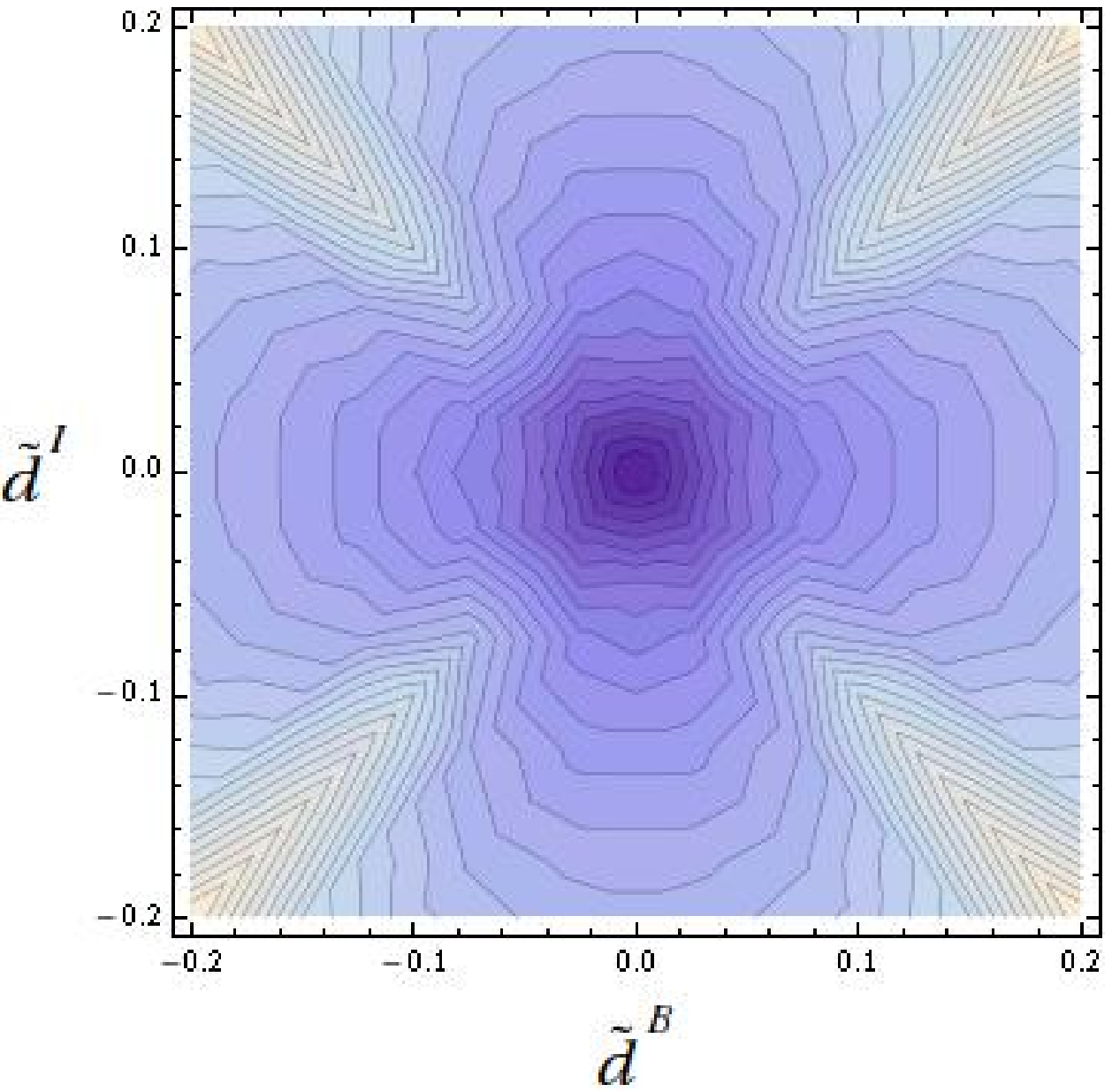}}
\caption{The diffusion coefficient $DT$ versus the mass parameter $m$ as defined in equation \eqref{dicmc} at $\dt^B=5\cdot 10^{-5}$ (a) and $\dt^B=20$ (b). The five different curves correspond to $\dt^I=0$ (black), $\dt^I=\frac{1}{4}\dt^B$ (green), $\dt^B=\frac{1}{2}\dt^B$ (blue), $\dt^B=\frac{3}{4}\dt^B$ (red) and $\dt^I=\dt^I$ (orange). (c) is the "enlarged" phase diagram, where we plot the position of the minimum of the diffusion coefficient versus the densities. (d) shows a contour plot of the "enlarged" phase diagram, which demonstrates the breaking of the rotational symmetry $O(2)$ down to the discrete symmetry $\Z4$.}\label{fig:diff}
\end{figure}

In this section we consider the effective diffusion coefficient computed from the membrane paradigm developed in \cite{Kovtun:2003wp}  and extended in~\cite{Myers:2007we}. The diffusion coefficient for the pure baryonic case is presented in \cite{Erdmenger:2007ja}. To include isospin density we use the following procedure. The finite baryon and isospin density enter the diffusion coefficient through the D$7$ embedding function~$\chi (\rho, \tilde d^B,\tilde d^I)$ which appears in the metric $G(\rho, \tilde d^B, \tilde d^I)$ \eqref{inducedmetric}. We obtain the explicit embedding function by solving its equation of motion~\eqref{eomchi} and then simply plug in the metric into the diffusion formula for the case of zero densities\footnote{Notice that a strict dervivation of the diffusion coefficent for the case of finite densities would change equation \eqref{membranepara} such that terms explicitly depending on the finite densities appear. We leave this task for further studies and restrict our analysis to equation \eqref{membranepara}.}
\begin{equation}\eqlabel{membranepara}
D=\frac{\sqrt{G}}{G_{11}\sqrt{-G_{00}G_{44}}}\Bigg|_{\rho=1}\int_{\rho=1}^\infty\frac{-G_{00}G_{44}}{\sqrt{-G}}\,.
\end{equation}
The calculation of $D$ makes use of the isospin depending embedding function $\chi(\rho,\dt^B,\dt^I)$. Thus, we call it the effective baryon diffusion coefficient. This procedure yields the plots given in figure~\ref{fig:diff} (a) and (b).\par
In the limit $m\to\infty$ and $m\to0$, the diffusion coefficient approaches always $D=1/(2\pi T)$, which coincides with the result for the diffusion coefficient of R-charges in $\caln=4$ SYM. For zero densities the diffusion coefficient $D$ vanishes at the phase transition because all the quarks are bound to mesons \cite{Myers:2007we}. If we increase the baryon and isospin densities, there are always free quarks available and the diffusion coefficient never vanishes \cite{Erdmenger:2007ja}. For densities larger than the critical one, the first order phase transition disappears. The only structure in $DT$ which survives in the large densities region is a minimum. We consider this minmum of the diffusion coefficient as a hydrodynamic crossover point. Including these crossover points in the thermodynamical phase diagram from fig.~\ref{fig:phasediagram} (a), we obtain fig.~\ref{fig:diff} (c), where the surface marks the phase transition and cross over point, respectively. As we expect, the rotational symmetry $O(2)$ is completely broken for large densities and in the case $N_f=2$ there is only the discrete symmetry $\Z4$ left.\par
It is important to notice is that the baryon density increases the diffusion coefficent as we see in fig.~\ref{fig:diff} (a), (b) and \cite{Erdmenger:2007ja}. Increasing the isospin density instead reduces the diffusuion coefficient (see fig.~\ref{fig:diff} (b)). Therefore, the effects of the baryon and the isospin density on this effective baryon diffusion coefficient are opposite.\par

\section{Mesons at finite chemical potential}
\label{sec:mesons}
	
	In this section we use the AdS/CFT correspondence to calculate the mesonic
	spectral function of the field theory by investigating the fluctuations of
	D$7$-branes. We then study the effects of chemical potentials on the meson
	spectrum.
	
	As in \cite{Myers:2007we}, we interpret resonances in the spectral functions
	of flavor currents as quasiparticle bound states, which are identified with
	mesons. A direct verification of this identification was given in
	\cite{Erdmenger:2007ja}. There the congruence of the spectra obtained from
	spectral functions and the analytically known meson spectra from
	\cite{Kruczenski:2003be} was shown in an appropriate limit.
	
	\subsection{Mesonic bound states and brane fluctuations}
		
		To describe mesons, we compute the two point correlation function of
		flavor currents $J(\vec x)$ of the field theory. The retarded Green
		function $G^R$ in momentum space may be written as
		\begin{equation}
			\label{eq:GreenRetarded}
			G^R(\omega, \bm q) = -\ii \int\!\dd^4x \; \ee^{\ii\,\vec k\vec x}\, \theta(x^0) \left<\left[J(\vec x), J(0)\right]\right>.
		\end{equation}
		We write $\vec k$ for the four-momentum vector with spatial
		three-momentum $\bm q$ and timelike component $\omega$, such that $\vec
		k=(\omega,\bm q)^\mathrm{T}$. The description of the spectrum of mesons
		is given in terms of the spectral function $\R(\omega,\bm q)$, defined
		as
		\begin{equation}
		\label{eq:specFuncFromG}
			\R(\omega, \bm q) = -2 \Im G^R(\omega, \bm q).
		\end{equation}
		Poles in the Green function give rise to peaks in the spectral function,
		which we will study below. The position, magnitude and width of these
		peaks are determined by the position and structure of the poles of
		$G^R$. They encode the mass and lifetime of the mesons
		\cite{Hoyos:2006gb,Hoyos:2007zz,Amado:2007yr}.
		
		To compute the current correlator in \eqref{eq:GreenRetarded}, we use
		the AdS propagator of the relevant supergravity field. The dual gravity
		field to the scalar, pseudo-scalar and vector flavor currents are the
		fluctuations of the probe D$7$-branes and the fluctuations of the gauge
		field on the branes. In terms of the coordinates introduced in
		section~\ref{sec:holographicSetup}, the scalar currents are dual to
		fluctuations of the scalar embedding function $\theta(\rho)$, the
		pseudo-scalar currents correspond to fluctuations of the angular
		embedding $\phi(\rho)$, and the vector current is mapped to fluctuations
		of the gauge field $A(\rho)$ on the branes.

		\subsubsection{Calculation of spectral functions}
			
			The computation is performed following the prescription of
			\cite{Son:2002sd,Herzog:2002pc}. Moreover, in this section we extend
			the results of \cite{Erdmenger:2007ja} and make use of the
			calculations performed there.
			
			The method of calculating spectral functions in a holographic model
			amounts to solving the equations of motion for a supergravity field
			$\tilde A(\rho,\bm q)$ in AdS space.  Then evaluate
			\begin{equation}
				G^R(\omega,\bm q) = \frac{N_fN_cT^2}{8}\,\lim_{\rho\to\infty}\left(\rho^3\frac{\partial_\rho \tilde A(\rho,\vec k)}{\tilde A(\rho,\vec k)}\right)
			\end{equation}
			to obtain the Green function and eventually make use of
			\eqref{eq:specFuncFromG}, which gives the spectral function in terms
			of the solution $\tilde A(\rho,\vec k)$,
			\begin{equation}
			\label{eq:specFuncFromA}
				\R(\omega,\vec k) = - \frac{N_f N_c T^2}{4}\; \mathrm{Im}\lim_{\rho\to\infty}\left(\rho^3 \frac{\partial_\rho \tilde A(\rho)}{\tilde A(\rho)} \right).
			\end{equation}
			Note that the formulae for $G^R$ and $\R$ above describe the result
			of the procedure described in \cite{Son:2002sd}, tailored to the
			field and brane configuration used in this paper.
			
			The remaining task is to derive and solve the equations of motion
			for the field $\tilde A$. We will now follow this procedure for the
			cases of non-vanishing baryonic or isospin chemical potential in the
			case of a theory with $N_f=2$ flavors. In particular we concentrate
			on the vector mesons, dual to fluctuations $\tilde A$ of the gauge
			field $A$ on the D$7$-brane. For simplicity, we restrict ourselves
			to the limit of vanishing spatial momentum $\bm q=0$. The vector
			mesons show the interesting feature that the spectral lines split up
			in the case of non-vanishing isospin chemical potential
			\cite{Erdmenger:2007ja}. In section~\ref{sec:spectrumIso} we will
			make use of this property to explore the phase diagram of
			fundamental matter in the D$3$/D$7$ setup.
	
	\subsection{Spectra at finite baryon density}
		
		\subsubsection{Equations of motion}
		
			In the case of pure baryonic chemical potential the relevant action
			is the DBI-action
			\eqref{actiondiagNf=2}, which now takes the form
			\begin{equation}
			\label{eq:DBIbaryonic}
				S = -T_{\text{D$7$}}\int\!\dd^8\xi\; \mathop\mathrm{Str} \sqrt{\left| \det \left( G + 2\pi \alpha'\, F\right) \right|}.
			\end{equation}
			We have introduced chemical potentials in
			section~\ref{sec:introchempot} and derived the equation of motion
			for the background gauge field $A$ and their solution in
			section~\ref{sec:eomGaugeFields}.
			
			To calculate the mesonic spectral functions, we now investigate
			small fluctuations of the background field configuration, which
			extremizes the action \eqref{eq:DBIbaryonic}. Therefore we add small
			fluctuations $\tilde A$ to the background $A$,
			\begin{align}
								 A &\mapsto A + \tilde A,\\
				\Rightarrow\quad F &\mapsto F + \tilde F.\hphantom{\quad\Rightarrow}
			\end{align}
			Note that according to section~\ref{sec:introchempot}, only the
			timelike component $A_0(\rho)$ of the background field is
			non-vanishing. The fluctuations, however, are gauged to have
			non-vanishing components along all Minkowski directions and depend
			on the Minkowski coordinates $\vec x$ and on the radial coordinate
			$\rho$, $\tilde A_{0,1,2,3}=\tilde A_{0,1,2,3}(\vec x,\rho)$ and
			$\tilde A_4\equiv 0$. We write
			\begin{equation}
				\mathcal G \equiv G + 2\pi\alpha' F
			\end{equation}
			and in this way cast the Lagrangian for the fluctuations into the
			form
			\begin{equation}
				\mathcal L = \sqrt{\left| \det \left( \mathcal G + 2\pi\alpha'\,\tilde F\right)\right|}.
			\end{equation}
			
			In general the non-linear structure of the Lagrangian induces
			couplings between vector mesons, scalar and pseudo-scalar mesons, as
			there will be couplings of fluctuations in $\mathcal G$ and those of
			$F=\dd A$. However, in the limit of $\bm q\to 0$, which we are
			working in, these couplings do not occur \cite{Mas:2008jz}. We
			assume the fluctuations $\tilde A$ to be small, justifying the
			consideration of the linearized equations of motion for the field
			components $\tilde A_\mu$,
			\begin{equation}
				\label{eq:eomFluct}
					0=  \partial_\nu\left[  \sqrt{\left|\det \mathcal G\right|}
						 \left( \mathcal G^{\mu\nu}\mathcal G^{\sigma\gamma}-\mathcal G^{\mu\sigma}\mathcal G^{\nu\gamma} - \mathcal G^{[\nu \sigma]} \mathcal G^{\gamma\mu} \right) \partial_{[\gamma}\tilde A_{\mu ]} \right],
			\end{equation}
			where we write upper indices on $\mathcal G$ to denote elements of
			$\mathcal G^{-1}$.
			
			As in \cite{Myers:2007we} we should consider gauge invariant
			combinations of field components only. For arbitrary momentum we are
			free to choose a reference frame in Minkowski space, such that $\vec
			k=(\omega,q,0,0)$, and we have gauge invariant longitudinal and
			transversal components
			\begin{equation}
				\label{eq:EfromA}
				E_\parallel = \omega \tilde A_x + q \tilde A_0,\qquad E_\perp = \omega \tilde A_{x,z}.
			\end{equation}
			In the limit $\bm q\to 0$, the equations of motion (and therefore
			also the according Green functions) for all components become
			identical, since there is no distinction between longitudinal and
			transversal modes.
			
			In momentum space,
			\begin{equation}
				\label{eq:fourierA}
				\tilde A_\mu(\rho,\vec{x}) = \int\!\! \frac{\dd^4 k}{(2\pi)^4}\, \ee^{\ii\vec{k}\vec{x}} \tilde A_\mu(\rho,\vec{k}) \,,
			\end{equation}
			the equations of motion \eqref{eq:eomFluct} for the gauge invariant
			fields are given by
			\begin{equation}
			\begin{split}
				\label{eq:eomEq0}
				0 =\,& E''+ \frac{\partial_\rho\left(\sqrt{|\det G|}G^{22}G^{44}\right)}{\sqrt{|\det G|} G^{22}G^{44}}\,
					  E'- \frac{G^{00}}{G^{44}}\, \varrho_H^2 \omega^2 E \\  
				 =\,& E'' + \partial_\rho \ln \left(\frac{\rho^3 f \left(1-\chi^2\right)^2}{\sqrt{1 - \chi^2 + \rho^2 {\chi'}^2 - 2 \frac{f}{\tilde f^2} (1-\chi^2)  (\partial_\rho A_0 )^2}}\right) E' \\
				&\hphantom{E''} + 8 \wn^2 \frac{\tilde f}{f^2} \frac{1-\chi^2 + \rho^2 {\chi'}^2 }{\rho^4 (1-\chi^2)}\,E,
			\end{split}
			\end{equation}
			with dimensionless $\wn = \omega/(2\pi T)$, and dimensionless
			coordinate $\rho$ as well as dimensionless fields $\tilde A\mapsto
			\frac{2\pi\alpha'}{\varrho_H}\tilde A$. A prime denotes a derivative
			with respect to $\rho$.

		\subsubsection{Numerical results and interpretation}
		\label{sec:resultsBaryonic}
			
			We solve the equation of motion \eqref{eq:eomEq0} numerically, using
			in-falling wave boundary conditions at the black hole horizon.
			Subsequently we make use of \eqref{eq:specFuncFromA} (where
			$A\mapsto E$) to compute the spectral function. Remember that this
			function depends parametrically on the particle density (or on the
			chemical potential, respectively), which influences the background
			field solutions $A$ and the embeddings $\chi$ in \eqref{eq:eomEq0}.
			
\begin{figure}
				\centering
				\psfrag{w}{\small $\wn$}
				\psfrag{R}{\small $\mathfrak{R}$}
				\psfrag{R0}{\footnotesize $\mathfrak{R}_0$}
				\psfrag{Rtemp}{\footnotesize $\mathfrak{R}(\wn,0)$}
				\psfrag{dt025}{\footnotesize $\tilde d = 0.25$}
				\psfrag{m5}{\footnotesize $m=5$}
				\includegraphics[width=.6\linewidth]{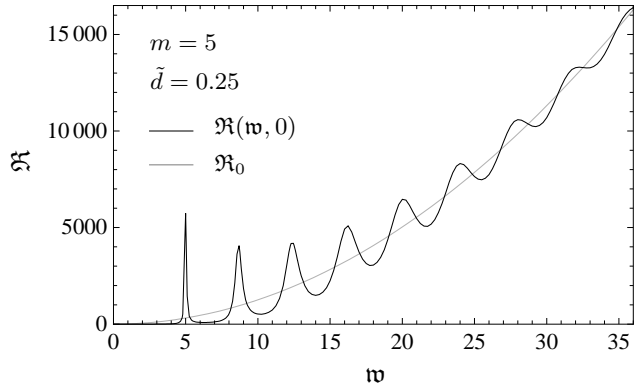}
				\caption{A spectral function oscillating around the asymptotic
					result $\R_0$. We plot spectral functions in units of
					$N_fN_cT^2/4$.}
				\label{fig:exampleSpecFunc}
			\end{figure}

			The large $\wn$ behavior of the spectral functions can be found
			analytically and is given by
			\begin{equation}
				\R_0(\wn,0) = N_fN_cT^2\pi\wn^2.
			\end{equation}
			An example for a spectral function and the oscillations around
			$\R_0$ can be found in fig.~\ref{fig:exampleSpecFunc}. One clearly
			sees the various excitations as peaks in the spectral function. Each
			peak corresponds to a quark-antiquark bound state with vanishing
			angular momentum.
			
			The position $\wn_n$ of a peak corresponds to a (dimensionless)
			meson mass $M_n=|\vec k|/(2\pi T)=\wn_n$. In \cite{Erdmenger:2007ja}
			it was observed that the meson masses obtained in this way indeed
			agree with the spectrum from \cite{Kruczenski:2003be}, found
			analytically in the supersymmetric zero temperature limit $\tilde
			d\to 0$, $m\to\infty$,
			\begin{equation}
			\label{eq:susyMesonSpectrum}
				\wn_n = \frac{M_n}{2\pi T}= m\,\sqrt{\frac{(n+2)(n+1)}{2}}\,,\qquad n=0,1,2,\ldots,
			\end{equation}
			where $m$ is the parameter defined in \eqref{dicmc}. At finite
			temperature and baryon density we will observe deviations from this
			spectrum. Nevertheless, at small densities and not too small $m$,
			the first peaks in the spectrum still come very close to this
			formula, see fig.~\ref{fig:exampleSpecFunc} with the first peak
			corresponding to $n=0$, the second to $n=1$ and so on.
			
			The dependence on temperature is related to the dependence on the
			quark mass, since the relevant parameter in the D$3$/D$7$-setup is
			the quotient $m$. The dependence on this parameter was intensively
			investigated in \cite{Erdmenger:2007ja}, extensions to finite
			momentum and the coupling to scalar and pseudo-scalar modes can be
			found in \cite{Myers:2008cj,Mas:2008jz}.
			
\begin{figure}
				\centering
				\psfrag{m5}{\footnotesize $m=5$}
				\psfrag{w}{$\wn$}
				\psfrag{R}{$\mathfrak{R}(\wn,0)$}
				\psfrag{dt1}{\footnotesize $\tilde d=1$}
				\psfrag{dt2}{\footnotesize $\tilde d=2$}
				\psfrag{dt20}{\footnotesize $\tilde d=20$}
				\psfrag{dt200}{\footnotesize $\tilde d=200$}
				\psfrag{dt2000}{\footnotesize $\tilde d=2000$}
				\includegraphics[width=.6\linewidth]{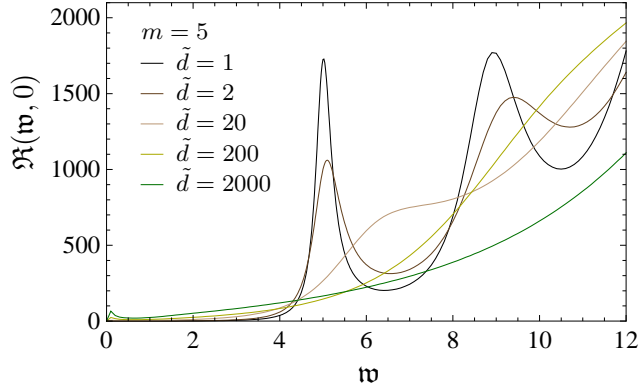}
				\caption{Spectral functions for various baryon densities
					$\tilde d$, again normalized to $N_fN_cT^2/4$. The
					black and brown curves for small $\tilde d$ show meson
					states around $\wn_0=5$ and higher excitations. At
					higher values of $\tilde d$ these excitations
					disappear, and at very high densities a new structure forms
					at small $\wn$.}
				\label{fig:variousDT}
			\end{figure}

			We work in the canonical ensemble and will now investigate the
			effects of variations in $\tilde d$. Spectral functions for various
			finite baryonic $\tilde d$ are shown in fig.~\ref{fig:variousDT}.
			Again, the peaks in these spectral functions indicate that quarks
			form bound states. At low baryon densities the positions of the
			peaks agree with the supersymmetric result
			\eqref{eq:susyMesonSpectrum}, e.\,g.\ the peaks at $\wn_0\approx 5$
			and $\wn_1\approx 8.7$ in the black curve correspond to the
			excitations with $n=0$ and $n=1$ in \eqref{eq:susyMesonSpectrum}.
			Increasing the quark density leads to a broadening of the peaks,
			which indicates decreasing stability of mesons at increasing baryon
			density \cite{Aharony:2007uu,Paredes:2008nf}. At the same time the
			positions of the peaks change, which indicates a dependence of the
			meson mass on the baryon density. Further increasing the quark
			densities leads to the formation of a new structure at $\wn<1$.
			We will discuss this structure together with the results at finite
			isospin density.
			
	\subsection{Spectra at finite isospin density}
	\label{sec:spectrumIso}
		
		\subsubsection{Equations of motion}
			
			To investigate isospin effects, we again start from the
			action~\eqref{actiondiagNf=2}. This time we choose $F$ as in
			\eqref{U(2)expand} with only $F^I\neq0$. Since the $\rho$-dependent
			time component of the gauge field generates the chemical potential
			on the field theory side, again we only have non-zero components
			$F^a_{40}=-F^a_{04}$ in the background. Now $a$ is an index labeling
			the $SU(2)$ components. So our gauge fields $A$ not only carry
			Lorentz-indices but additionally $SU(2)$ gauge indices, as in
			$A^a_\mu$ with $a=1,2,3$. In general the field strength tensor
			therefore has elements $F^a_{\mu\nu} = 2\partial_{[\mu} A^a_{\nu]} +
			f^{abc}A^b_\mu A^c_\nu$, with structure constants
			$f^{abc}=\varepsilon^{abc}$.

			In the non-Abelian field strength tensor, the term quadratic in the
			gauge field describes a self interaction of the gauge field. The
			coupling constant for this interaction may be determined by a
			redefinition of the gauge field, such that the kinetic term of the
			effective four-dimensional theory has the canonical form. In
			appendix~\ref{sec:mesoncoupling} we show that the redefinition is
			given by
			\begin{equation}
				A\mapsto \frac{c}{\sqrt{\lambda}}A\,,
			\end{equation}
			where the dimensionless constant $c$ depends on the geometry of the
			D$7$ worldvolume directions along $\rho$ and on the $S^3$, which are
			transverse to the directions of the D$3$. In particular, $c$ is
			independent of the 't~Hooft coupling $\lambda$. Determining the
			exact value of $c$ is left to further work in terms of the ideas
			presented in appendix~\ref{sec:mesoncoupling}. In the following we
			set $c=\frac{4\pi}{\sqrt 2}$. The field strength tensor in the
			redefined fields is given by
			\begin{equation}
			  F^a_{\mu\nu}=2\partial_{[\mu} A^a_{\nu]}+\frac{c}{\sqrt{\lambda}}f^{abc}A^b_\mu A^c_\nu
			\end{equation}
			
			We make use of the derivations performed in \cite{Erdmenger:2007ja}
			(where the above redefinition was implicitly used) and assume the
			isospin chemical potential to be oriented the direction labeled by
			$a=3$, as motivated in~\eqref{U(2)expand}. Therefore the only
			non-vanishing background component is $F^3_{40}=-F^3_{04}$, which
			coincides with the background for the case of a baryonic chemical
			potential.
			
			In dimensionless quantities, as in the previous subsection, the
			linearized equations of motion in the case of vanishing spatial
			momentum for the components $\tilde A^a_\mu$ of gauge field
			fluctuations $\tilde A$ can be derived as
			\begin{equation}
			\begin{aligned}
			\label{eq:eomIsoFluct}  
				0=\;&\partial_\kappa \left[ \sqrt{\left|\det \mathcal G\right|}
				\left( \mathcal G^{\nu\kappa} \mathcal G^{\sigma\mu} - \mathcal G^{\nu\sigma} \mathcal G^{\kappa\mu} \right)
				\check F_{\mu\nu}^a \right] \\ 
				& - \sqrt{\left|\det \mathcal G\right|}
				 \frac{{\varrho_H}^2}{2\pi\alpha'} \, A_0^3 f^{ab3} \left( \mathcal G^{\nu 0} \mathcal G^{\sigma\mu} 
				 - \mathcal G^{\nu\sigma} \mathcal G^{0\mu} \right)\check F_{\mu\nu}^b,
			\end{aligned}
			\end{equation}
			where $\check{F}^a_{\mu\nu}=2\partial_{[\mu}\tilde
			A^a_{\nu]}+c/\sqrt{\lambda}f^{ab3} A_0^3( \delta_{0\mu} \tilde A_\nu^b+
			\delta_{0\nu} \tilde A_\mu^b)\,\varrho^2_H/(2\pi\alpha')$ contains
			all field strength contributions linear in $\tilde A$.
			
			To solve these equations of motion, we first introduce the gauge
			invariant fields $E^a$ as in~\eqref{eq:EfromA} and then perform
			another change of basis to express the equations of motion in terms
			of the fields
			\begin{equation}
				\label{eq:flavorTrafo}
				X =E^1+ \ii E ^2,\qquad	Y =E^1- \ii E ^2, \qquad E^3.
			\end{equation}
			In this basis the equations of motion decouple and may be written as
			\begin{alignat}{3}
			\label{eq:eomX}
				0=&\;{X}'' &&+\frac{\partial_\rho\left(\sqrt{\left| \det \mathcal{G} \right|} \mathcal{G}^{44} \mathcal{G}^{22}\right)}{\sqrt{\left|\det \mathcal{G}\right|} \mathcal{G}^{44} \mathcal{G}^{22}} {X}' &&-
				4 \frac{\varrho_H^4}{R^4} \frac{\mathcal{G}^{00}}{\mathcal{G}^{44}} \, \left (\wn- \mathfrak{m}\right )^2 X\, , \\    
			\label{eq:eomY}
				0=&\;{Y}'' &&+\frac{\partial_\rho\left(\sqrt{\left|\det \mathcal{G}\right|} \mathcal{G}^{44} \mathcal{G}^{22}\right)}{\sqrt{\left|\det \mathcal{G}\right|} \mathcal{G}^{44} \mathcal{G}^{22}} {Y}' &&-
				4 \frac{\varrho_H^4}{R^4}\frac{\mathcal{G}^{00}}{\mathcal{G}^{44}}\, \left (\wn+ \mathfrak m\right )^2 Y\, , \\   
			\label{eq:eomE3}
				0=&\;{E^3}'' &&+\frac{\partial_\rho\left(\sqrt{\left|\det \mathcal{G}\right|} \mathcal{G}^{44} \mathcal{G}^{22}\right)}{\sqrt{\left|\det \mathcal{G}\right|} \mathcal{G}^{44} \mathcal{G}^{22}} {E^3}' &&-
				4 \frac{\varrho_H^4}{R^4} \frac{\mathcal{G}^{00}}{\mathcal{G}^{44}} \, \wn^2 E^3\, ,
			\end{alignat}
			where we introduced
			\begin{equation}
			\label{eq:dimLessm}
				\mn=\frac{\sqrt{2}\,c}{4\pi}\, A^3_0(\rho).
			\end{equation}
			Here $A^3_0(\rho)$ is the dimensionless background gauge field. The
			quantity $\mn$ is related to the dimensionful background field
			$A_0^3(\rho)$ by
			$\mn=\frac{c}{\sqrt{\lambda}}\frac{A^3_0(\rho)}{2\pi T}$.
			
			As a result of the introduction of an isospin chemical potential
			instead of a baryonic one, we end up with three equations of motion
			for the decoupled fields $X$, $Y$ and $E^3$. These equations
			now will have three distinct solutions because of the differences in
			the last terms of \eqref{eq:eomX} to \eqref{eq:eomE3}. Since the
			difference is a shift in $\wn$ one would expect the solutions $X$
			and $Y$ to be shifted versions of $E^3$. However, the shift
			parameter $\mn$ does depend on $\rho$. The equation of motion for
			$E^3$ is identical to the baryonic case and as such was discussed in
			the previous subsection.
			
			The three solutions $X$, $Y$ and $E^3$ constitute the isospin
			triplet of mesons which may be constructed out of the isospin $1/2$
			quarks of the field theory. This is similar to the $\rho$-meson in
			QCD. The mode $E^3$ coincides with the solution in case of a pure
			baryonic chemical potential, while the other two solutions have
			peaks in the spectral function at lower and higher values of $\wn$
			\cite{Erdmenger:2007ja}. The magnitude of this splitting of the
			spectral lines is determined by the chemical potential.
			
			In the limit of zero frequency $\wn\to 0$, equations \eqref{eq:eomX}
			and \eqref{eq:eomY} coincide and will result in identical solutions
			$X$ and $Y$. In this limit the solution $E^3$, though, differs from
			$X$ and $Y$, by means of the last term. So for small frequencies
			$\wn$, we expect differences between the solutions $E^3$ and $X$,
			$Y$.

		\subsubsection{Numerical results and interpretation}
		
			We now investigate the effects of finite isospin density on the
			spectrum. After solving \eqref{eq:eomX} to \eqref{eq:eomE3}
			numerically, the correlators may eventually be evaluated as given by
			\eqref{eq:specFuncFromA}, where $A$ has to be replaced by $E^3$, $X$
			or $Y$. The peaks in the spectral functions again correspond to
			mesons.
			
\begin{figure}
				\centering
				\psfrag{w}{\small $\wn$}
				\psfrag{R}{\small $\R$}
				\psfrag{m3}{\small $m=3$}
				\psfrag{dt8}{\small $\tilde d=8$}
				\psfrag{dt14}{\small $\tilde d=14$}
				\includegraphics[width=.6\linewidth]{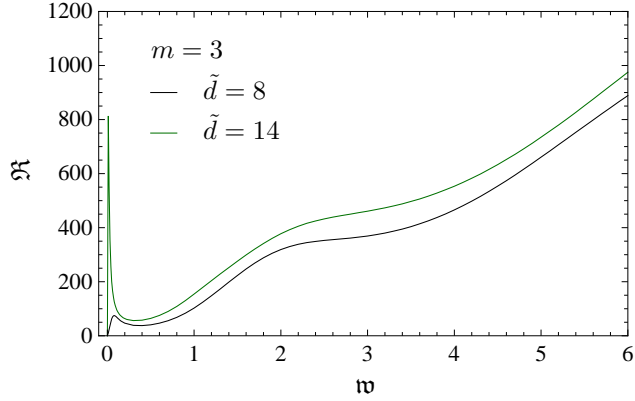}
				\caption{Spectral function calculated from \eqref{eq:eomY}.
					A new peak forms at high densities.}
				\label{fig:exampleIso}
			\end{figure}

			An interesting feature at finite isospin chemical potential is the
			formation of a new peak in the spectral function in the regime of
			small $\wn$ at high density/high chemical potential, see
			fig.~\ref{fig:exampleIso}. Notice that compared to the baryonic case
			the density, at which the new peak forms, is about two orders of
			magnitude smaller. As in the baryonic case, the excitations related
			to the supersymmetric spectrum broaden, the corrsponding mesons
			become unstable.
			
			One should keep in mind that the structure of the spectral function
			is determined by the pole structure of the retarded correlator
			\eqref{eq:GreenRetarded}. The poles of this function are located in
			the complex $\omega$-plane at positions $\Omega_n \in \mathbb{C}$.
			The spectral functions show the imaginary part of the correlator at
			real valued $\omega$. Any pole in the vicinity of the real axis will
			therefore introduce narrow high peaks in the spectral function,
			while poles far from the real axis have less influence and merely
			introduce small and broad structures.
			
\begin{figure}
				\centering
				\psfrag{w}{\small $\wn$}
				\psfrag{R}{\small $\R_X$}
				\psfrag{m3}[br]{\footnotesize $m=3$}
				\psfrag{dt146}{\footnotesize $\tilde d=14.6$}
				\psfrag{dots}{\footnotesize $\vdots$}
				\psfrag{dt158}{\footnotesize $\tilde d=15.8$}
				\includegraphics[width=.6\linewidth]{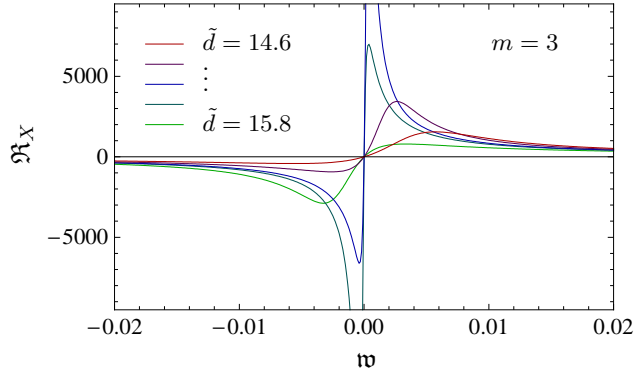}
				\caption{Plot of the spectral function around $\wn=0$. At a
					value of $\tilde d = 15.35$ a pole appears at $\wn=0$.}
				\label{fig:poleExample}
			\end{figure}

			In reminiscence of the real valued \emph{normal modes} of solutions
			for oscillators the complex $\Omega_n$ are called \emph{quasinormal
			modes}. The imaginary part of the quasinormal modes describes
			damping, as long as $\Im \Omega_n < 0$. The figures shown in this
			section indicate a dependence of the position of the quasinormal
			modes on the chemical potential/particle density. From
			fig.~\ref{fig:exampleIso} we deduce that a quasinormal mode
			approaches the origin of the complex $\omega$ plane as the particle
			density is increased. We observe a pole at $\wn=0$ for a certain
			particle density $\tilde d_{\text{crit}}$, the value depends on $m$.
			An impression of the variation in the spectral function is given in
			fig.~\ref{fig:poleExample}.

\begin{figure}
				\centering
				\psfrag{Imw}{\footnotesize $\Im \wn$}
				\psfrag{Rew}{\footnotesize $\Re \wn$}
				\psfrag{smile}{$\ddot\smile$}
				\psfrag{frown}{$\ddot\frown$}
				\includegraphics[height=.25\textheight]{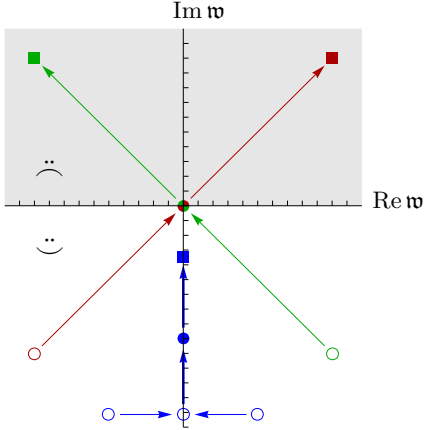}
				\caption{A sketch of the positions and movements of the
					quasinormal modes under changes of $\tilde d$. Color indicates
					the function: \textcolor{red}{$\text{red}=Y$},
					\textcolor{green}{$\text{green}=X$},
					\textcolor{blue}{$\text{blue}=E^3$}.
					The symbols indicate the range of $\tilde d$: $\circ<\tilde d_{\text{crit}}$,
					$\bullet=\tilde d_{\text{crit}}$, $\text{\tiny$\blacksquare$}>\tilde d_{\text{crit}}$.
					Poles in the gray region introduce instabilities.}
				\label{fig:qnmSketch}
			\end{figure}

\begin{figure}
				\centering
				\includegraphics[width=\linewidth]{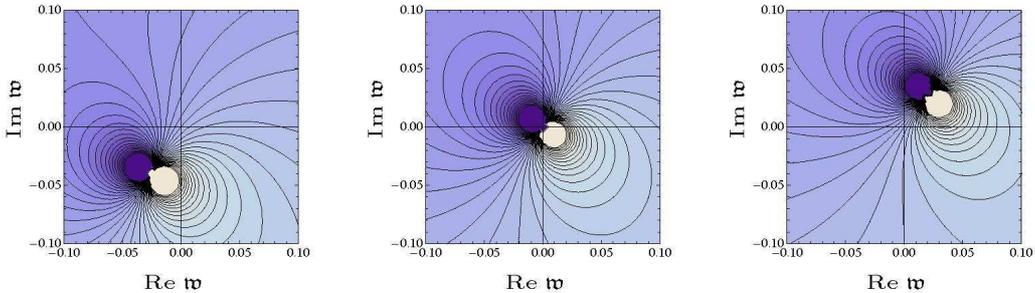}
				\caption{Contour plot of the spectral function for $Y$ around
					$\wn=0$ in the complex $\wn$-plane. Left: $\tilde
					d=10<\tilde d_{\text{crit}}$, center: $\tilde
					d=15.352=\tilde d_{\text{crit}}$, right: $\tilde
					d=20.704>\tilde d_{\text{crit}}$. The three graphs where
					generated for $m=3$, dark shading indicates small values of
					$\R$, light shading indicates large values. A pole in the
					upper half plane introduces an instability.}
				\label{fig:threeExamples}
			\end{figure}

			In fig.~\ref{fig:qnmSketch} we qualitatively sketch the result from
			the investigation of the behavior of the quasinormal modes closest
			to the origin of the complex $\wn$-plane. These modes do \emph{not}
			produce the peaks corresponding to the spectrum
			\eqref{eq:susyMesonSpectrum}.
			At low densities all quasinormal modes are located in the lower half
			plane. When increasing the isospin density, the modes of the
			solutions $X$ and $Y$ to \eqref{eq:eomX} and \eqref{eq:eomY} move
			towards the origin of the frequency plane. At the same time two
			quasinormal modes of $E^3$ move towards each other and merge on the
			negative imaginary axis, then travel along the axis towards the
			origin as one single pole. At the critical value of $\tilde d$ the
			modes from $X$ and $Y$ meet at the origin, the quasinormal modes
			from $E^3$ still reside in the lower half plane. This observation
			matches the discussion below equations \eqref{eq:eomX}
			to~\eqref{eq:eomE3}, where we expected $X$ and $Y$ to behave
			similarly at small $\wn$, while $E^3$ should differ from this
			behavior. Upon further increasing the isospin density, the modes
			$\Omega$ from $X$ and $Y$ enter the upper half plane, maintaining
			their distinct directions. The sign change in $\Im\Omega$ from
			$\Im\Omega<0$ to $\Im\Omega>0$ indicates that a damped resonance
			changes into a self-enhancing one, and thus introduces an
			instability to the system. Figure fig.~\ref{fig:threeExamples}
			illustrates the transition of a quasinormal mode of $Y$ from the
			lower half plane to the upper half plane. The $E^3$-mode does not
			enter the upper half plane at any value of $\tilde d$ we
			considered. Compare this to the values of $\tilde d$ in
			fig.~\ref{fig:variousDT} and fig.~\ref{fig:exampleIso} at which the
			pole induces visible structures at small $\wn$. A comparable
			movement of poles in a different but related setup was found in
			\cite{Gubser:2008wv}. There the quasinormal modes of correlation
			functions of electromagnetic currents were investigated as a
			function of temperature.
			
			\medskip
			 
			In the following we interpret the observation of decaying mesons and
			the emergence of a new peak in the spectral function in terms of
			field theory quantities. In particular we speculate on a new phase
			in the phase diagram for fundamental matter in the D$3$/D$7$ setup.
			
			In the far UV, the field theory dual to our setup is supersymmetric,
			thus containing scalars as well as fermions, both of which
			contribute to the bound states we identified with mesons, even when
			supersymmetry is eventually broken. The meson decay at non-vanishing
			particle densities may be explained by the change
			of the shape of the potential for the scalars in the field theory
			upon the introduction of a non-vanishing density. As
			outlined in appendix~\ref{sec:scalarPotential}, a chemical potential
			may lead to an instability of the theory, since it induces a runaway
			potential for the scalar fields at small field values
			\cite{Harnik:2003ke}. Nevertheless, interactions of $\phi^4$-type
			lead to a Mexican hat style potential for larger field values. In
			this way the theory is stabilized at finite density $\tilde d$ while
			the scalar fields condensate. This squark condensate presumably
			contributes to the vev of the scalar flavor current,
			\begin{equation}
				\tilde d \propto \left< J^0 \right> \propto \left< \bar \psi\,\gamma^0\,\psi\right> + \left< \phi \;\partial^0\phi \right>. 
			\end{equation}
			In the AdS/CFT context the presence of an upside-down potential for
			the squark vev has been shown in \cite{Apreda:2005yz} using an
			instanton configuration in the dual supergravity background.
			
			The occurrence of a pole in the upper half plane of complex frequencies
			at finite $\tilde d_{\text{crit}}$
			indicates an instability of the
			theory. A comparable
			observation was made in~\cite{Aharony:2007uu}, where in fact the
			vector meson becomes unstable by means of negative values for its
			mass beyond some critical chemical potential. The
			difference between this work and \cite{Aharony:2007uu} is that our
			model includes scalar modes in addition to the fundamental fermions.
			Nevertheless, in both models an instability occurs at a
			critical value of the chemical potential.
			The theory may still be stabilized dynamically by vector
			condensation \cite{Buchel:2006aa}. In this case the system would
			enter a new phase of condensed vectors at densities larger than
			$\tilde d_{\text{crit}}$, in accordance with the expectation from QCD
                        calculations \cite{Lenaghan:2001sd,Sannino:2002wp,Voskresensky:1997ub}.
			
\begin{figure}
				\centering
				\psfrag{invm}{\small \hspace{-5ex}$1/m\:\propto\:T/M_q$}
				\psfrag{muM}{\small $\mu^I/M_q$}
				\psfrag{dt20p5}[bc]{\scriptsize \textcolor{black!65}{$\;\tilde d=20.5$}}
				\psfrag{bla1}{\footnotesize Minkowski}
				\psfrag{bla3}{\footnotesize black hole}
				\psfrag{bla2}{\footnotesize unstable}
				\includegraphics[width=.6\linewidth]{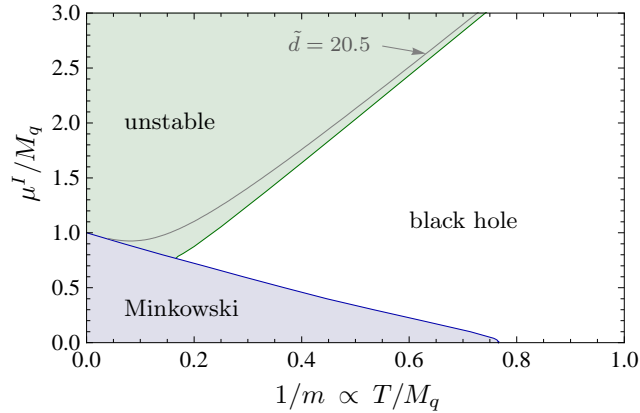}
				\caption{The phase diagram for fundamental matter in the
					D$3$/D$7$ setup. In the blue shaded region D$7$-branes have
					the topology of Minkowski embeddings. The white and green
					regions are described by black hole embeddings. However, in
					the green region we interpret our observations as an
					instability of the black hole embeddings in the D$3$/D$7$
					setup.}
				\label{fig:phaseDiagram}
			\end{figure}

			We perform the analysis of the pole structure at $\wn=0$ for
			various $m$, and interpret the phenomenon of the transition of poles
			into the upper half plane at finite critical particle density as a
			sign of the transition to an unstable phase. We relate the critical
			particle density $\tilde d_{\text{crit}}$ to the according chemical
			potential $\tilde \mu^I$ by \eqref{mu} and use the pairs of $m$ and
			critical dimensionful $\mu$ to trace the line of the phase
			transition in the phase diagram of fundamental matter in the
			D$3$/D$7$ setup. The result is drawn as a green line in
			fig.~\ref{fig:phaseDiagram}. The picture shows the $(\mu^I,T)$-plane
			of the phase diagram and contains three regions, drawn as blue
			shaded, white, and green shaded, as well as a blue and a green line,
			separating the different regions.
			
			The blue shaded region marks the range of parameters, in which
			fundamental matter is described by D$7$-branes with Minkowski
			embeddings. The blue line, delimiting the the blue region, marks the
			line of phase transitions to the black hole phase, where fundamental
			matter is described by D$7$-branes which have black hole embeddings.
			This phase transition was discussed in
			section~\ref{phasediagramgrand}. Using the symmetries
			\eqref{symNf=2}, this phase transition line can be mapped to the
			line of phase transitions between Minkowski and black hole
			embeddings, present at finite baryon chemical potential
			\cite{Mateos:2007vc,Karch:2007br,Erdmenger:2007cm}.
			
			The green line and green shaded region in the phase diagram in
			fig.~\ref{fig:phaseDiagram} mark the observation made in this
			section. The green line marks the values of $\tilde d_{\text{crit}}$
			at which the pole in the spectral function appears at $\wn=0$.
			Beyond the green line we enter the green shaded unstable region.
			
			We observe that the green line asymptotes to a straight line at high
			temperatures. Within the values computed by us, this line agrees
			with the asymptotic behavior of the contour of particle density with
			$\tilde d\approx 20.5$, drawn as a gray line in the phase diagram.
			We thus speculate on a finite critical particle density beyond which
			the black hole phase is unstable. This interpretation is supported
			by analogous studies of the phase diagram of $\N =4$
			super-Yang-Mills theory with R-symmetry chemical potentials, where a
			similar line in the phase diagram was discovered
			\cite{Yamada:2006rx,Cvetic:1999ne}. Up to now it is unclear whether
			the new phase indeed is unstable or if it only indicates that
			black hole embeddings are not capable of describing this state of
			matter.
			
			Note that the location of the green phase transition line in
			fig.~\ref{fig:phaseDiagram} as well as the results shown in
			fig.~\ref{fig:threeExamples} and fig.~\ref{fig:qnmSketch} are
			obtained from the analysis of poles in the spectral functions. These
			functions in turn are obtained as solutions to
			equations~\eqref{eq:eomX} to~\eqref{eq:eomE3}, which do depend on
			the so far unknown factor $c$ in \eqref{eq:dimLessm}. The
			computation of this factor is left to future work. It will determine
			the exact position of the green phase transition line in
			fig.~\ref{fig:phaseDiagram}. This will answer the question whether
			there is a triple point in the phase diagram and if the blue and
			green shaded regions meet at a common border.  Moreover, other
			poles than the ones investigated here may have influence on the
			stability of this system.

\section{Conclusion} 

In this work we have considered a probe of~$N_f$ D7 branes embedded in the AdS
Schwarzschild black hole background with an $U(N_f)$ gauge field. This
configuration is conjectured to be dual to a thermal quantum field theory with
isospin and baryon charge. We have considered both the canonical and grand
canonical ensemble. Our detailed study of this background provides a basis for
further studies. For instance it is now possible to investigate the mesonic
excitation spectrum or hydrodynamic properties in detail.

By including isospin density, we find four new phase boundaries coincident with
the fixed points of the accidental symmetries between baryon and isospin
density, which our setup displays. Along the lines of equal densities~$ |\tilde
d^B | = |\tilde d^I |$ and near~$T=0$, we find a sharp transition in the
chemical potentials. This situation is reminiscent of a corresponding transition
found in two-color QCD~\cite{Splittorff:2000mm}. For~$T=0$ the chemical
potential is discontinuous and a one-to-one correspondence between the
thermodynamic conjugate variables~$\tilde d$ and~$\tilde \mu$ can not be
established. This inconsistency between the two ensembles signals an instability
of the setup in that phase region. We expect that vector mesons condense in this
region as is also expected in QCD~\cite{Splittorff:2000mm}. A further detailed
study of this new phase requires the investigation of scalar and pseudoscalar
fluctuations as well. Furthermore, an analysis of the possible condensates 
is required, cf.~\cite{Aharony:2007uu}. This is a promising task for future
investigation. It would also be interesting to study the heat capacity and the
quasi-particle spectrum at low temperature as in \cite{Karch:2008fa} in the
more general setup provided in this paper. This study may give a better
understanding of the new quantum liquid found in \cite{Karch:2008fa}.

Here we have restricted to vector fluctuations about the finite isospin density
background. Analyzing these fluctuations, we found an unstable region in the
$(\mu^I,\,T)$ phase diagram which is bounded by a line of constant density. We
expect that mesons condense in this unstable phase. The emerging phase structure
appears to be a generic feature of holographic duals with chemical potentials as
a similar behavior has been observed for spinning branes~\cite{Gubser:1998jb}
and R-charged black hole backgrounds~\cite{Cvetic:1999ne,Yamada:2007gb}. It
would be interesting to shed more light on the connection between the new phase 
found in this work and in the backgrounds mentioned above.

Considering the effective baryon diffusion, we find competing behavior between
baryon and isospin density. For example at a given
temperature,~increasing~$\tilde d^B$ increases the effective baryon diffusion
while increasing~$\tilde d^I$ suppresses that diffusion. Note that this
effective baryon density collects both, the effect of finite baryon and of
finite isospin density on the effective baryon diffusion. It would be very
interesting to separate these two contributions, for example by a modification
of the membrane paradigm.

All the numerical results mentioned here were obtained for~$N_f=2$. However,
they may also be generated for arbitrary~$N_f\ll N_c$ from the general
expressions we have provided here.

\section*{Acknowledgments}
\addcontentsline{toc}{section}{Acknowledgments}

We are grateful to Martin Ammon and Karl Landsteiner for valuable discussions.
We thank  Alex Buchel, Junji Jia and Jonathan Shock for helpful comments. This
work was supported in part by \emph{The Cluster of Excellence for Fundamental Physics
-- Origin and Structure of the Universe} and by \emph{The European Union Excellence Grant MEXT-CT-2003-509661}.

\begin{appendix}
\section{Simplification of the DBI action}\label{appaction}

In this section we give some details for the simplifications done in the
equations (\ref{actiondiag}), (\ref{newfields}), (\ref{action}). We argued that
the square root in the action (\ref{actiondiag}) is diagonal in flavor space and
the DBI action decouple for the fields defined in (\ref{newfields}). To prove
this, we first have to calculate the square of the non-vanishing field strength
component $(F_{40})^2$. For this calculation we need some multiplication
relations for the matrices $\lambda^i$ given by
\begin{equation}
\begin{split}
\lambda^1\lambda^i&=\lambda^i\quad i=1,\dots,N_f\\
(\lambda^i)^2&=\diag(1,\dots,\overbrace{(N_f-1)^2}^{i\text{-th position}},\dots,1)\quad i=2,\dots,N_f\\
\lambda^i\lambda^j&=\diag(1,\dots,\overbrace{-(N_f-1)}^{i\text{-th position}},\dots,\overbrace{-(N_f-1)}^{j\text{-th position}},\dots,1)\quad i\not=j
\end{split}
\end{equation}
and the fact that all matrices $\lambda^i$ commute with each other since they
are the generators of the Cartan subalgebra. The $i$-th diagonal component of
the matrix $(F_{40})^2$ is then for $i=1$
\begin{equation}
\begin{split}
\left((F_{40})^2\right)_{11}&=\left[\sum_{j=1}^{N_f}(F^{I_j}_{40})^2(\lambda^j)^2+2\sum_{j=2}^{N_f}F^{I_1}_{40}F^{I_j}_{40}\lambda^1\lambda^j+2\sum_{j=2}^{N_f}\sum_{\substack{k=2\\k\not=j}}^{N_f}F^{I_j}_{40}F^{I_k}_{40}\lambda^j\lambda^k\right]_{11}\\
&=\left(\sum_{j=1}^{N_f}F^{I_j}_{40}\right)^2=(\del_\vrho X_1)^2
\end{split}
\end{equation}
and for $i\not=1$
\begin{equation}
\begin{split}
\left((F_{40})^2\right)_{ii}=&\Bigg[\sum_{\substack{j=1\\j\not=i}}^{N_f}(F^{I_j}_{40})^2(\lambda^j)^2+(F^{I_i}_{40})^2(\lambda^i)^2+2\sum_{\substack{j=1\\j\not=i}}F^{I_i}_{40}F^{I_j}_{40}\lambda^i\lambda^j\\
&+2\sum_{j=1}^{N_f}\sum_{\substack{k=1\\k\not=i}}^{N_f}F^{I_j}_{40}F^{I_k}_{40}\lambda^j\lambda^k\Bigg]_{ii}\\
=&\left(\sum_{\substack{j=1\\j\not=i}}^{N_f}F^{I_j}_{40}-(N_f-1)F^{I_i}_{40}\right)^2=(\del_\vrho X_i)^2\,,
\end{split}
\end{equation}
where we used the new fields as defined in equation (\ref{newfields}) in the last equalities. With this choice of the new fields and since $F_{40}$ is diagonal in the flavor space, each component of the diagonal is given by the expression known from the pure baryonic case \cite{Kobayashi:2006sb}
\begin{equation}
\begin{split}
\left(\sqrt{\lambda^1+(2\pi\alpha')^2 G^{00}G^{44}(F_{40})^2}\right)_{ii}&=\sqrt{1+(2\pi\alpha')^2 G^{00}G^{44}((F_{40})^2)_{ii}}\\
&=\sqrt{1+(2\pi\alpha')^2G^{00}G^{44}(\del_\vrho X_i)^2}\,.
\end{split}
\end{equation}
Taking the trace and simplifying the expressions of the metric factors, we get equation (\ref{action}).

\section{Effect of the accidental symmetries for $N_f=3$}\label{app:symNf=3}
In this section we give the action of the accidental symmetries discussed in section \ref{sym} in the physical basis for $N_f=3$:
\tiny
\begin{equation}
\begin{aligned}
d_1&\leftrightarrow d_2: && d^B\mapsto d^B && d^{I_2}\mapsto-(d^{I_2}+d^{I_3}) &&d^{I_3}\mapsto d^{I_3}\\
d_1&\leftrightarrow d_3: && d^B\mapsto d^B && d^{I_2}\mapsto d^{I_2} && d^{I_3}\mapsto-(d^{I_2}+d^{I_3})\\
d_2&\leftrightarrow d_3: && d^B\mapsto d^B && d^{I_2}\mapsto d^{I_3} && d^{I_3}\mapsto d^{I_2}\\
d_1&\leftrightarrow-d_1:\qquad && d^B\mapsto\frac{1}{3}(d^B-2d^{I_2}-2d^{I_3})\qquad && d^{I_2}\mapsto\frac{1}{3}(-2d^B+d^{I_2}-2d^{I_3})\qquad && d^{I_3}\mapsto\frac{1}{3}(-2d^B-2d^{I_2}+d^{I_3})\\
d_2&\leftrightarrow-d_2: && d^B\mapsto\frac{1}{3}(d^B+2d^{I_2})&& d^{I_2}\mapsto\frac{1}{3}(4d^B-d^{I_2}) && d^{I_3}\mapsto\frac{1}{3}(-2d^B+2d^{I_2}+d^{I_3})\\
d_3&\leftrightarrow-d_3: && d^B\mapsto\frac{1}{3}(d^B+2d^{I_3})&& d^{I_2}\mapsto\frac{1}{3}(-2d^B+d^{I_2}+2d^{I_3}) && d^{I_3}\mapsto\frac{1}{3}(4d^B-d^{I_3})\,.
\end{aligned}
\end{equation}
\normalsize
Since the transformation matrix to the physical basis for $N_f=3$ is not an orthogonal matrix, there is no induced approximate $O(3)$ symmetry in the physical basis. This is in general also true for all cases with $N_f\ge 3$.

\section[Chemical potentials in field theories:\\ Runaway potential and Bose-Einstein Condensation]{Chemical potentials in field theories: Runaway potential and Bose-Einstein Condensation}
\label{sec:scalarPotential}
In our setup we consider a theory which is supersymmetric in the far UV. Its fundamental matter consists of complex scalars  (squarks) and fermionic fields (quarks). In this section we describe the effect of the chemical potential on the field theory Lagrangian and on the vacuum as \eg in \cite{Harnik:2003ke}. We consider a theory with one complex scalar $\phi$ and one fermionic field $\psi$ with the same mass $M_q$ coupled to an $U(1)$ gauge field $A_\nu$. The time component of the $U(1)$ gauge field has a non-zero vev which induces the chemical potential $\mu$,
\begin{equation}
A_\nu=\mu\delta_{\nu 0}\,.
\end{equation}
The Lagrangian is given by
\begin{equation}
\call=-D_\nu\phi^* D^\nu\phi-M_q^2\phi^*\phi-\bar\psi(\dSlash{D}+M_q)\psi-\frac{1}{4}F_{\mu\nu}F^{\mu\nu}\,,
\end{equation}
where $D_\nu=\del_\nu-\ii A_\nu$ is the covariant derivative and $F_{\mu\nu}=\del_\mu A_\nu-\del_\nu A_\mu$ the field strength tensor. Expanding the Lagrangian around the non-zero vev of the gauge field, the Lagrangian becomes
\begin{equation}
\call=-\del_\nu\phi^*\del^\nu\phi-(M_q^2-\mu^2)\phi^*\phi+\mu J^s_0-\bar\psi(\dslash{\del}+M_q)\psi+\mu J_0^F\,,\eqlabel{fieldl}
\end{equation}
where $J^S_\mu=\ii[(\del_\mu\phi^*)\phi-\phi^*(\del_\mu\phi)]$, $\left(J^\mu\right)^F=-\ii\bar\psi\gamma^\mu\psi$ are conserved currents. These conserved currents are the population density for the scalar $N_S$ and fermionic fields $N_F$, such that the linear terms in the Lagrangian are $\mu N_S$ and $\mu N_F$.\par
The mass term $-(M_q^2-\mu^2)\phi^2$ of the Lagrangian (\ref{fieldl}) introduces an instability if $\mu>M_q$ since the corresponding potential $V=(M_q^2-\mu^2)\phi^2+\cdots$ is not bounded from below. In some systems this runaway potential is stabilized by higer interactions  and becomes a Mexican hat potential such that the scalar condense and the scalar density becomes non-zero. This condensation is known as Bose-Einstein-Condensation.

\section{Coupling constant for vector meson interaction}
\label{sec:mesoncoupling}

In this section we show how the coupling constant for the interaction of vector
mesons can be computed, extending the ideas presented in \cite{Kruczenski:2003be}. This coupling constant in the effective four-dimensional
meson theory can be determined by redefinition of the gauge fields such that the
kinetic term has canonical form. This coupling constant depends on the
geometry of the extra dimensions.

First we consider the eight-dimensional theory determined by the DBI action
$S^{(2)}_{\text{DBI}}$ expanded to second order in the fluctuations $\tilde A$,
\begin{equation}
	S^{(2)}_{\text{DBI}}=\frac{T_{D7}(2\pi\alpha')^2}{4}\int\!\dd^8\xi\sqrt{-G}\;G^{\mu\mu'}G^{\nu\nu'}F_{\mu'\nu}F_{\nu'\mu}\,,
\end{equation}
where $G$ contains the background fields and we simplify the analysis by
considering only Abelian gauge fields. Defining the dimensionless coordinate
$\rho=\varrho/R$ and integrating out the contribution of the $S^3$, we obtain
\begin{equation}
	S^{(2)}_{\text{DBI}}=\frac{T_{D_7}(2\pi\alpha')^2\vol(S_3)R^4}{4}\int\!\dd^4 x\int\!\dd\rho\sqrt{-G}\;G^{\mu\mu'}G^{\mu\nu'}F_{\mu'\nu}F_{\nu'\mu}\,.
\end{equation}
To obtain a four-dimensional effective theory we have to integrate over the
coordinate $\rho$. This contribution depends on the geometry induced
by the $\rho$ dependence of the metric factors. However, we expect that it is
independent of the 't Hooft coupling $\lambda$. We parametrize this contribution
by $c'$. The kinetic term of the effective theory is then given by
\begin{equation}
  S^{(2)}_{\text{DBI}}=\frac{T_{D7}(2\pi\alpha')^2\vol(S_3)R^4 c'}{4}\int\!\dd^4 x \;F_{\mu\nu}F^{\mu\nu}\,,
\end{equation}
where the prefactor may be written as
\begin{equation}
  \frac{T_{D7}(2\pi\alpha')^2\vol(S_3)R^4 c'}{4}=\frac{\lambda}{g_{\text{YM}}^2 c^2}\,,
\end{equation}
where the numerical values independent of the 't Hooft coupling are grouped into
the coefficient $c$. From this we can read off that a rescaling of the form
\begin{equation}
  A\mapsto\frac{c}{\sqrt{\lambda}}A
\end{equation}
casts the Lagrangian into canonical form with a prefactor of $1/g_{\text{YM}}^2$.
\end{appendix}

\providecommand{\href}[2]{#2}\begingroup\raggedright\endgroup


\begin{thebibliography}{10}

\bibitem{Maldacena:1997re}
J.~M. Maldacena, \emph{ {The large N limit of superconformal field theories and
  supergravity}}, Adv. Theor. Math. Phys. {\bf 2} (1998) 231--252,
\href{http://www.slac.stanford.edu/spires/find/hep/www?texkey=Maldacena:1997re%
}{arXiv:hep-th/9711200}.

\bibitem{Aharony:1999ti}
O.~Aharony, S.~S. Gubser, J.~M. Maldacena, H.~Ooguri, and Y.~Oz, \emph{ {Large
  N field theories, string theory and gravity}}, Phys. Rept. {\bf 323} (2000)
  183--386,
\href{http://www.slac.stanford.edu/spires/find/hep/www?texkey=Aharony:1999ti}{%
arXiv:hep-th/9905111}.

\bibitem{Witten:1998zw}
E.~Witten, \emph{ {Anti-de Sitter space, thermal phase transition, and
  confinement in gauge theories}}, Adv. Theor. Math. Phys. {\bf 2} (1998)
  505--532,
\href{http://www.slac.stanford.edu/spires/find/hep/www?texkey=Witten:1998zw}{a%
rXiv:hep-th/9803131}.

\bibitem{Son:2007vk}
D.~T. Son and A.~O. Starinets, \emph{ {Viscosity, Black Holes, and Quantum
  Field Theory}}, Ann. Rev. Nucl. Part. Sci. {\bf 57} (2007) 95--118,
\href{http://www.slac.stanford.edu/spires/find/hep/www?texkey=Son:2007vk}{arXi%
v:0704.0240}.

\bibitem{Policastro:2001yc}
G.~Policastro, D.~T. Son, and A.~O. Starinets, \emph{ {The shear viscosity of
  strongly coupled N = 4 supersymmetric Yang-Mills plasma}}, Phys. Rev. Lett.
  {\bf 87} (2001) 081601,
\href{http://www.slac.stanford.edu/spires/find/hep/www?texkey=Policastro:2001y%
c}{arXiv:hep-th/0104066}.

\bibitem{Policastro:2002se}
G.~Policastro, D.~T. Son, and A.~O. Starinets, \emph{ {From AdS/CFT
  correspondence to hydrodynamics}}, JHEP {\bf 09} (2002) 043,
\href{http://www.slac.stanford.edu/spires/find/hep/www?texkey=Policastro:2002s%
e}{arXiv:hep-th/0205052}.

\bibitem{Policastro:2002tn}
G.~Policastro, D.~T. Son, and A.~O. Starinets, \emph{ {From AdS/CFT
  correspondence to hydrodynamics. II: Sound waves}}, JHEP {\bf 12} (2002) 054,
\href{http://www.slac.stanford.edu/spires/find/hep/www?texkey=Policastro:2002t%
n}{arXiv:hep-th/0210220}.

\bibitem{Herzog:2002fn}
C.~P. Herzog, \emph{ {The hydrodynamics of M-theory}}, JHEP {\bf 12} (2002)
  026,
\href{http://www.slac.stanford.edu/spires/find/hep/www?texkey=Herzog:2002fn}{a%
rXiv:hep-th/0210126}.

\bibitem{Herzog:2003ke}
C.~P. Herzog, \emph{ {The sound of M-theory}}, Phys. Rev. {\bf D68} (2003)
  024013,
\href{http://www.slac.stanford.edu/spires/find/hep/www?texkey=Herzog:2003ke}{a%
rXiv:hep-th/0302086}.

\bibitem{Kovtun:2003wp}
P.~Kovtun, D.~T. Son, and A.~O. Starinets, \emph{ {Holography and
  hydrodynamics: Diffusion on stretched horizons}}, JHEP {\bf 10} (2003) 064,
\href{http://www.slac.stanford.edu/spires/find/hep/www?texkey=Kovtun:2003wp}{a%
rXiv:hep-th/0309213}.

\bibitem{Buchel:2003tz}
A.~Buchel and J.~T. Liu, \emph{ {Universality of the shear viscosity in
  supergravity}}, Phys. Rev. Lett. {\bf 93} (2004) 090602,
\href{http://www.slac.stanford.edu/spires/find/hep/www?texkey=Buchel:2003tz}{a%
rXiv:hep-th/0311175}.

\bibitem{Kovtun:2004de}
P.~Kovtun, D.~T. Son, and A.~O. Starinets, \emph{ {Viscosity in strongly
  interacting quantum field theories from black hole physics}}, Phys. Rev.
  Lett. {\bf 94} (2005) 111601,
\href{http://www.slac.stanford.edu/spires/find/hep/www?texkey=Kovtun:2004de}{a%
rXiv:hep-th/0405231}.

\bibitem{Buchel:2004di}
A.~Buchel, J.~T. Liu, and A.~O. Starinets, \emph{ {Coupling constant dependence
  of the shear viscosity in N=4 supersymmetric Yang-Mills theory}}, Nucl. Phys.
  {\bf B707} (2005) 56--68,
\href{http://www.slac.stanford.edu/spires/find/hep/www?texkey=Buchel:2004di}{a%
rXiv:hep-th/0406264}.

\bibitem{Benincasa:2005iv}
P.~Benincasa, A.~Buchel, and A.~O. Starinets, \emph{ {Sound waves in strongly
  coupled non-conformal gauge theory plasma}}, Nucl. Phys. {\bf B733} (2006)
  160--187,
\href{http://www.slac.stanford.edu/spires/find/hep/www?texkey=Benincasa:2005iv%
}{arXiv:hep-th/0507026}.

\bibitem{Mas:2006dy}
J.~Mas, \emph{ {Shear viscosity from R-charged AdS black holes}}, JHEP {\bf 03}
  (2006) 016,
\href{http://www.slac.stanford.edu/spires/find/hep/www?texkey=Mas:2006dy}{arXi%
v:hep-th/0601144}.

\bibitem{Son:2006em}
D.~T. Son and A.~O. Starinets, \emph{ {Hydrodynamics of R-charged black
  holes}}, JHEP {\bf 03} (2006) 052,
\href{http://www.slac.stanford.edu/spires/find/hep/www?texkey=Son:2006em}{arXi%
v:hep-th/0601157}.

\bibitem{Maeda:2006by}
K.~Maeda, M.~Natsuume, and T.~Okamura, \emph{ {Viscosity of gauge theory plasma
  with a chemical potential from AdS/CFT}}, Phys. Rev. {\bf D73} (2006) 066013,
\href{http://www.slac.stanford.edu/spires/find/hep/www?texkey=Maeda:2006by}{ar%
Xiv:hep-th/0602010}.

\bibitem{Herzog:2006se}
C.~P. Herzog, \emph{ {Energy loss of heavy quarks from asymptotically AdS
  geometries}}, JHEP {\bf 09} (2006) 032,
\href{http://www.slac.stanford.edu/spires/find/hep/www?texkey=Herzog:2006se}{a%
rXiv:hep-th/0605191}.

\bibitem{CasalderreySolana:2006rq}
J.~Casalderrey-Solana and D.~Teaney, \emph{ {Heavy quark diffusion in strongly
  coupled N = 4 Yang Mills}}, Phys. Rev. {\bf D74} (2006) 085012,
\href{http://www.slac.stanford.edu/spires/find/hep/www?texkey=CasalderreySolan%
a:2006rq}{arXiv:hep-ph/0605199}.

\bibitem{Buchel:2008wy}
A.~Buchel, \emph{ {Shear viscosity of CFT plasma at finite coupling}}, Phys.
  Lett. {\bf B665} (2008) 298--304,
\href{http://www.slac.stanford.edu/spires/find/hep/www?texkey=Buchel:2008wy}{a%
rXiv:0804.3161}.

\bibitem{Myers:2008yi}
R.~C. Myers, M.~F. Paulos, and A.~Sinha, \emph{ {Quantum corrections to
  eta/s}},
\href{http://www.slac.stanford.edu/spires/find/hep/www?texkey=Myers:2008yi}{ar%
Xiv:0806.2156}.

\bibitem{Karch:2002sh}
A.~Karch and E.~Katz, \emph{ {Adding flavor to AdS/CFT}}, JHEP {\bf 06} (2002)
  043,
\href{http://www.slac.stanford.edu/spires/find/hep/www?texkey=Karch:2002sh}{ar%
Xiv:hep-th/0205236}.

\bibitem{Babington:2003vm}
J.~Babington, J.~Erdmenger, N.~J. Evans, Z.~Guralnik, and I.~Kirsch, \emph{
  {Chiral symmetry breaking and pions in non-supersymmetric gauge / gravity
  duals}}, Phys. Rev. {\bf D69} (2004) 066007,
\href{http://www.slac.stanford.edu/spires/find/hep/www?texkey=Babington:2003vm%
}{arXiv:hep-th/0306018}.

\bibitem{Kirsch:2004km}
I.~Kirsch, \emph{ {Generalizations of the AdS/CFT correspondence}}, Fortsch.
  Phys. {\bf 52} (2004) 727--826,
\href{http://www.slac.stanford.edu/spires/find/hep/www?texkey=Kirsch:2004km}{a%
rXiv:hep-th/0406274}.

\bibitem{Mateos:2007vn}
D.~Mateos, R.~C. Myers, and R.~M. Thomson, \emph{ {Thermodynamics of the
  brane}}, JHEP {\bf 05} (2007) 067,
\href{http://www.slac.stanford.edu/spires/find/hep/www?texkey=Mateos:2007vn}{a%
rXiv:hep-th/0701132}.

\bibitem{Kruczenski:2003uq}
M.~Kruczenski, D.~Mateos, R.~C. Myers, and D.~J. Winters, \emph{ {Towards a
  holographic dual of large-N(c) QCD}}, JHEP {\bf 05} (2004) 041,
\href{http://www.slac.stanford.edu/spires/find/hep/www?texkey=Kruczenski:2003u%
q}{arXiv:hep-th/0311270}.

\bibitem{Kobayashi:2006sb}
S.~Kobayashi, D.~Mateos, S.~Matsuura, R.~C. Myers, and R.~M. Thomson, \emph{
  {Holographic phase transitions at finite baryon density}}, JHEP {\bf 02}
  (2007) 016,
\href{http://www.slac.stanford.edu/spires/find/hep/www?texkey=Kobayashi:2006sb%
}{arXiv:hep-th/0611099}.

\bibitem{Mateos:2007vc}
D.~Mateos, S.~Matsuura, R.~C. Myers, and R.~M. Thomson, \emph{ {Holographic
  phase transitions at finite chemical potential}}, JHEP {\bf 11} (2007) 085,
\href{http://www.slac.stanford.edu/spires/find/hep/www?texkey=Mateos:2007vc}{a%
rXiv:0709.1225}.

\bibitem{Nakamura:2006xk}
S.~Nakamura, Y.~Seo, S.-J. Sin, and K.~P. Yogendran, \emph{ {A new phase at
  finite quark density from AdS/CFT}}, J. Korean Phys. Soc. {\bf 52} (2008)
  1734--1739,
\href{http://www.slac.stanford.edu/spires/find/hep/www?texkey=Nakamura:2006xk}%
{arXiv:hep-th/0611021}.

\bibitem{Nakamura:2007nx}
S.~Nakamura, Y.~Seo, S.-J. Sin, and K.~P. Yogendran, \emph{ {Baryon-charge
  Chemical Potential in AdS/CFT}}, Prog. Theor. Phys. {\bf 120} (2008) 51--76,
\href{http://www.slac.stanford.edu/spires/find/hep/www?texkey=Nakamura:2007nx}%
{arXiv:0708.2818}.

\bibitem{Ghoroku:2007re}
K.~Ghoroku, M.~Ishihara, and A.~Nakamura, \emph{ {D3/D7 holographic Gauge
  theory and Chemical potential}}, Phys. Rev. {\bf D76} (2007) 124006,
\href{http://www.slac.stanford.edu/spires/find/hep/www?texkey=Ghoroku:2007re}{%
arXiv:0708.3706}.

\bibitem{Kim:2006gp}
K.-Y. Kim, S.-J. Sin, and I.~Zahed, \emph{ {Dense hadronic matter in
  holographic QCD}},
\href{http://www.slac.stanford.edu/spires/find/hep/www?texkey=Kim:2006gp}{arXi%
v:hep-th/0608046}.

\bibitem{Horigome:2006xu}
N.~Horigome and Y.~Tanii, \emph{ {Holographic chiral phase transition with
  chemical potential}}, JHEP {\bf 01} (2007) 072,
\href{http://www.slac.stanford.edu/spires/find/hep/www?texkey=Horigome:2006xu}%
{arXiv:hep-th/0608198}.

\bibitem{Son:2002sd}
D.~T. Son and A.~O. Starinets, \emph{ {Minkowski-space correlators in AdS/CFT
  correspondence: Recipe and applications}}, JHEP {\bf 09} (2002) 042,
\href{http://www.slac.stanford.edu/spires/find/hep/www?texkey=Son:2002sd}{arXi%
v:hep-th/0205051}.

\bibitem{Herzog:2002pc}
C.~P. Herzog and D.~T. Son, \emph{ {Schwinger-Keldysh propagators from AdS/CFT
  correspondence}}, JHEP {\bf 03} (2003) 046,
\href{http://www.slac.stanford.edu/spires/find/hep/www?texkey=Herzog:2002pc}{a%
rXiv:hep-th/0212072}.

\bibitem{Hoyos:2006gb}
C.~Hoyos-Badajoz, K.~Landsteiner, and S.~Montero, \emph{ {Holographic Meson
  Melting}}, JHEP {\bf 04} (2007) 031,
\href{http://www.slac.stanford.edu/spires/find/hep/www?texkey=Hoyos:2006gb}{ar%
Xiv:hep-th/0612169}.

\bibitem{Hoyos:2007zz}
C.~Hoyos-Badajoz, K.~Landsteiner, and S.~Montero, \emph{ {Quasinormal modes and
  meson decay rates}}, Fortsch. Phys. {\bf 55} (2007)
760--764.

\bibitem{Amado:2007yr}
I.~Amado, C.~Hoyos-Badajoz, K.~Landsteiner, and S.~Montero, \emph{ {Residues of
  Correlators in the Strongly Coupled N=4 Plasma}}, Phys. Rev. {\bf D77} (2008)
  065004,
\href{http://www.slac.stanford.edu/spires/find/hep/www?texkey=Amado:2007yr}{ar%
Xiv:0710.4458}.

\bibitem{Myers:2007we}
R.~C. Myers, A.~O. Starinets, and R.~M. Thomson, \emph{ {Holographic spectral
  functions and diffusion constants for fundamental matter}}, JHEP {\bf 11}
  (2007) 091,
\href{http://www.slac.stanford.edu/spires/find/hep/www?texkey=Myers:2007we}{ar%
Xiv:0706.0162}.

\bibitem{Mateos:2007yp}
D.~Mateos and L.~Patino, \emph{ {Bright branes for strongly coupled plasmas}},
  JHEP {\bf 11} (2007) 025,
\href{http://www.slac.stanford.edu/spires/find/hep/www?texkey=Mateos:2007yp}{a%
rXiv:0709.2168}.

\bibitem{Myers:2008cj}
R.~C. Myers and A.~Sinha, \emph{ {The fast life of holographic mesons}},
\href{http://www.slac.stanford.edu/spires/find/hep/www?texkey=Myers:2008cj}{ar%
Xiv:0804.2168}.

\bibitem{Mas:2008jz}
J.~Mas, J.~P. Shock, J.~Tarrio, and D.~Zoakos, \emph{ {Holographic Spectral
  Functions at Finite Baryon Density}},
\href{http://www.slac.stanford.edu/spires/find/hep/www?texkey=Mas:2008jz}{arXi%
v:0805.2601}.

\bibitem{Erdmenger:2007ja}
J.~Erdmenger, M.~Kaminski, and F.~Rust, \emph{ {Holographic vector mesons from
  spectral functions at finite baryon or isospin density}}, Phys. Rev. {\bf
  D77} (2008) 046005,
\href{http://www.slac.stanford.edu/spires/find/hep/www?texkey=Erdmenger:2007ja%
}{arXiv:0710.0334}.

\bibitem{Kruczenski:2003be}
M.~Kruczenski, D.~Mateos, R.~C. Myers, and D.~J. Winters, \emph{ {Meson
  spectroscopy in AdS/CFT with flavour}}, JHEP {\bf 07} (2003) 049,
\href{http://www.slac.stanford.edu/spires/find/hep/www?texkey=Kruczenski:2003b%
e}{arXiv:hep-th/0304032}.

\bibitem{Yamada:2006rx}
D.~Yamada and L.~G. Yaffe, \emph{ {Phase diagram of N = 4 super-Yang-Mills
  theory with R- symmetry chemical potentials}}, JHEP {\bf 09} (2006) 027,
\href{http://www.slac.stanford.edu/spires/find/hep/www?texkey=Yamada:2006rx}{a%
rXiv:hep-th/0602074}.

\bibitem{Splittorff:2000mm}
K.~Splittorff, D.~T. Son, and M.~A. Stephanov, \emph{ {QCD-like theories at
  finite baryon and isospin density}}, Phys. Rev. {\bf D64} (2001) 016003,
\href{http://www.slac.stanford.edu/spires/find/hep/www?texkey=Splittorff:2000m%
m}{arXiv:hep-ph/0012274}.

\bibitem{0034-4885-50-7-001}
K.~Binder, \emph{ Theory of first-order phase transitions}, Reports on Progress
  in Physics {\bf 50} (1987), no.~7, 783--859.

\bibitem{Aharony:2007uu}
O.~Aharony, K.~Peeters, J.~Sonnenschein, and M.~Zamaklar, \emph{ {Rho meson
  condensation at finite isospin chemical potential in a holographic model for
  QCD}}, JHEP {\bf 02} (2008) 071,
\href{http://www.slac.stanford.edu/spires/find/hep/www?texkey=Aharony:2007uu}{%
arXiv:0709.3948}.

\bibitem{Erdmenger:2007ap}
J.~Erdmenger, M.~Kaminski, and F.~Rust, \emph{ {Isospin diffusion in thermal
  AdS/CFT with flavor}}, Phys. Rev. {\bf D76} (2007) 046001,
\href{http://www.slac.stanford.edu/spires/find/hep/www?texkey=Erdmenger:2007ap%
}{arXiv:0704.1290}.

\bibitem{Apreda:2005yz}
R.~Apreda, J.~Erdmenger, N.~Evans, and Z.~Guralnik, \emph{ {Strong coupling
  effective Higgs potential and a first order thermal phase transition from
  AdS/CFT duality}}, Phys. Rev. {\bf D71} (2005) 126002,
\href{http://www.slac.stanford.edu/spires/find/hep/www?texkey=Apreda:2005yz}{a%
rXiv:hep-th/0504151}.

\bibitem{Karch:2007br}
A.~Karch and A.~O'Bannon, \emph{ {Holographic Thermodynamics at Finite Baryon
  Density: Some Exact Results}}, JHEP {\bf 11} (2007) 074,
\href{http://www.slac.stanford.edu/spires/find/hep/www?texkey=Karch:2007br}{ar%
Xiv:0709.0570}.

\bibitem{Karch:2005ms}
A.~Karch, A.~O'Bannon, and K.~Skenderis, \emph{ {Holographic renormalization of
  probe D-branes in AdS/CFT}}, JHEP {\bf 04} (2006) 015,
\href{http://www.slac.stanford.edu/spires/find/hep/www?texkey=Karch:2005ms}{ar%
Xiv:hep-th/0512125}.

\bibitem{Landau:1980sm}
L.~D. Landau and E.~M. Lifshitz, \emph{ { Course of Theoretical Physics, Volume
  5, Statistical Physics }},. Pergamon Press (1980) 544p.

\bibitem{spinGlass}
{A. H. Marshall}, {B. Chakraborty}, and {S. Nagel}, \emph{ Numerical studies of
  the compressible Ising spin glass}, Europhysics Letters {\bf 74} (may, 2006)
  699--705.

\bibitem{PhysRev.48.55}
E.~A. Uehling, \emph{ Polarization Effects in the Positron Theory}, Phys. Rev.
  {\bf 48} (Jul, 1935) 55--63.

\bibitem{Cvetic:1999ne}
M.~Cvetic and S.~S. Gubser, \emph{ {Phases of R-charged black holes, spinning
  branes and strongly coupled gauge theories}}, JHEP {\bf 04} (1999) 024,
\href{http://www.slac.stanford.edu/spires/find/hep/www?texkey=Cvetic:1999ne}{a%
rXiv:hep-th/9902195}.

\bibitem{Voskresensky:1997ub}
D.~N. Voskresensky, \emph{ {On the possibility of the condensation of the
  charged rho meson field in dense isospin asymmetric baryon matter}}, Phys.
  Lett. {\bf B392} (1997)
262--266.

\bibitem{Lenaghan:2001sd}
J.~T. Lenaghan, F.~Sannino, and K.~Splittorff, \emph{ {The superfluid and
  conformal phase transitions of two- color QCD}}, Phys. Rev. {\bf D65} (2002)
  054002,
\href{http://www.slac.stanford.edu/spires/find/hep/www?texkey=Lenaghan:2001sd}%
{arXiv:hep-ph/0107099}.

\bibitem{Sannino:2002wp}
F.~Sannino, \emph{ {General structure of relativistic vector condensation}},
  Phys. Rev. {\bf D67} (2003) 054006,
\href{http://www.slac.stanford.edu/spires/find/hep/www?texkey=Sannino:2002wp}{%
arXiv:hep-ph/0211367}.

\bibitem{Son:2000by}
D.~T. Son and M.~A. Stephanov, \emph{ {QCD at finite isospin density: From pion
  to quark antiquark condensation}}, Phys. Atom. Nucl. {\bf 64} (2001)
  834--842,
\href{http://www.slac.stanford.edu/spires/find/hep/www?texkey=Son:2000by}{arXi%
v:hep-ph/0011365}.

\bibitem{Son:2000xc}
D.~T. Son and M.~A. Stephanov, \emph{ {QCD at finite isospin density}}, Phys.
  Rev. Lett. {\bf 86} (2001) 592--595,
\href{http://www.slac.stanford.edu/spires/find/hep/www?texkey=Son:2000xc}{arXi%
v:hep-ph/0005225}.

\bibitem{Ebert:2008tp}
D.~Ebert, K.~G. Klimenko, A.~V. Tyukov, and V.~C. Zhukovsky, \emph{ {Pion
  condensation of quark matter in the static Einstein universe}},
\href{http://www.slac.stanford.edu/spires/find/hep/www?texkey=Ebert:2008tp}{ar%
Xiv:0804.0765}.

\bibitem{He:2005nk}
L.-y. He, M.~Jin, and P.-f. Zhuang, \emph{ {Pion superfluidity and meson
  properties at finite isospin density}}, Phys. Rev. {\bf D71} (2005) 116001,
\href{http://www.slac.stanford.edu/spires/find/hep/www?texkey=He:2005nk}{arXiv%
:hep-ph/0503272}.

\bibitem{Paredes:2008nf}
A.~Paredes, K.~Peeters, and M.~Zamaklar, \emph{ {Mesons versus quasi-normal
  modes: undercooling and overheating}}, JHEP {\bf 05} (2008) 027,
\href{http://www.slac.stanford.edu/spires/find/hep/www?texkey=Paredes:2008nf}{%
arXiv:0803.0759}.

\bibitem{Gubser:2008wv}
S.~S. Gubser and S.~S. Pufu, \emph{ {The gravity dual of a p-wave
  superconductor}},
\href{http://www.slac.stanford.edu/spires/find/hep/www?texkey=Gubser:2008wv}{a%
rXiv:0805.2960}.

\bibitem{Harnik:2003ke}
R.~Harnik, D.~T. Larson, and H.~Murayama, \emph{ {Supersymmetric color
  superconductivity}}, JHEP {\bf 03} (2004) 049,
\href{http://www.slac.stanford.edu/spires/find/hep/www?texkey=Harnik:2003ke}{a%
rXiv:hep-ph/0309224}.

\bibitem{Buchel:2006aa}
A.~Buchel, J.~Jia, and V.~A. Miransky, \emph{ {Dynamical stabilization of
  runaway potentials at finite density}}, Phys. Lett. {\bf B647} (2007)
  305--308,
\href{http://www.slac.stanford.edu/spires/find/hep/www?texkey=Buchel:2006aa}{a%
rXiv:hep-th/0609031}.

\bibitem{Erdmenger:2007cm}
J.~Erdmenger, N.~Evans, I.~Kirsch, and E.~Threlfall, \emph{ {Mesons in
  Gauge/Gravity Duals - A Review}}, Eur. Phys. J. {\bf A35} (2008) 81--133,
\href{http://www.slac.stanford.edu/spires/find/hep/www?texkey=Erdmenger:2007cm%
}{arXiv:0711.4467}.

\bibitem{Karch:2008fa}
A.~Karch, D.~T. Son, and A.~O. Starinets, \emph{ {Zero Sound from Holography}},
\href{http://www.slac.stanford.edu/spires/find/hep/www?texkey=Karch:2008fa}{ar%
Xiv:0806.3796}.

\bibitem{Gubser:1998jb}
S.~S. Gubser, \emph{ {Thermodynamics of spinning D3-branes}}, Nucl. Phys. {\bf
  B551} (1999) 667--684,
\href{http://www.slac.stanford.edu/spires/find/hep/www?texkey=Gubser:1998jb}{a%
rXiv:hep-th/9810225}.

\bibitem{Yamada:2007gb}
D.~Yamada, \emph{ {Metastability of R-charged black holes}}, Class. Quant.
  Grav. {\bf 24} (2007) 3347--3376,
\href{http://www.slac.stanford.edu/spires/find/hep/www?texkey=Yamada:2007gb}{a%
rXiv:hep-th/0701254}.

\end{thebibliography}

\end{document}